\documentclass[twocolumns]{aa} 

\usepackage{graphicx}
\usepackage{epstopdf}
\usepackage{deluxetable}
\usepackage{times,epsfig} 
\usepackage{mathrsfs}  
\usepackage{natbib}
\usepackage{amsmath}
\usepackage{amssymb}
\bibpunct{(}{)}{;}{a}{}{,} 
\usepackage{caption}

\usepackage[english]{babel}
\usepackage{longtable}
\usepackage{url}
\usepackage{hyperref}

\authorrunning{Taddia et al.}
\titlerunning{PTF and CCCP long-rising SNe~II}
\begin{document}

\title{Long-rising Type II supernovae from PTF and CCCP}

\author{F. Taddia\inst{1}
\and J. Sollerman\inst{1}
\and C. Fremling\inst{1}
\and K. Migotto\inst{1}
\and A. Gal-Yam\inst{2}
\and S. Armen\inst{3}
\and G. Duggan\inst{4}
\and M. Ergon\inst{1}
\and A.~V. Filippenko\inst{5}
\and C. Fransson\inst{1}
\and G. Hosseinzadeh\inst{6,7}
\and M.~M. Kasliwal\inst{8}
\and R.~R. Laher\inst{9}
\and G. Leloudas\inst{2,10}
\and D.~C. Leonard\inst{3}
\and R. Lunnan\inst{4}
\and F.~J. Masci\inst{11}
\and D.-S. Moon\inst{12}
\and J.~M. Silverman\inst{13}
\and P.~R. Wozniak\inst{14}}

\institute{The Oskar Klein Centre, Department of Astronomy, Stockholm University, AlbaNova, 10691 Stockholm, Sweden.\\ (\email{francesco.taddia@astro.su.se})
\and Department of Particle Physics \& Astrophysics, Weizmann Institute of Science, Rehovot 76100, Israel.
\and Department of Astronomy, San Diego State University, San Diego, CA
92182-1221, USA.
 \and Astronomy Department, California Institute of Technology, Pasadena, California 91125, USA.
\and Department of Astronomy, University of California, Berkeley, CA 94720-3411, USA.
 \and Las Cumbres Observatory Global Telescope, 6740 Cortona Dr, Suite 102, Goleta, CA 93117, USA.
 \and Department of Physics, University of California, Santa Barbara, CA 93106-9530, USA.
 \and The Observatories, Carnegie Institution for Science, 813 Santa Barbara Street, Pasadena, CA 91101, USA. 
 \and Spitzer Science Center, California Institute of Technology, M/S 314-6, Pasadena, CA 91125, USA.
\and Dark Cosmology Centre, Niels Bohr Institute, University of Copenhagen, Juliane Maries Vej 30, 2100 Copenhagen, Denmark.
\and Infrared Processing and Analysis Center, California Institute of Technology, MS 100-22, Pasadena, CA 91125, USA.
\and Department of Astronomy and Astrophysics, University of Toronto, Toronto, ON M5S 3H4, Canada.
\and Department of Astronomy, University of Texas, Austin, TX 78712, USA.
\and Los Alamos National Laboratory, MS D436, Los Alamos, NM 87545, USA.}

\date{Received  XX / Accepted XX}

\abstract
{Supernova (SN) 1987A was a peculiar hydrogen-rich event with a long-rising ($\sim84$~d) light curve, stemming from the explosion of a compact blue supergiant star. Only a few similar events have been presented in the literature in recent decades.}
{We present new data for a sample of six long-rising Type II SNe (SNe~II), three of which were discovered and observed by the Palomar Transient Factory (PTF) and three observed by the Caltech Core-Collapse Project (CCCP). Our aim is to enlarge this small family of long-rising SNe~II, characterizing their differences in terms of progenitor and explosion parameters. We also study the metallicity of their environments.}
{Optical light curves, spectra, and host-galaxy properties of these SNe are presented and analyzed. Detailed comparisons with known SN~1987A-like events in the literature are shown, with particular emphasis on the absolute magnitudes, colors, expansion velocities, and host-galaxy metallicities. Bolometric properties are derived from the multiband light curves. By modeling the early-time emission with scaling relations derived from the SuperNova Explosion Code (SNEC) models of MESA progenitor stars, we estimate the progenitor radii of these transients. The modeling of the bolometric light curves also allows us to estimate other progenitor and explosion parameters, such as the ejected $^{56}$Ni mass, the explosion energy, and the ejecta mass.}
{We present PTF12kso, a long-rising SN~II that is estimated to have the largest amount of ejected $^{56}$Ni mass measured for this class. PTF09gpn and PTF12kso are found at the lowest host metallicities observed for this SN group. The variety of early light-curve luminosities depends on the wide range of progenitor radii of these SNe, from a few tens of R$_{\odot}$ (SN~2005ci) up to thousands (SN~2004ek) with some intermediate cases between 100~R$_{\odot}$ (PTF09gpn) and 300~R$_{\odot}$ (SN~2004em).}
{We confirm that long-rising SNe~II with light-curve shapes closely resembling that of SN~1987A 
generally arise from blue supergiant (BSG) stars. However, some of them, such as SN~2004em, likely have progenitors with larger radii ($\sim300$~R$_{\odot}$, typical of yellow supergiants) and can thus be regarded as intermediate cases between normal SNe~IIP and SN~1987A-like SNe. Some extended red supergiant (RSG) stars such as the progenitor of SN~2004ek can also produce long-rising SNe~II if they synthesized a large amount of $^{56}$Ni in the explosion. Low host metallicity is confirmed as a characteristic of the SNe arising from compact BSG stars.}


\maketitle
\section{Introduction}
\label{sec:intro}
Supernova (SN) 1987A marked the explosion of a blue supergiant (BSG) star (Sanduleak $-69^\circ$202, \citealp{gilmozzi87}), and it showed a peculiar long-rising light curve (84 days from explosion to peak) and a hydrogen-rich (Type II SN) spectrum. Given its extraordinary proximity [merely 50~kpc, in the Large Magellanic Cloud (LMC)], SN~1987A is to date the best-studied SN. 
It changed our understanding of SN explosions (see the reviews by \citealp{arnett89} and \citealp{mccray93}), and it continues to trigger scientific activity; it is
the only SN that can be spatially resolved in the optical and was the first one whose transition to a young SN remnant could be observed \citep{larsson11}.

In the last decade, a handful of SNe resembling SN~1987A have been presented in the literature. Most of these SN~1987A-like SNe were collected by \citet{pastorello12}. 
The best-observed objects are SNe~1998A \citep{pastorello05}, 2000cb \citep{kleiser11}, 2006V, 2006au \citep{taddia12}, and 2009E \citep{pastorello12}. 
Well-sampled optical light curves of the long-rising ($>40$~d) SNe~II 2004ek, 2004em, and 2005ci by the Caltech Core-Collapse Project (CCCP; \citealp{galyam07_CCCP}) were shown by \citet{arcavi12}, and here we present their full datasets, including unpublished spectroscopy. \citet{gonzales15} also discussed slow risers in their sample of SNe~II. Some of their SNe lack 
spectral classifications and might be of Type~IIn;
however, at least one spectroscopically confirmed SN~1987A-like event is presented (SNLS-07D2an), with a rise time of $>60$~d. 
The first SN~1987A-like event was likely SN~1909A \citep{young88}, but for this object there is no spectral classification.

The analysis of the light curves and spectra of the SNe discussed by \citet{pastorello12} led to the conclusion
that they arise from the explosions of compact ($R<100$~R$_{\odot}$) BSG stars, with initial masses of about $M_{\rm ZAMS} = 20$~M$_{\odot}$, explosion energies of a few $10^{51}$~erg, and $^{56}$Ni masses of $\sim10^{-1}$~M$_{\odot}$. 

As BSG stars are not expected to explode in the standard stellar evolution models, following the explosion of Sanduleak $-69^\circ$202 there were many attempts to explain this surprising fact (see \citealp{pod92} for a review). Models showed that a low metallicity (like that measured in the LMC) and/or fast rotation could make a single $M_{\rm ZAMS}=20$~M$_{\odot}$ star explode in the blue part of the Hertzsprung-Russell (HR) diagram. 
We \citep{taddia13} presented a detailed study of the metallicity at the locations of the known SN~1987A-like events, finding that these SNe indeed tend to occur at lower (LMC-like) metallicity than normal SNe~IIP. Alternatively, a binary scenario may also explain BSG explosions, and in the specific case of SN~1987A \citep{pod92} it may also explain the complex shape of the circumstellar medium (CSM).

Given the rarity of long-rising SNe~II, it is important to extend the number of well-observed objects, in order to study the variety of their properties.
The observations from the Palomar Transient Factory (PTF; \citealp{rahmer08,rau09}) and from the CCCP that we present here are 
therefore significant
additions to this SN subgroup.
 
The PTF discoveries are particularly important as they are not biased toward bright (i.e., metal-rich) host galaxies \citep{arcavi10}. Our small PTF sample of SN~1987A-like events allowed us to find these SNe at unprecedented low metallicity, with important implications for the progenitor scenario. We notice that other rare and peculiar core-collapse (CC) SNe have been observed in low or very low metallicity environments (e.g., the Type Ic-BL PTF12gzk; \citealp{benami12}). Other recent examples are the class of hydrogen-poor superluminous SNe (SLSNe; \citealp{quimby11,galyam12,lunnan14,leloudas15}), and that of the SN impostors \citep{taddia15}.

The PTF survey and its continuation (the intermediate PTF, iPTF; \citealp{kulkarni13}) not only extends the domain of SN discovery to low-luminosity hosts, but thanks to the high cadence allows regular discoveries of very young transients (e.g.,  iPTF13ast; \citealp{galyam14}). Indeed, some of our PTF (but also CCCP) objects were likely discovered and observed soon after explosion. Therefore, some of them show possible signatures of the shock-breakout cooling tail in their optical light curves \citep{chevalier92, chevalier08, rabinak11, piro13}. This feature is rarely observed and contains important information to understand the nature of the SN progenitor. First, it allows an estimate of the explosion epoch with small uncertainty, which is crucial for deriving the properties of the progenitor from light-curve and spectral modeling. In addition, especially for objects that are radioactively powered at peak (like our long-rising SNe~II), the modeling of the early-time light curve can constrain the radius of the SN progenitor.  

 \begin{figure}
 \centering
\includegraphics[width=9cm,angle=0]{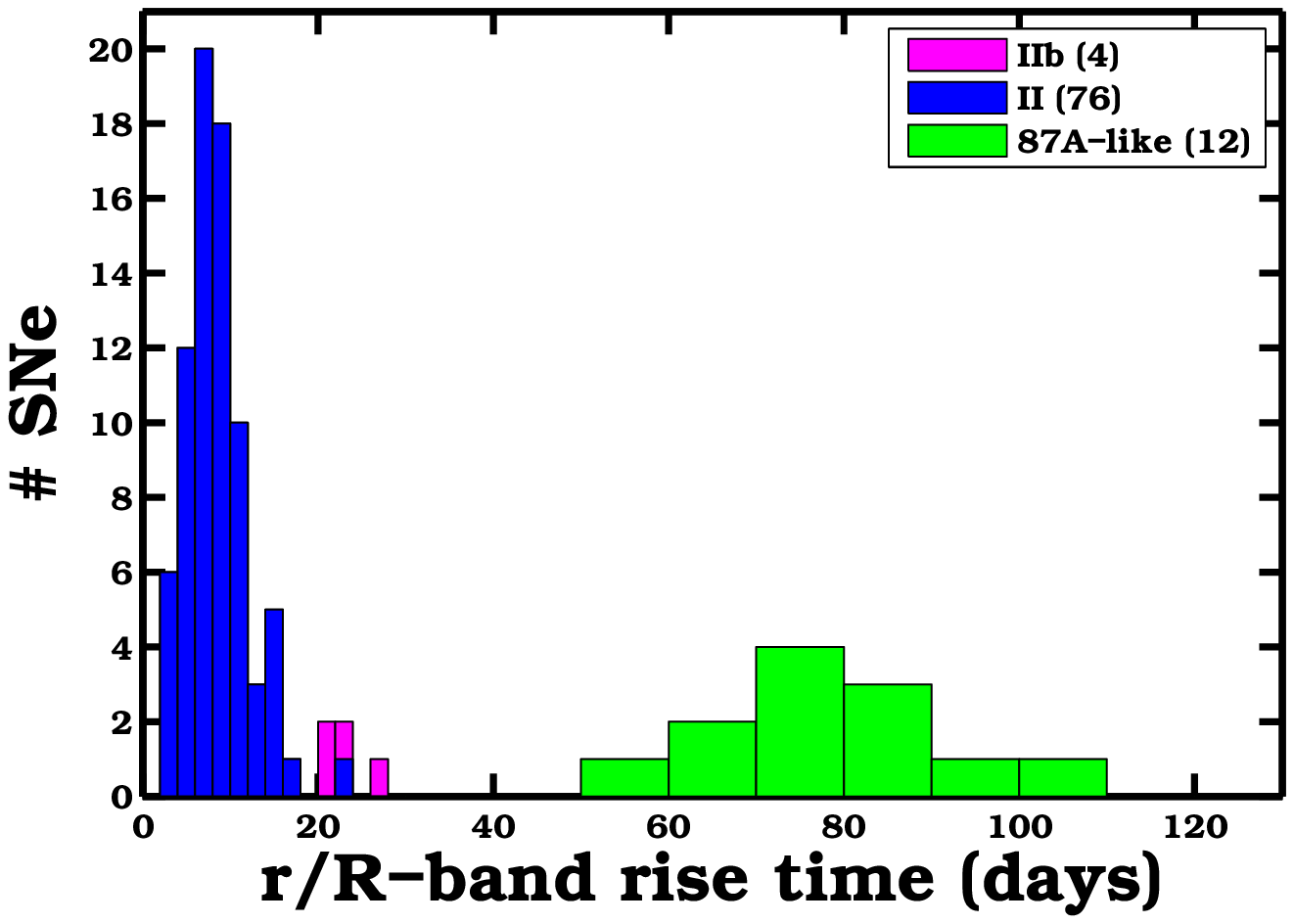}
\includegraphics[width=9cm,angle=0]{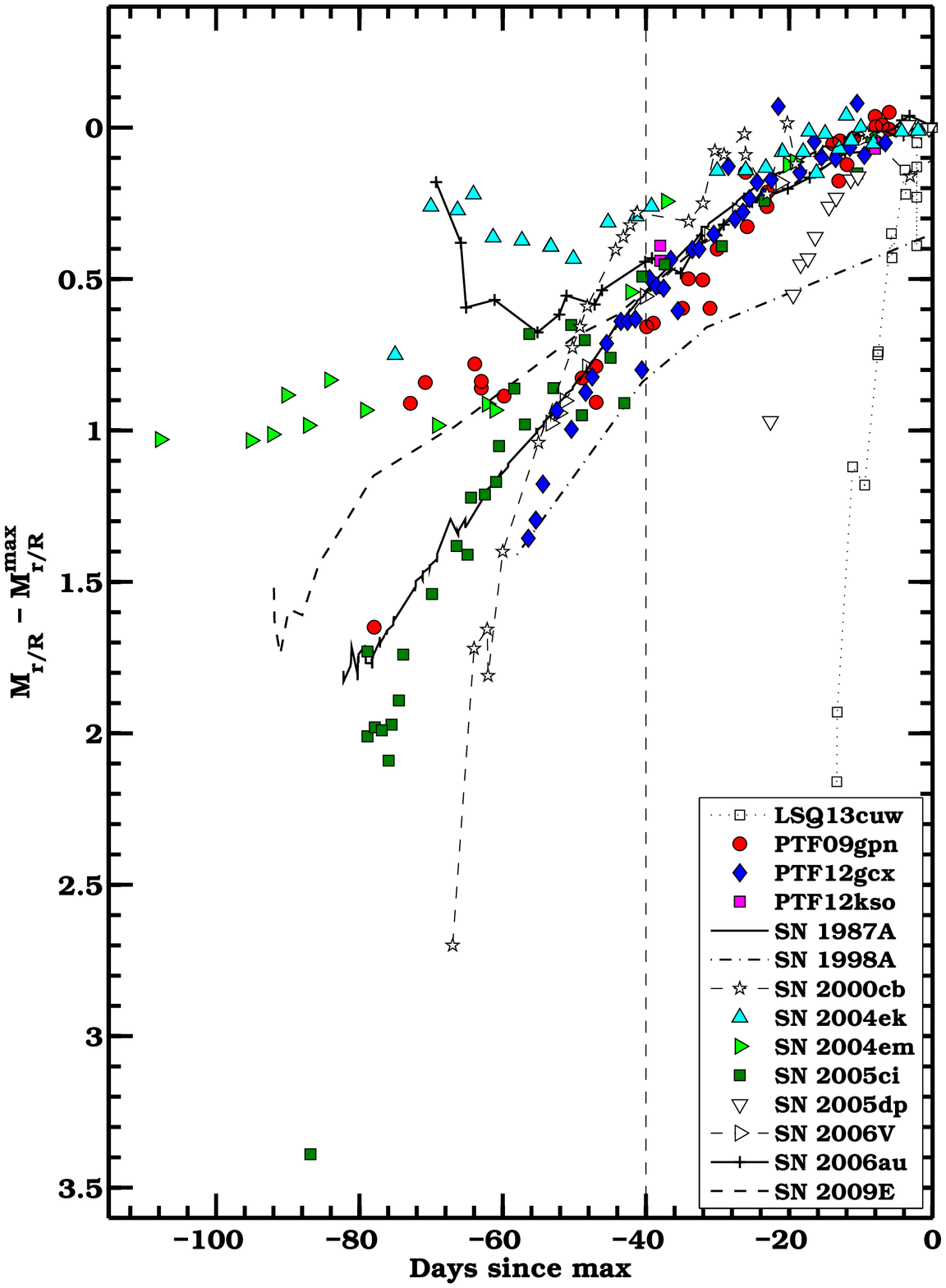}
  \caption{\label{rise} Top panel: distribution of $r/R$-band rise times 
  for SNe~IIP, IIL, IIb, and SN~1987A-like SNe from the literature \citep{gall15,taddia15} and from this work. (See Filippenko 1997 for a review of SN classification.) We define as long-rising SNe~II or SN~1987A-like events those with rise time $>40$~d. These SNe are powered mainly by radioactive decay at peak, whereas normal SN~II peaks are dominated by shock energy. Bottom panel: $r/R$-band light curves of slow-rising SNe~II (from this work, when colored, and from the literature), aligned to the peak epoch and scaled to match their peaks. All of them show rise times $>40$~d, but display a variety of shapes before peak. We also show SN~2005dp \citep{arcavi12} and LSQ13cuw \citep{gall15}, which have relatively long rise times compared to normal SNe~IIP, but still $<40$~d.}
 \end{figure} 
 
Here we present new observations of six long-rising SNe~II, which appear rather similar to SN~1987A. 
The paper is structured as follows. In Sec.~\ref{sec:sample} we describe the SN sample, and in Sec.~\ref{sec:data} we discuss data acquisition and reduction. Section~\ref{sec:host} includes the analysis of the host galaxies. We describe and analyze SN photometry and spectroscopy in Sec.~\ref{sec:phot} and Sec.~\ref{sec:spectra}, respectively. Bolometric properties are derived in Sec.~\ref{sec:boloprop}. Data are modeled in Sec.~\ref{sec:models}, in order to derive the progenitor parameters. We discuss the results in Sec.~\ref{sec:discussion} and give our conclusions in Sec.~\ref{sec:conclusion}.

\section{The supernova sample}
\label{sec:sample}

We define a SN~II as ``long-rising" based on the epoch of the (second) peak of its $r/R$-band light curve. In Fig.~\ref{rise} (top panel) we show the distribution of $r/R$-band rise times of SNe~II, with data for normal SNe~II (IIP and IIL) and SNe~IIb from \citet{gall15}, \citet{rubin15}, and \citet{taddia15}, as well as data for long-rising SNe~II from the light curves discussed in Sec.~\ref{sec:phot} and displayed in the bottom panel of Fig.~\ref{rise}. SNe showing rise times $>40$~d are considered ``long risers" in this work. Interestingly, among SNe~II (excluding SNe~IIn, which are powered by CSM interaction), there is a lack of SNe rising in 30--50~d. This is likely due to the fact that the explosion energy and the ejecta mass (up to $\sim25$~M$_{\odot}$) of SNe~II appear to be linearly correlated \citep{utrobin10}, and their ratios imply long rise times for compact progenitor stars (see Eq.~\ref{eq:trise}). 

SNe~II with rise times of $\sim20$--25~d are presented by \citet{rubin15} (e.g., iPTF14bas), and in that work these are named ``extremely slow risers.'' However, we include in our definition of slow risers only those SNe with rise times $>40$~d, to be sure to include only objects that are mainly powered by radioactive decay at peak, like SN~1987A. In SNe~IIP and IIL, a rise of only a few weeks in the optical could instead be due to a temperature effect, with most of the flux in the ultraviolet (UV) part of the spectrum until $\sim20$--25~d. This is why we exclude from our sample another CCCP SN~II with relatively slow rise (24~d), SN~2005dp (\citealp{arcavi12}). The rising part of the $R$-band light curve of this transient is shown in the bottom panel of Fig.~\ref{rise}, and it is clearly distinct from the family of slow risers resembling SN~1987A. For SNe~IIb, the peak is powered by radioactive decay as in SN~1987A, but their shorter rise times are due to the lower ejecta mass as compared to SN~1987A-like SNe. SNe~IIb have lost their H envelopes almost entirely prior to exploding.

As shown in the bottom panel of Fig.~\ref{rise}, the slow risers can exhibit a variety of light-curve shapes at early epochs. Their rise can be only a few tenths of a magnitude from the early minimum to the second peak (e.g., PTF09gpn), or it can be several magnitudes (e.g., SN~1987A). We include both of these types of objects in the category of long-rising SNe~II or SN~1987A-like events. Their common feature is the late-time peak, which we will show also appears in their bolometric light curves; it is therefore likely to be powered mainly by the decay chain of $^{56}$Ni. We will propose that the variety of the early-time light curves is a result of the explosion of progenitors with different radii (see Sec.~\ref{sec:hydro} and the results by \citealp{young04}).

Here we present data for six SNe~II with long-rising light curves.
Three of them were discovered and followed by PTF (PTF09gpn, PTF12gcx, PTF12kso), and three were observed by CCCP (SNe 2004ek, 2004em, 2005ci). For the PTF objects we present unpublished light curves and spectra. For the CCCP objects we present unpublished spectra and make use of the light curves already included in \citet{arcavi12}.
In Table~\ref{tab:sample} the basic information on the targets is given, including SN coordinates, host-galaxy name and type, redshift, distance, and extinction. The redshifts are obtained from NED\footnote{The NASA/IPAC Extragalactic Database, \href{http://ned.ipac.caltech.edu}{http://ned.ipac.caltech.edu} } for the CCCP SNe and for PTF12gcx, and from the host-galaxy emission lines in the case of the other two PTF objects. We adopted redshift-independent distances (from NED) when available (i.e., in the case of the CCCP targets), whereas we used the known redshift and the WMAP5 cosmological parameters (H$_{0} = 70.5$~km~s$^{-1}$~Mpc$^{-1}$, $\Omega_{M}=0.27$, $\Omega_{\Lambda}=0.73$; \citealp{komatsu09}) to compute the distances for the hosts without redshift-independent estimates (i.e., for the PTF SNe).
The Galactic extinction is obtained from NED \citep{schlafly11}, the host-galaxy extinction (non-negligible only in the case of SN~2004ek) is obtained from the \ion{Na}{i}~D equivalent width (EW) via the relation $E(B-V)=0.16$\,EW(\ion{Na}{i}~D) \citep{taubenberger06} and assuming $R_V=3.1$. Here we neglect the correction for time dilation as these SNe are all in low-redshift galaxies.

\begin{figure}[b]
 \centering
\includegraphics[width=9cm,angle=0]{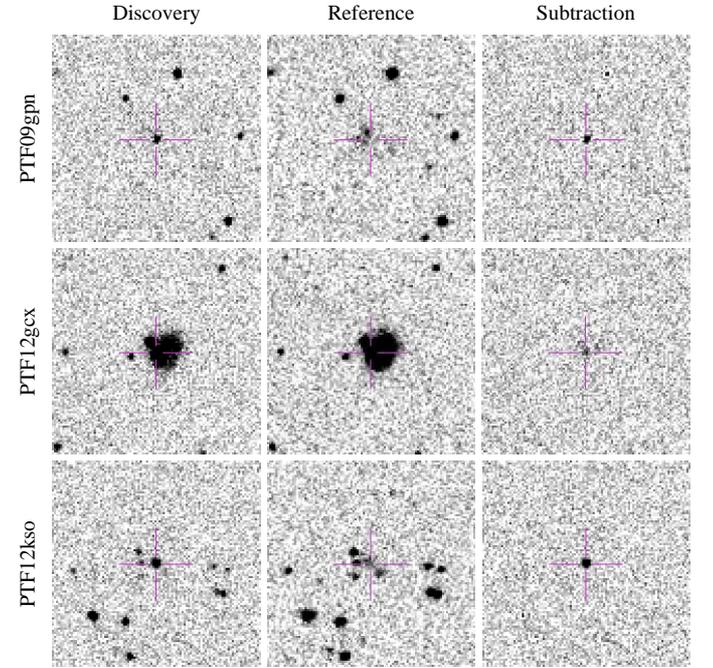}
  \caption{Discovery, reference, and subtracted P48 $r$-band images for the three PTF SNe. North is up and east is to the left. The field of view is 1.67\arcmin $\times$ 1.67\arcmin. \label{discoimage}}
 \end{figure} 

\begin{figure*}
 \centering
 $\begin{array}{cc}
\includegraphics[width=4cm,angle=0]{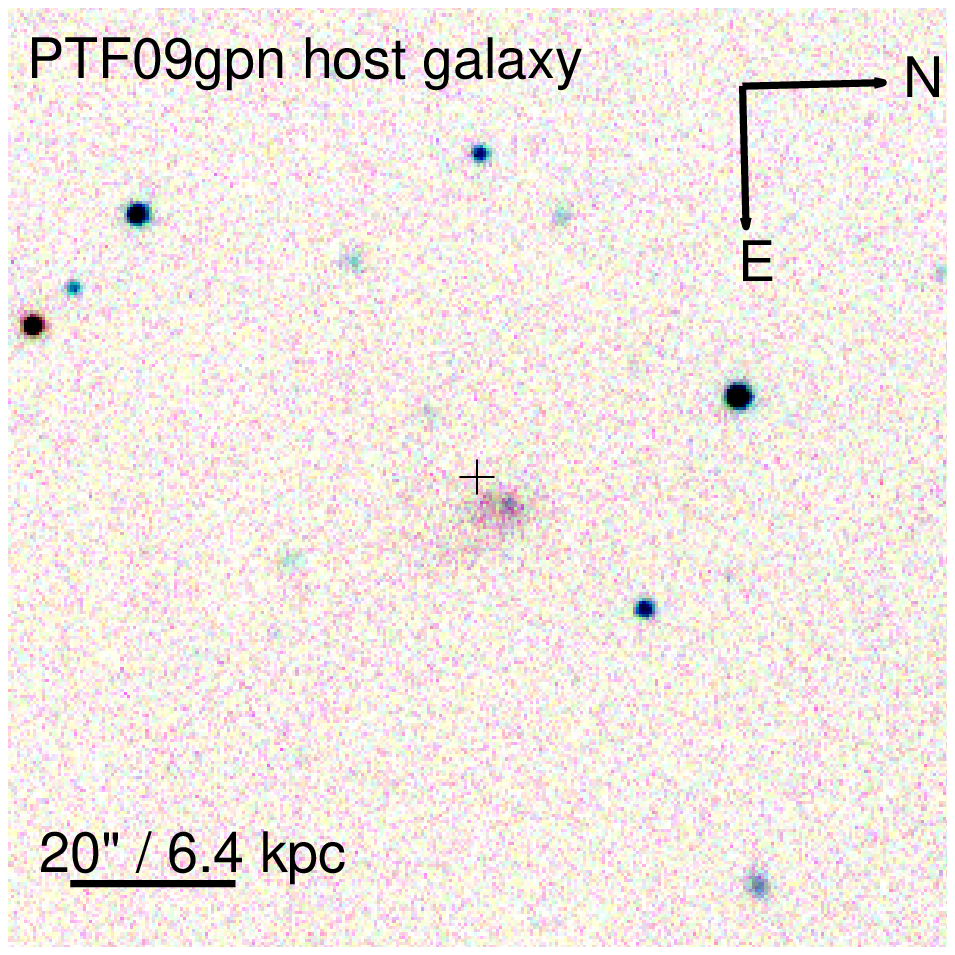}&
\includegraphics[width=7.17cm,angle=0]{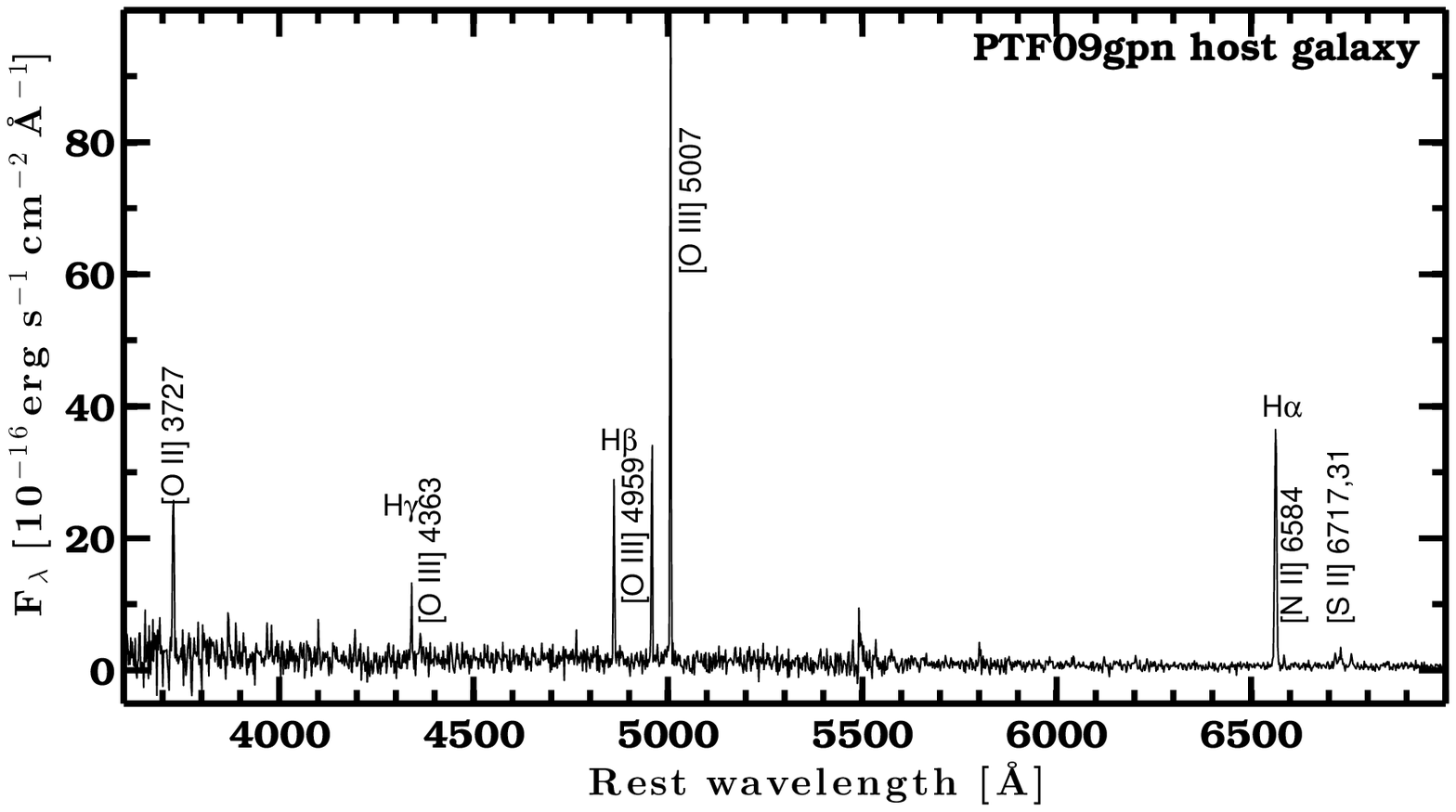}\\
\includegraphics[width=4cm,angle=0]{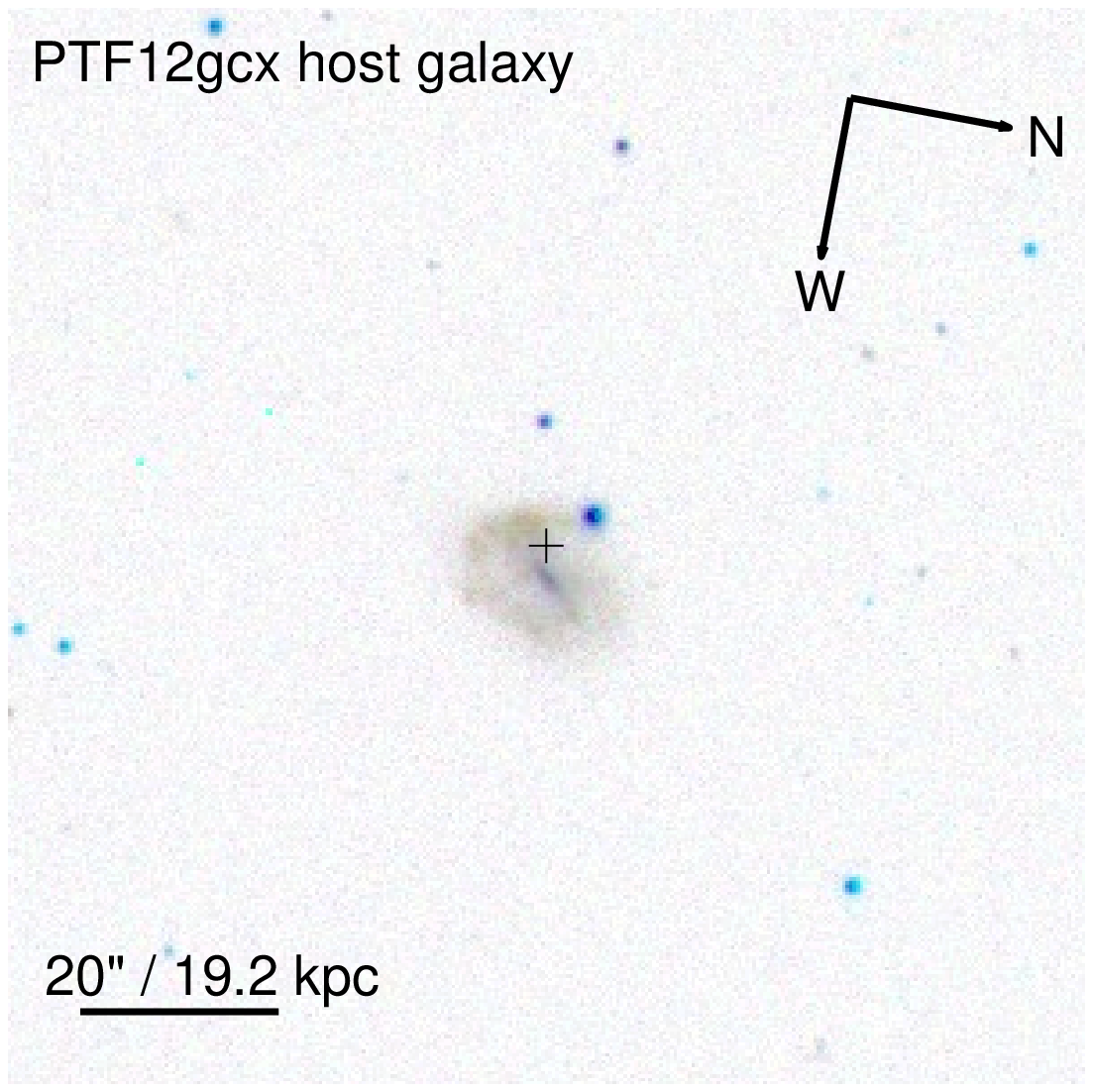}&
\includegraphics[width=7.17cm,angle=0]{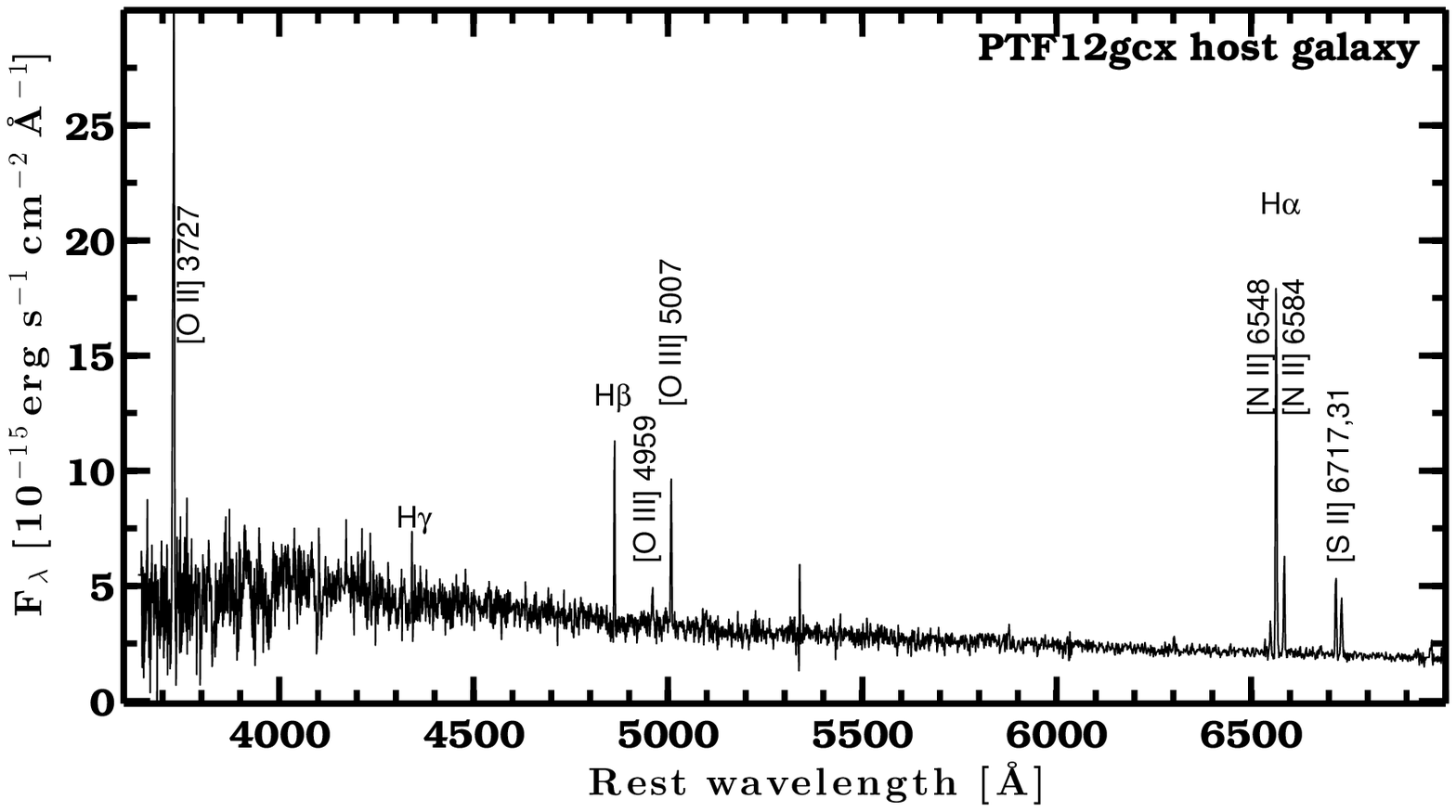}\\
\includegraphics[width=4cm,angle=0]{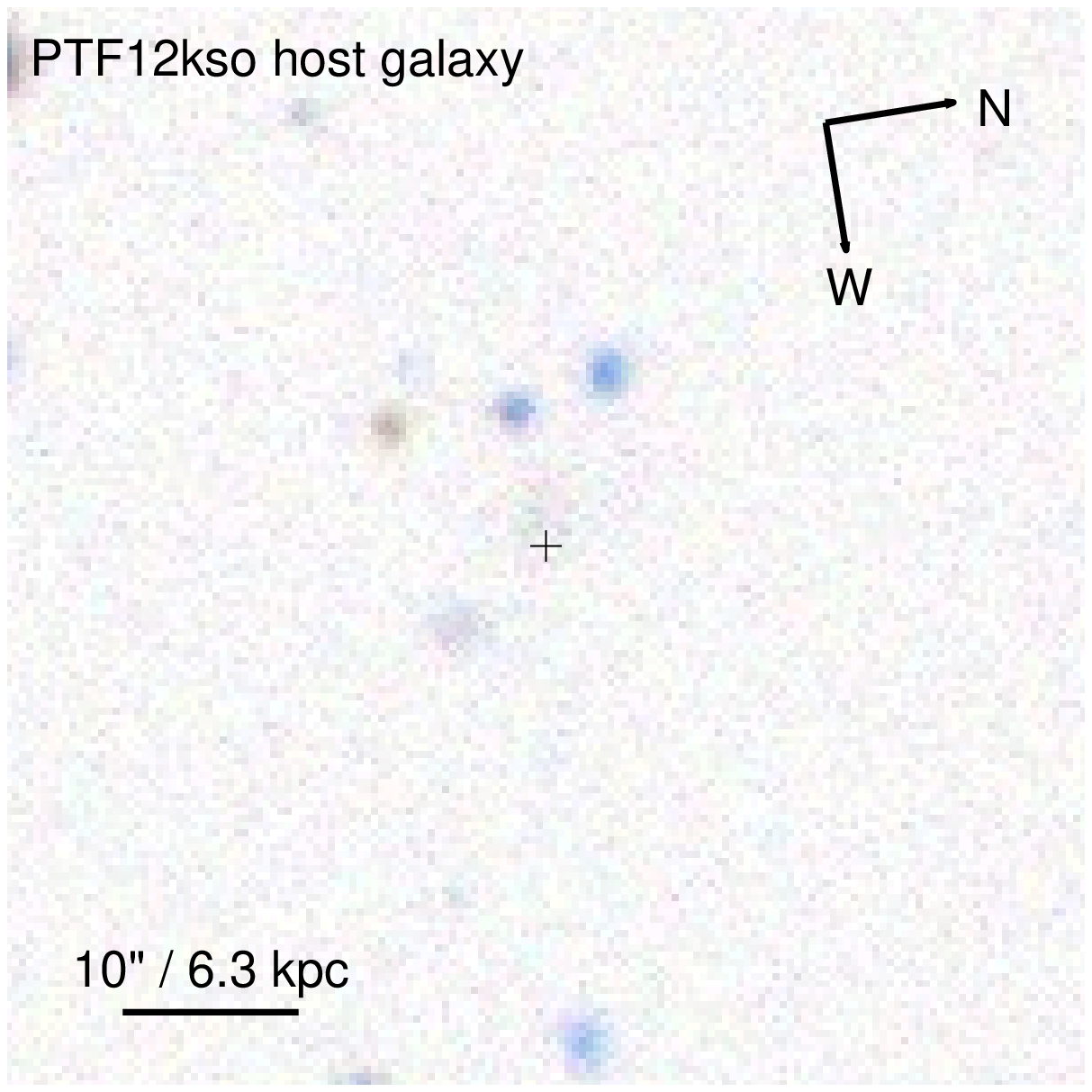}&
\includegraphics[width=7.17cm,angle=0]{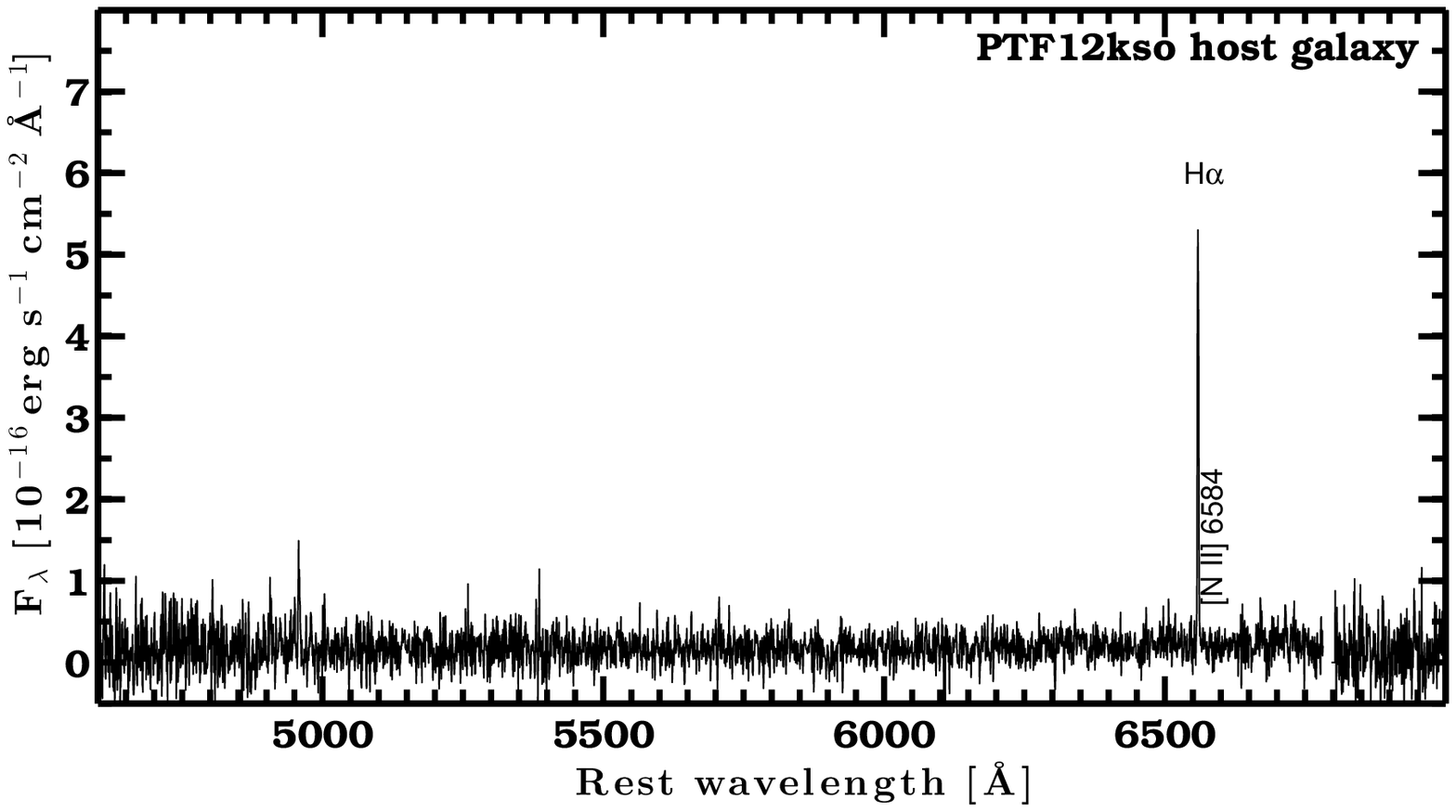}\\
\end{array}$
  \caption{\label{galaimage} Color images and spectra of the three host galaxies of our long-rising SNe~II from PTF. The hosts of PTF09gpn and PTF12kso are particularly faint and metal poor. In the color images the position of the SN is marked by a black ``+", and the scale as well as the orientation are reported. The color image of the host of PTF09gpn was obtained by combining three images in $gri$ filters that we obtained with P60. The images of the other hosts are from SDSS. In the host-galaxy spectra, shown in the rest frame, we identified the most prominent emission lines. Some of them were used to estimate the metallicity.}
 \end{figure*} 

\section{Data acquisition and reduction}
\label{sec:data}

\subsection{Discovery and explosion epochs}
In Table~\ref{tab:disco} we report the last nondetection date and limiting magnitude, the discovery date and magnitude (along with their references), and the assumed explosion epoch for each SN.
 It is important to constrain the explosion date for each SN, in order to compare the different SNe at similar phases and to derive the progenitor parameters. 
The last nondetection and the discovery epochs can be used to constrain the explosion epoch, if they occur close in time and the nondetection magnitude limit is sufficiently deep. Alternatively, a power-law (PL) fit to the early-time light curves 
can be used \citep[e.g.,][]{cao13}. 

For SNe 2004em, 2005ci, PTF09gpn, and PTF12gcx, we used a PL fit to the early light curves to estimate the explosion epoch. For SN~2004em we fit the first six epochs of the bolometric light curve (see Sec.~\ref{sec:boloprop}). For SN~2005ci, PTF09gpn, and PTF12gcx, we fit the first 11, 3, and 7 epochs of the $r/R$-band light curves (see Sec.~\ref{sec:phot}), respectively. The uncertainty in the explosion date is determined by the fit uncertainty and/or by the difference between the discovery and the last nondetection epochs.

For SN~2004ek we derive the explosion date from the average between discovery and the last nondetection, and the uncertainty from the interval between these two epochs. We adopt the same method for PTF12kso as we do not have observations of this SN at early times that would allow us to fit a PL. 

In the rest of the paper the phase of each object will be defined in days since explosion.

\subsection{CCCP supernova data}

The CCCP $BVRI$ SN light curves were obtained from \citet{arcavi12} via Wiserep \citep{yaron12}. Additional data for SNe~2004ek and 2005ci were obtained from \citet[][unfiltered]{kleiser11} and \citet[][$BVRI$]{tse08}, respectively. The $BVRI$ photometry from \citet{tse08} was shifted by $+$0.1, $+$0.2, $+$0.2, and $+$0.1 mag (respectively), in order to match the CCCP photometry of SN~2004ek. We report all the magnitudes in Table~\ref{tab:photCCCP}.

Five spectra of each of SNe~2004ek and 2004em were obtained with the Double Spectrograph (DBSP; \citealp{oke82}) mounted on the 200-inch Hale telescope at Palomar Observatory (P200), the Low Resolution Imaging Spectrometer (LRIS; \citealp{oke95}) mounted on the Keck~1 10-m telescope, and the Kast spectrograph (Miller \& Stone 1993) on the 3-m Shane reflector at Lick Observatory. Three spectra of SN~2005ci were obtained with P200+DBSP.
For a detailed log of the CCCP spectral observations, see Table~\ref{tab:spectra}. The spectra were reduced in a standard manner, including (i) bias and flatfield corrections, (ii) one-dimensional (1D) spectral extraction from the 2D frames after tracing and choosing proper apertures and background regions, (iii) wavelength calibration with spectra of calibration lamps, and (iv) flux calibration with spectrophotometric standard stars observed during the same night. 

\subsection{PTF supernova data}

In Fig.~\ref{discoimage} we show the discovery images, along with reference and subtracted images, of the three PTF SNe. For these SNe 
$r$-band photometry was obtained with the 48-inch Samuel Oschin Telescope at Palomar Observatory (P48), equipped with the 96~Mpixel mosaic camera CFH12K \citep{rahmer08} and a Mould $R$ filter \citep{ofek12}. Additional $g$-band photometry of PTF12kso was obtained with the same telescope and instrument.
PTF09gpn, PTF12gcx, and PTF12kso were also imaged in the $Bgri$ bands with the automated Palomar 60-inch telescope (P60; \citealp{cenko06}). PTF12gcx was not detected in the $B$ band, whereas for PTF09gpn $z$-band photometry was also obtained. For PTF12kso, a large number of photometric epochs  with the 1-m telescopes of the Las Cumbres Observatory Global Telescope Network (LCOGT; \citealp{brown13}) in the $griz$ bands were also obtained. For the same SN additional $BVRI$ photometry was obtained at Mount Laguna Observatory (MLO; \citealp{smith69}) using the telescope, camera, and reduction technique described by \citet{smith15}.

 Point-spread-function (PSF) photometry was obtained on template-subtracted P48 and P60 images using the Palomar Transient Factory Image Differencing and Extraction (PTFIDE) pipeline\footnote{\href{http://spider.ipac.caltech.edu/staff/fmasci/home/miscscience/ptfide-v4.5.pdf}{http://spider.ipac.caltech.edu/staff/fmasci/home/miscscience/ptfide-v4.5.pdf}}$^{,}$\footnote{\href{ http://web.ipac.caltech.edu/staff/fmasci/home/miscscience/forcedphot.pdf}{ http://web.ipac.caltech.edu/staff/fmasci/home/miscscience/forcedphot.pdf}} and the P60 pipeline presented by \citet{fremling16}. The photometry was calibrated against Sloan Digital Sky Survey (SDSS; \citealp{ahn14}) stars in the SN field (or observed the same night in other fields, in the case of PTF09gpn). The magnitudes are reported in Table~\ref{tab:photPTF}. For the LCOGT images of PTF12kso, we estimated the galaxy contribution by fitting a low-order surface and then performed PSF photometry. 

A single optical spectrum of PTF09gpn was obtained using the P200+DBSP. Keck-1+LRIS was used to acquire an optical spectrum of the host galaxy. 
 For PTF12gcx, three spectra were obtained with three different facilities: Keck~2$+$DEep Imaging Multi-Object Spectrograph (DEIMOS; \citealp{faber03}), P200+DBSP, and Lick 3-m+Kast. Four spectra of PTF12kso were obtained with P200+DBSP and Keck~2+DEIMOS, and one with the Copernico 1.82m+AFOSC (obtained from the Asiago Supernova Classification program\footnote{\href{http://graspa.oapd.inaf.it/asiago_class.html}{http://graspa.oapd.inaf.it/asiago\_class.html}}).
  Spectra of the three host galaxies were obtained with P200+DBSP (PTF09gpn), from the SDSS archive (PTF12gcx), and with Keck~2+DEIMOS (PTF12kso).
 The spectra were reduced in a standard manner as outlined for the CCCP spectra. Table~\ref{tab:spectra} reports a log of the PTF spectral observations.

\begin{figure*}
 \centering
 $\begin{array}{ccc}
\includegraphics[width=6cm,angle=0]{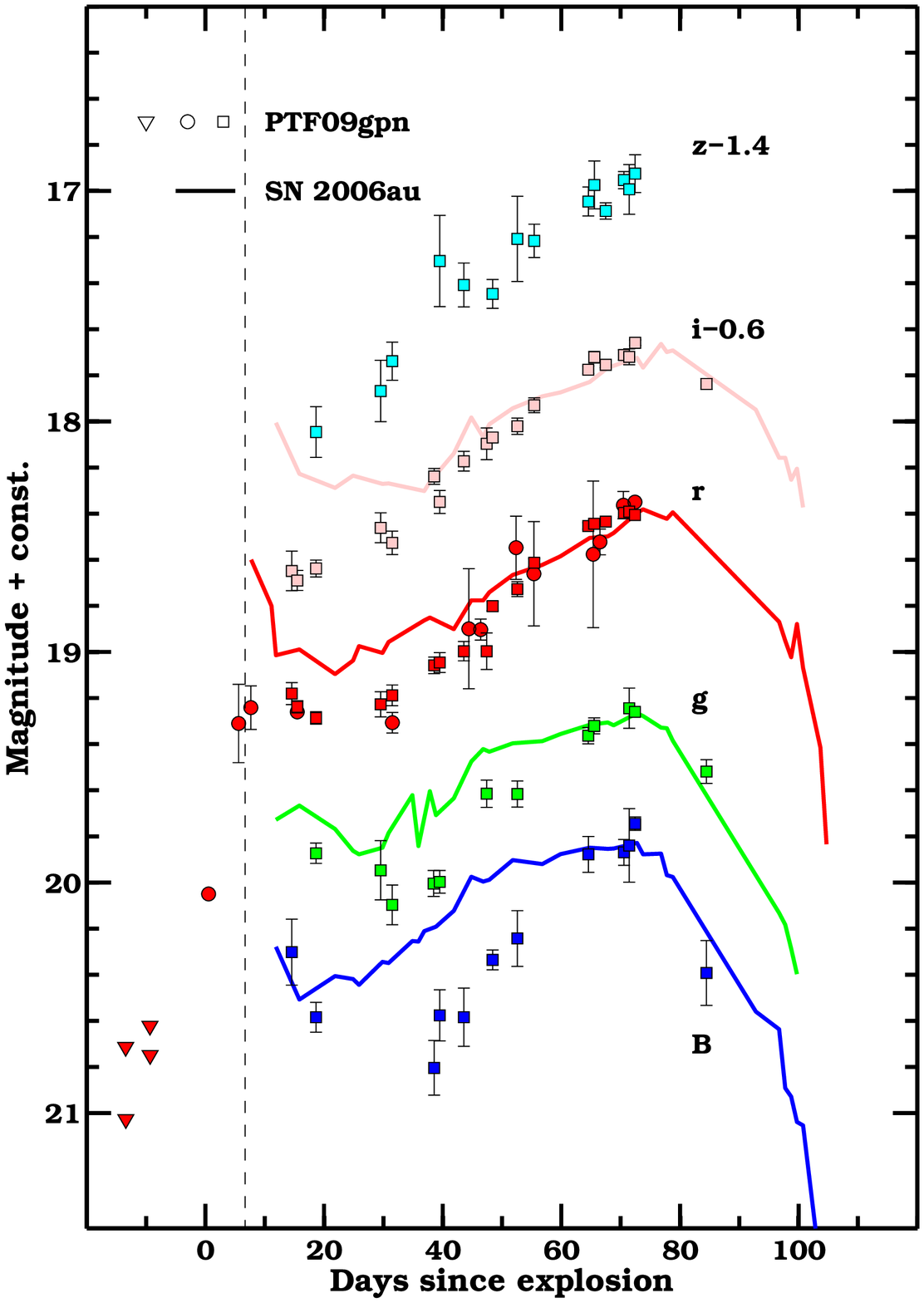}&
\includegraphics[width=6cm,angle=0]{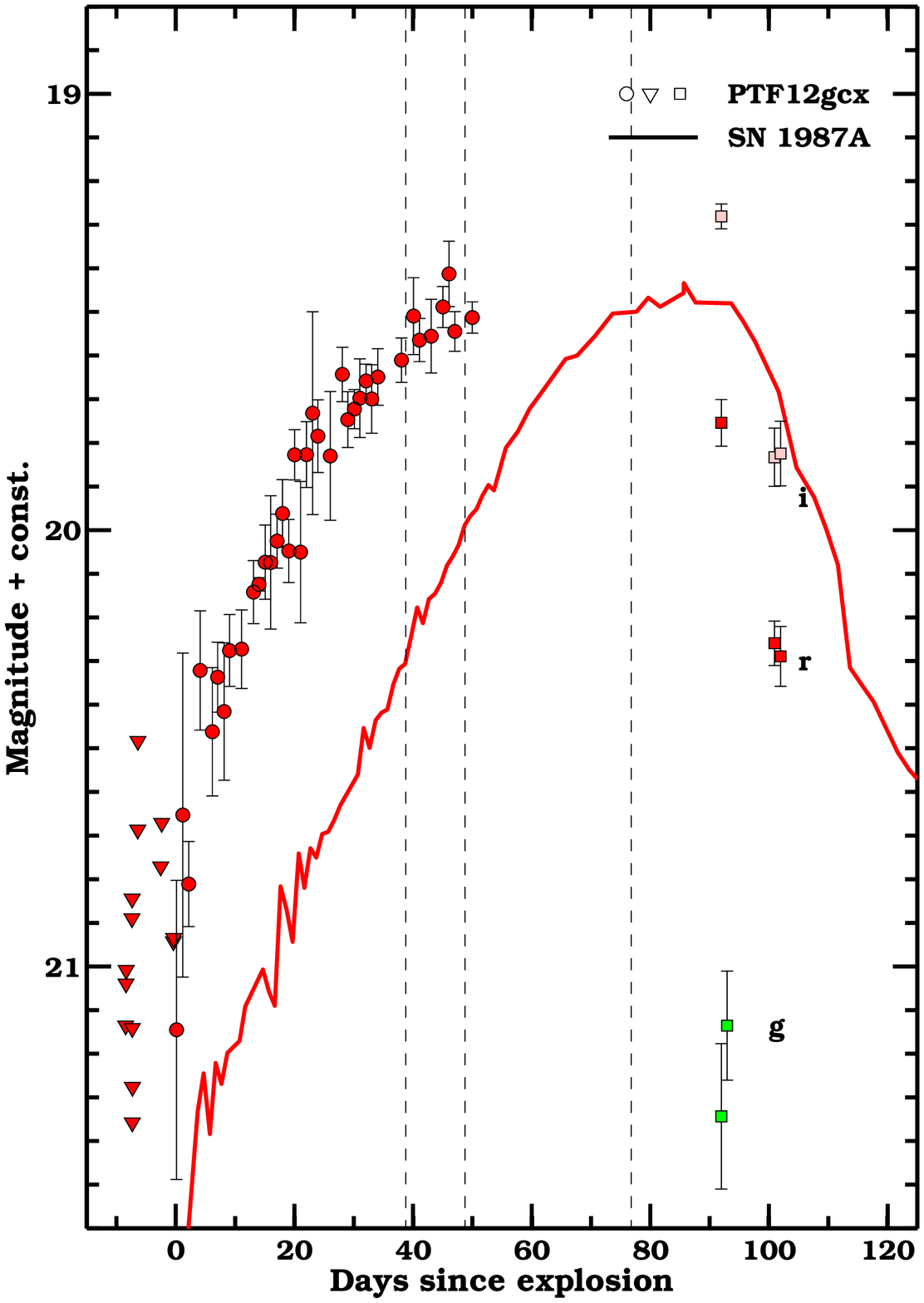}&
\includegraphics[width=6cm,angle=0]{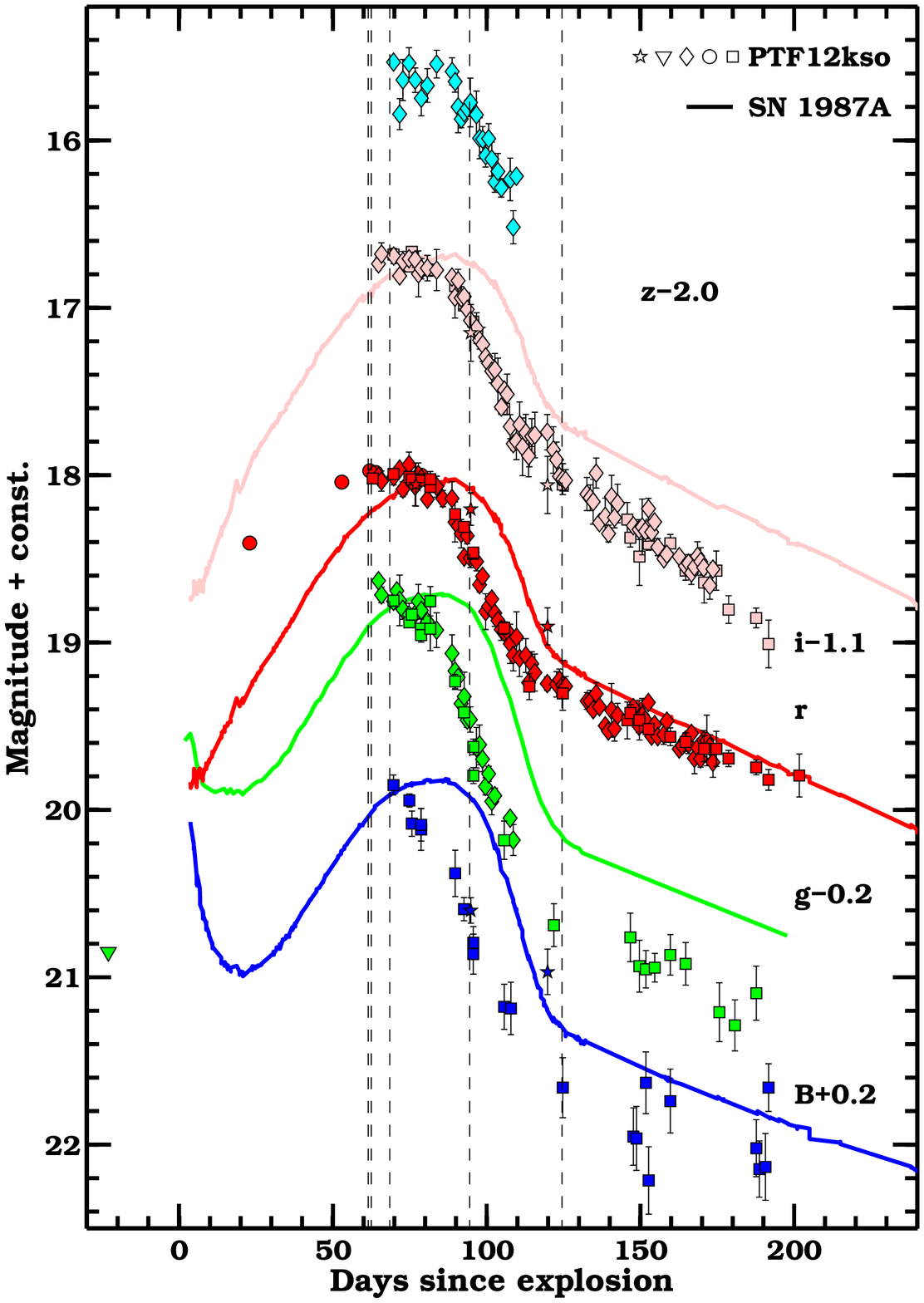}\\
\end{array}$
  \caption{Optical light curves of PTF09gpn, PTF12gcx, and PTF12kso. The circles show detections obtained with the P48 telescope and the squares indicate observations with the P60 telescope in $Bgriz$. Diamonds and stars mark the LCOGT and MLO photometry (obtained on the Johnson-Cousins $BVRI$ photometric system; see Table~\ref{tab:photPTF}) for PTF12kso. Triangles indicate prediscovery $3\sigma$ limiting magnitudes from P48 observations. The corresponding light curves of SN~2006au \citep[from][]{taddia12} and SN~1987A (solid lines) are overplotted and shifted in magnitude to match those of our objects. The vertical dashed lines mark the spectroscopic epochs. \label{lcPTF}}
 \end{figure*}

\section{Host galaxies}
\label{sec:host}

We have already examined the host galaxies of the CCCP SN~1987A-like SNe in \citet{taddia13} along with those of the entire literature sample of long-rising SNe~II. There we found that SN~1987A-like SN locations are more metal-poor than those of normal SNe~II. These peculiar SNe are located either in the metal-poor outskirts of bright galaxies or in dwarf hosts. Here we present and analyze spectra and images of the PTF objects, which are shown in Fig.~\ref{galaimage}.

PTF09gpn was located in a faint host galaxy characterized by an apparent magnitude of 19.13 in the $r$ band \citep{arcavi10}. A color image of the galaxy is shown in Fig.~\ref{galaimage} (top-left panel). The host-galaxy spectrum (top-right panel of Fig.~\ref{galaimage}) reveals several emission lines (Balmer lines, 
[\ion{O}{iii}]~$\lambda\lambda$4363, 4959, 5007, [\ion{O}{ii}]~$\lambda$3727, [\ion{N}{ii}]~$\lambda$6584, 
[\ion{S}{ii}]~$\lambda\lambda$6717, 6731; see Fig.~\ref{galaimage}).

The absolute magnitude of the host galaxy in the $r$ band is $M_{r}=-15.35$~mag, which is that of a dwarf 
galaxy. By using the value of the absolute magnitude in the $r$~band and Eq.~1 in  \citet{arcavi10}, we estimated the metallicity to be 0.13~Z$_{\odot}$, or 12$+$log(O/H) = 7.80 (where 12$+$log(O/H)$_{\odot}$ = 8.69~dex; \citealp{asplund09}). This would place PTF09gpn at the lowest end of the known metallicity distribution of SN~1987A-like events \citep{taddia13}.

 To provide a more direct measurement of the host metallicity, we made use of the emission-line fluxes in the host-galaxy spectrum. We first modeled the spectrum to account for the Balmer absorption lines due to the underlying stellar population, using the STARLIGHT code \citep{cid05}\footnote{\href{http://astro.ufsc.br/starlight/}{http://astro.ufsc.br/starlight/}} as in \citet{taddia15}. Next we measured the Balmer-line fluxes and those of several other lines after subtracting the continuum; see Table~\ref{line}. We then computed the total extinction, which turned out to be negligible as measured from the Balmer-line decrement \citep{osterbrock89}. Different diagnostics, namely R23 \citep{m91}, N2, and O3N2 \citep{pp04}, were used to determine 12$+$log(O/H).
R23 indicate an oxygen abundance of 12$+$log(O/H)$_{\odot}$ = $7.77\pm0.15$ (see Table~\ref{line}), confirming that PTF09gpn is characterized by one of the lowest metallicities measured so far. N2 and O3N2 indicate a somewhat higher metallicity (12$+$log(O/H)$_{\odot}$ = $8.06\pm0.18/0.14$), but they rely on the poor detection of $[\ion{N}{ii}]$~$\lambda6584$ (2.5$\sigma$). Also, it is known that there are offsets between different strong-line diagnostics \citep{kewley08}.

The bright [\ion{O}{iii}] emission makes the spectrum of this host similar to those of the so-called ``green pea" (G.P.) galaxies \citep{cardamone09} and H-poor SLSN hosts (\citealp{leloudas15}), which often show a preference for extreme emission-line galaxies (\citealp{amorin15}). The [\ion{O}{iii}]~$\lambda$5007/[\ion{O}{ii}]~$\lambda$3727 ratio is 4.2, consistent with those of G.P. galaxies \citep{jaskot13}. The diameter of this galaxy is $\sim10$\arcsec, which corresponds to $\sim3.2$~kpc, also consistent with the size of G.P. galaxies.

The H$\alpha$ luminosity is $L$(H$\alpha$) = $2.4\times10^{40}$~erg~s$^{-1}$. 
This gives an estimate of the star-formation rate (SFR) by applying the
formula of \citet{kennicutt12}:log$_{10}$(SFR) = log$_{10}L$(H$\alpha$)$-$41.27.
We obtain SFR = 0.07~M$_{\odot}$~yr$^{-1}$. The spectrum was calibrated against the $r$-band magnitude of the galaxy.

From the STARLIGHT fit we also obtained a value for the stellar mass ($M_{*} = 1.24\times10^{8}$~M$_{\odot}$), which implies a specific star-formation rate (sSFR) of $5.6\times10^{-10}$~yr$^{-1}$.

PTF12gcx was hosted by a spiral galaxy, SDSS J154417.02$+$095743.8, 
characterized by SDSS apparent 
magnitudes of $ugriz=$ 18.08(0.04), 16.92(0.01), 16.51(0.01), 16.25(0.01), 16.06(0.02) mag. A color image of the galaxy from SDSS is shown in Fig.~\ref{galaimage} (center-left panel). The spectrum of its nucleus (see Fig.~\ref{galaimage}, center-right panel) reveals the same emission lines as shown by the host of PTF09gpn, with the exception of [\ion{O}{iii}]~$\lambda$4363 and H$\delta$, which are too faint to be detected in this case.

The absolute magnitude of the host galaxy in the $r$ band is $M_{r} = 20.08$~mag, which is that of a bright spiral galaxy. 
Again, by using the value of the absolute magnitude in the $r$~band and \citet{arcavi10}, we estimated a metallicity of 0.75~Z$_{\odot}$, or 12$+$log(O/H) = 8.56. This places PTF12gcx at the highest end of the known metallicity distribution of SN~1987A-like events \citep{taddia13}, together with SNe~1998A, 2004em, and 2006au.

We made use of the emission-line fluxes (see Table~\ref{line}) in the host-galaxy spectrum and strong-line diagnostics to further constrain the metallicity. 
The SN exploded relatively close to the host center (at 4$\arcsec$ separation), so the SDSS spectrum (obtained with a 3$\arcsec$-wide fiber) can provide a good estimate of the metallicity at the position of PTF12gcx. N2 and O3N2 indicate 12$+$log(O/H)$_{\odot}$ = $8.52\pm0.18$ and 12$+$log(O/H)$_{\odot}$ = $8.56\pm0.14$ (respectively), confirming that PTF12gcx is at relatively high metallicity.  R23 indicates a slightly lower oxygen abundance of 12$+$log(O/H)$_{\odot}$ = $8.27\pm0.15$ (see Table~\ref{line}).

The H$\alpha$ luminosity is $L$(H$\alpha$) = $2.5\times10^{40}$~erg~s$^{-1}$, which corresponds to a SFR of 0.13~M$_{\odot}$~yr$^{-1}$ \citep{kennicutt12}. The spectrum was calibrated against the SDSS $r$-band magnitude of the galaxy.
With $M^{*} = 6.4\times10^{9}$~M$_{\odot}$, we obtain sSFR~$= 2.0\times10^{-11}$~yr$^{-1}$.

PTF12kso was hosted by an extremely faint galaxy, barely detected in the SDSS images (see Fig.~\ref{galaimage}, bottom-left panel). Aperture photometry on stacked P48 $r$-band images reveals a host apparent magnitude of $r =$ 20.64~mag. The absolute magnitude of $-15.12$ mag implies a metallicity of $Z=0.0024=0.12$~Z$_{\odot}$, or 12$+$log(O/H) = 7.76, which makes the host of PTF12kso one of the most metal-poor SN~1987A-like SN hosts observed so far.

The spectrum of this faint host galaxy exhibits strong H$\alpha$ emission and a faint continuum. We use the flux of H$\alpha$ and the limit on the flux of [\ion{N}{ii}] to estimate an oxygen abundance of 12$+$log(O/H) $\lesssim8.04$ with the N2 method. 
The SFR obtained from the H$\alpha$ luminosity is SFR = 0.015~M$_{\odot}$~yr$^{-1}$. The spectral continuum is too faint and noisy to obtain a reliable $M^{*}$ estimate from the STARLIGHT fit.

In summary, two out of our three new metallicity measurements push the oxygen-abundance distribution of the host galaxies of SN~1987A-like SNe shown in \citet{taddia13} to even lower values, making the SN~1987A-like SNe one of the most metal-poor populations of CC~SNe, together with SN impostors (\citealp{taddia15met}) and H-poor SLSN hosts (\citealp{leloudas15}).

 \begin{figure*}
 \centering
 $\begin{array}{ccc}
\includegraphics[width=6cm,angle=0]{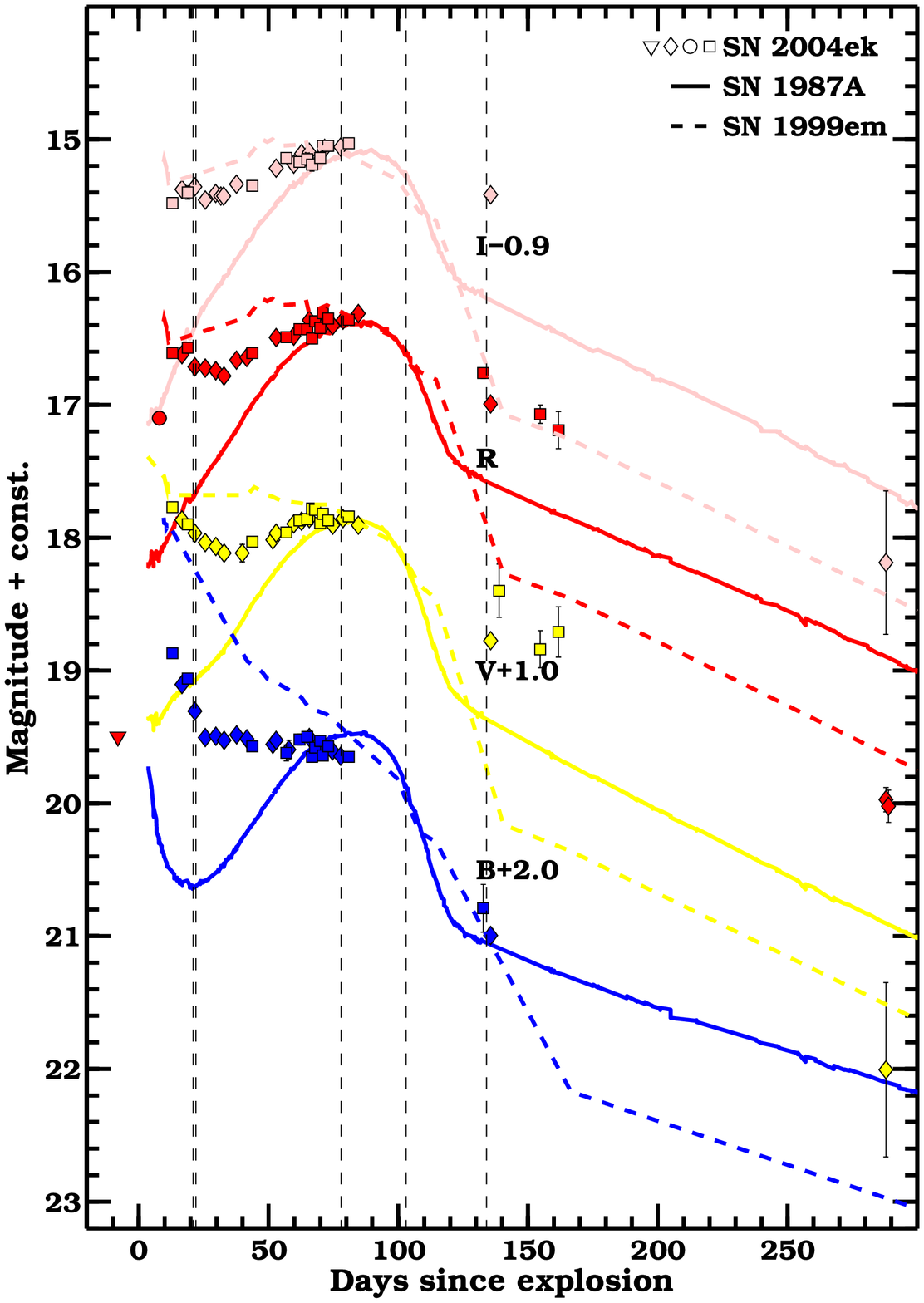}&
\includegraphics[width=6cm,angle=0]{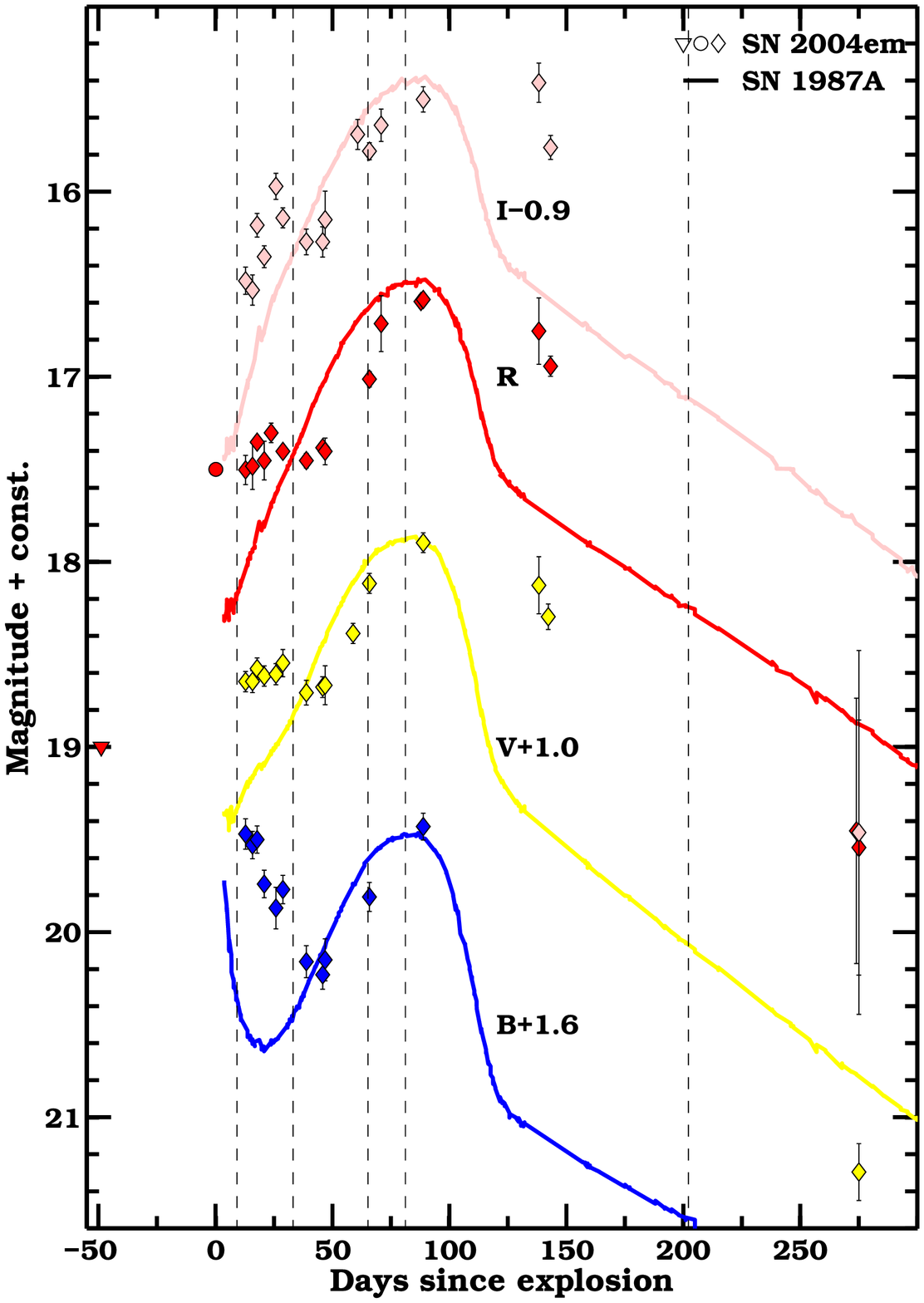}&
\includegraphics[width=6cm,angle=0]{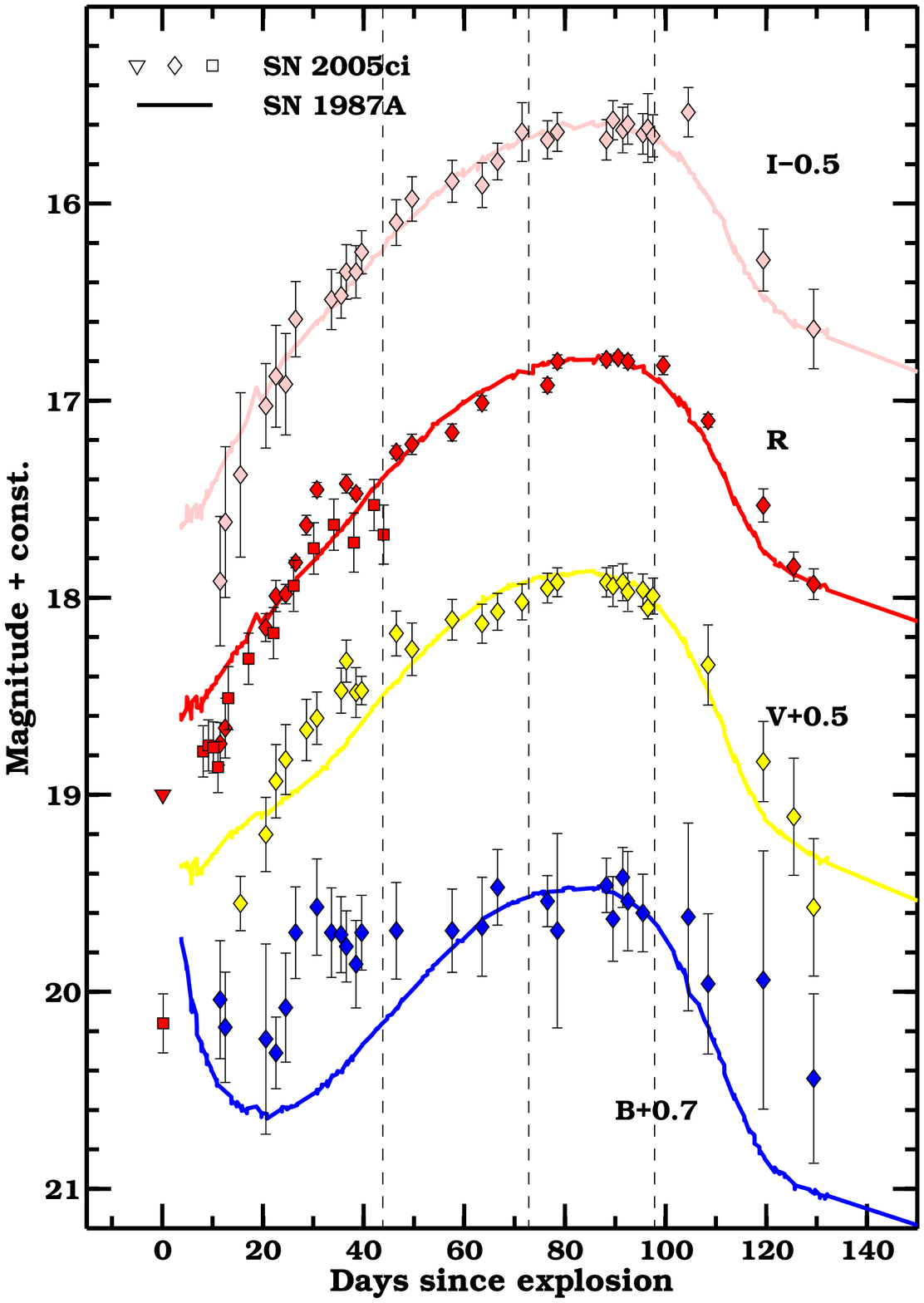}\\
\end{array}$
  \caption{Optical light curves of SNe~2004ek, 2004em, and 2005ci. Diamonds indicate CCCP photometry, squares indicate photometry from \citet{tse08} for SN~2004ek and from \citet{kleiser11} for SN~2005ci. For SNe~2004ek and 2004em, the circle indicates the discovery magnitude from \citet{disco04ek} and \citet{disco04em}. The $BVRI$ photometry from \citet{tse08} was shifted by $+$0.1, $+$0.2, $+$0.2, and $+$0.1 mag (respectively), in order to match the CCCP photometry of SN~2004ek. Triangles indicate limits from prediscovery nondetections from the literature (see Table~\ref{tab:disco}). The corresponding light curves of SNe~1987A (solid lines) and 1999em (from \citealp{Elmhamdi03}, dashed lines) are overplotted and shifted in magnitude to match the peaks of our objects. The vertical dashed lines mark the spectroscopic epochs. \label{lcCCCP}}
 \end{figure*}

\section{Supernova light curves and colors}
\label{sec:phot}

\subsection{Light-curve shape}

\subsubsection{PTF supernovae}

PTF09gpn was imaged in the $Bgriz$ bands, and the light curves (with the exception of the late $r$-band epochs) are reported in 
Fig.~\ref{lcPTF} (left-hand panel).
The $r$-band light curve of PTF09gpn covers the first $\sim350$~d. The $r$-band light curve sampling is 
rather frequent  (2.4~d average cadence) until $\sim72$~d. Several epochs were taken at late times (300 to 350~d).
In the other bands the follow-up observations started $\sim2$ weeks after discovery, with an average cadence of 4.1, 3.6, 5.7, and 5.7~d for $z$, $i$, $g$, and $B$ (respectively). In these filters the last photometric epoch was taken at $\sim82$~d ($Bgi$) and $\sim72$~d ($z$).
The $r$-band light curve shows a rise of $\sim0.8$~mag in the first 10~d. Then it settles onto an early plateau between days 10 and 30, and it finally starts a long rise to peak ($\sim0.9$~mag brighter than the plateau) that occurs at $\sim75$~d. 
All of the other bands reach maximum around the same time, and between days 10 and 30 they also exhibit an early-time plateau, preceded by a decline in the case of $B$ and $g$. Since the follow-up observations started 2 weeks later for these filters than in the $r$ band, no early-time rise could be detected. The overplotted and scaled light curves of SN~2006au (from \citealp{taddia12}) clearly show the close resemblance of PTF09gpn to this SN~1987A-like SN.

PTF12gcx was imaged in the $r$ band with P48. 
The light curve (central panel of Fig.~\ref{lcPTF}) is characterized by frequent sampling, with the SN being imaged every 1.3 nights on average during the rise ($\sim1.4$~mag from discovery to peak). Three epochs (the last at $\sim100$~d) were taken after peak, which occurred at day 57. After peak the decline is of $\sim1$~mag in $\sim40$~d.
The light-curve shape is very similar to that of SN~1987A, which is also shown in the figure. A difference is that PTF12gcx rises faster in the first days after discovery and thus peaks earlier than SN~1987A. A few epochs with P60 in $gri$
were also taken after the peak.

The light curves of PTF12kso are shown in the right-hand panel of Fig.~\ref{lcPTF}.
This SN was imaged in the $Bgriz$ bands. The first observation in $r$ with P48 occurred at $\sim20$~d. Then, after a  gap of $\sim20$~d in the light-curve coverage, at $\sim40$~d intense follow-up observations started in all the bands with P60 and in $griz$ with LCOGT. The SN was observed and detected by LCOGT almost every night from a few days before peak (which occurred at 61~d in $r$) until $\sim170$~d ($\sim110$~d in $g$ and $z$). P60 took additional images up to 200~d. After peak the light curves decline by $\sim1$~mag in $\sim60$~d, but then after 120~d the decline rate becomes
smaller by about 50\%, with the light curve setting onto a tail very similar to that observed in SN~1987A.

Rise times and peak magnitudes for the PTF SNe are summarized in Table~\ref{tab:rise}.

\subsubsection{CCCP supernovae}

\begin{figure*}
 \centering
\includegraphics[width=17cm,angle=0]{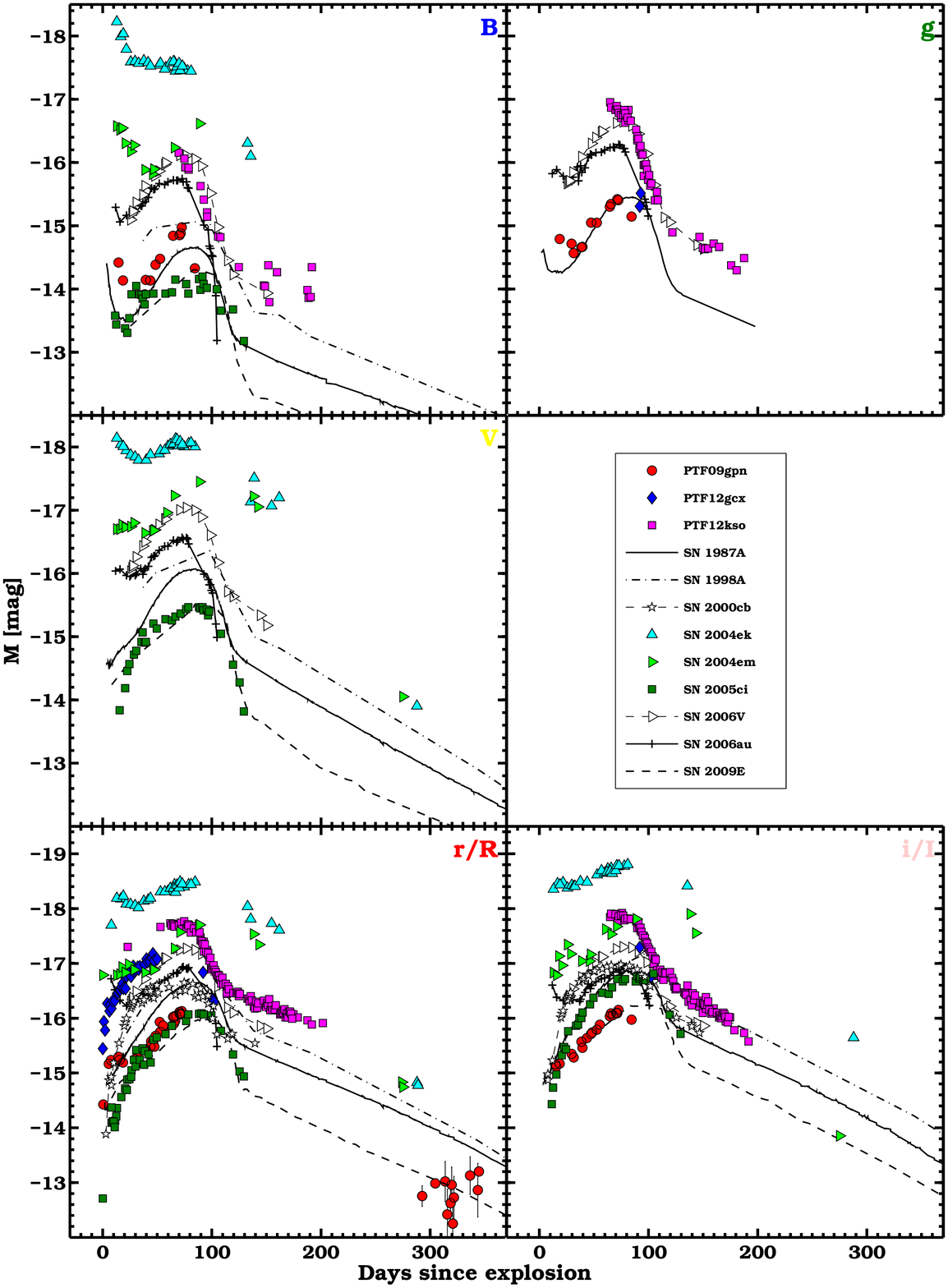}
  \caption{$B,$ $g$, $V$, $r/R$, and $i/I$-band absolute magnitudes for our PTF and CCCP SN~1987A-like SNe (colored symbols) and for similar events (black lines and symbols) from the literature (\citealp{kleiser11}; \citealp{pastorello05}; \citealp{pastorello12}; \citealp{taddia12}; and references therein).\label{abs_lc}}
 \end{figure*}

SN~2004ek was observed in $BVRI$ by \citet{arcavi12} and \citet{tse08}. The light curves are shown in Fig.~\ref{lcCCCP} (top-left panel). The monitoring started at $\sim13$~d and, before maximum (which occurred at $\sim90$~d in $VRI$),
the light-curve sampling was rather frequent, with the SN observed every 2.7 nights on average. After peak, a few epochs were obtained between  $\sim130$ and $\sim160$~d. The last observations occurred at almost 290~d in all bands except $B$, for which the last epoch was obtained at $\sim130$~d. The light curves of SN~2004ek show an early decline up to $\sim25$--32~d in all the bands except $I$, which shows an early plateau of comparable length. The decline is faster in the bluer bands, with $B$ and $R$ exhibiting a decline of $\sim0.5$ and 0.2~mag from the early peak, respectively. Following this phase, the $VRI$ bands show a slow rise to peak ($\sim0.4$ mag in $\sim55$ days), which occurred at $\sim80$ days. At the same epochs, the $B$ band shows a plateau. The early, long-lasting declining phase and the slow rise to maximum shown by SN~2004ek make this SN a very peculiar object --- sort of an intermediate case between SN~1987A and a normal SN~IIP such as SN~1999em, as we show in Fig.~\ref{lcCCCP}. 
The post-maximum decline up to 160~d is also quite slow in all of the redder bands,  as compared with SNe~1987A and 1999em.
However, the last epoch obtained at 290~d shows a difference in magnitude from peak comparable to that of SN~1999em and larger than that of SN~1987A.

SN~2004em has $BVRI$ light-curve coverage and shape that are rather similar to those of SN~2004ek (see Fig.~\ref{lcCCCP}, top-right panel). 
For this SN the monitoring began at $\sim2$ weeks, and most of the epochs were obtained before maximum (which occurred at $\sim110$~d). Before maximum the SN was observed once per week on average. A few additional epochs were obtained at $\sim140$~d, and the last one at 275~d.
This SN shows a 0.6~mag decline in the $B$-band light curve in the first $\sim40$ days. At the same epochs, the $V$ and $R$ bands appear constant, whereas the $I$ band first rises by 0.5~mag for $\sim2$ weeks, and then also shows a short plateau up to $\sim40$ days. After the early phase, all of the light curves rise to peak. The rise to peak is steeper than in the case of SN~2004ek, reaching about 1~mag over 50~d in $R$. After peak, the decline is slower than for SN~1987A, as shown in the figure. The last epoch shows that the difference in magnitude from peak to tail is similar to that of SN~1987A.

The light curves of SN~2005ci are rather similar to those of SN~1987A, as shown in Fig.~\ref{lcCCCP} (bottom-left panel). This SN was observed by \citet{arcavi12} in $BVRI$ and unfiltered ($\sim R$) by \citet{kleiser11}. The CCCP follow-up observations started at 11~d, with an average cadence of 4 nights and the last epoch taken at 130~d. The unfiltered light curve has the first epoch only a few hours after explosion, and covers the first 44~d.
The rise to peak, which occurs at $\sim87$~d in $R$, 
is faster than for SN~1987A in the first $\sim30$--40 days in all bands. Thereafter, the slope becomes less steep ($\sim0.5$ mag in $\sim50$~d in $R$).
The $B$ band shows hints of an early plateau before day 20. Its shape is flatter than that of SN~1987A.
The post-peak decline ($\sim1$~mag in $\sim40$~d in the $R$ band) is similar to that of SN~1987A.

Rise times and peak magnitudes for the CCCP SNe are summarized in Table~\ref{tab:rise}.

\subsection{Absolute magnitudes}

In Fig.~\ref{abs_lc} we plot the absolute magnitudes in $B$, $g$, $V$, $r/R/unf.$, and $i/I$ filters of all our SNe (colored symbols) and of other well-observed long-rising SNe~II resembling SN~1987A (black symbols and lines). The literature data were taken from 
\citet{kleiser11}, \citet{pastorello05}, \citet{pastorello12}, \citet{taddia12}, and reference therein.
For our SNe we adopted the distances and extinctions listed in Table~\ref{tab:sample}. For the SNe in the literature we obtained redshift-independent distances from NED when available; otherwise we adopted WMAP5 cosmological parameters and a redshift from NED, to be consistent with the distances computed for our objects. The Galactic extinctions were obtained from NED. The explosion epochs and the host-galaxy extinctions were obtained from the literature.

The only band in which all our objects and those from the literature have a frequently sampled and extended light curve is $r/R/unf$. The comparison in this band reveals that the magnitude of the (second) peak ranges from $-16.0$~mag (as in the case of SNe~2005ci, 2009E, and PTF09gpn) to $-18.5$~mag (SN~2004ek). One PTF SN (PTF12kso) and two CCCP objects (SNe 2004ek and 2004em) are more luminous than all of the other SN~1987A-like events previously published (among those SNe, SN~2006V was hitherto the most luminous), peaking at $M_{r/R}<-17.3$~mag. PTF09gpn and SN~2005ci are instead among the least luminous SN~1987A-like SNe, being only marginally brighter than SN~2009E. 
On the tail (i.e., after $\sim130$~d), the SNe which have brighter peaks also have brighter tails. However, at very late epochs ($\sim290$~d), the absolute magnitudes of SNe~2004ek and 2004em are slightly brighter (up to 0.4~mag) than those of SN~1998A, which at peak was up to 1.5~mag fainter. 
Compared to the $r/R$ band, the $B$ and $V$ bands exhibit wider ranges of peak magnitudes (3.5 and 2.5~mag, respectively).

At early phases, when most of the flux is in the bluer bands, $B$ ranges from $-18.2$~mag (SN~2004ek) down to $-13.3$~mag (SN~2005ci). In between, we find SNe~1987A, 2004em, 2006au, and PTF09gpn, showing early declines.

The large range of absolute magnitudes has important implications for the progenitor parameters, in particular for the $^{56}$Ni mass and the progenitor radius (see Sec.~\ref{sec:models}).

\subsection{Colors}

\begin{figure}
 \centering
\includegraphics[width=9cm,angle=0]{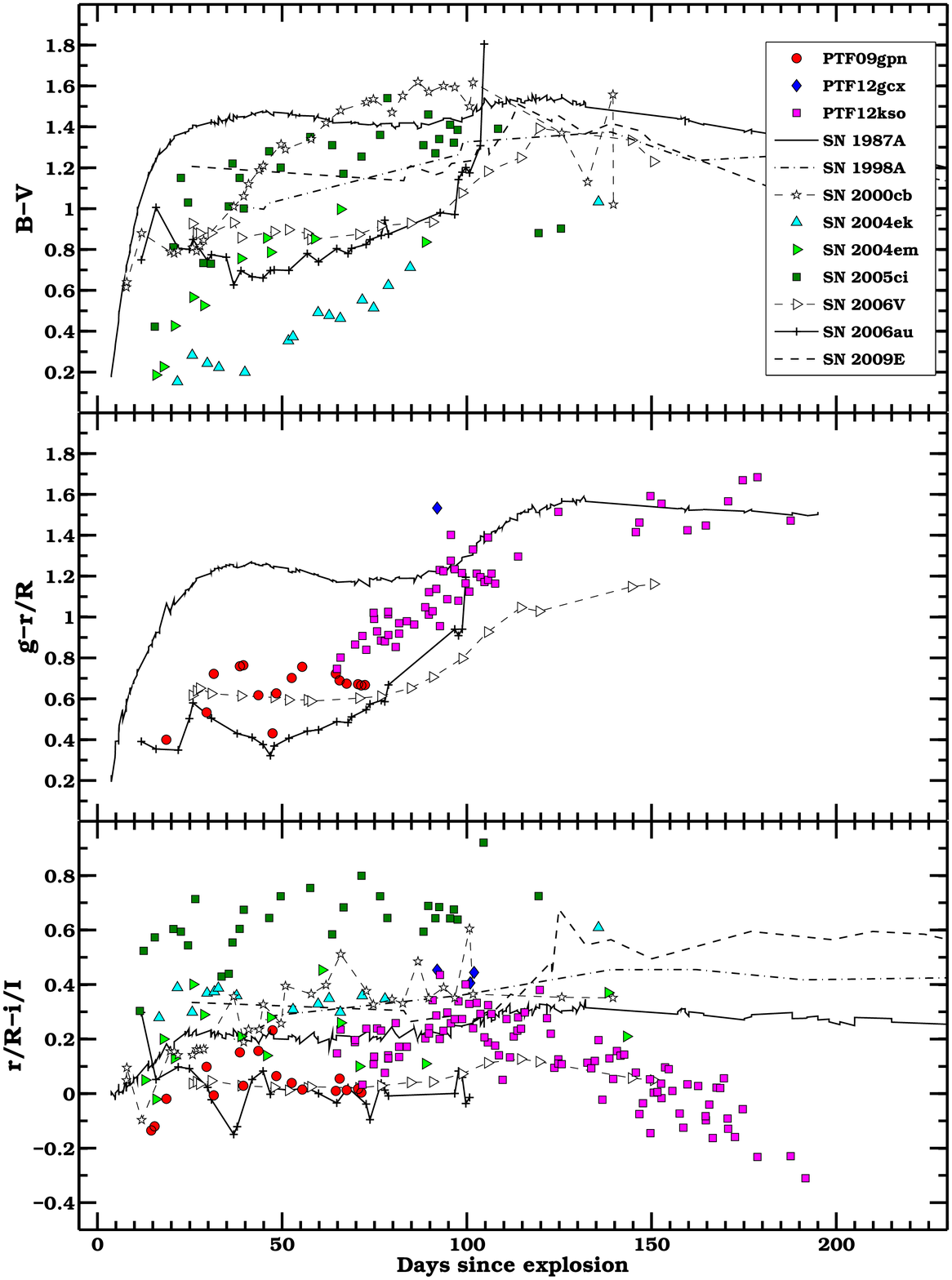}
  \caption{Color comparison among our PTF and CCCP SN~1987A-like SNe (colored symbols) and similar events (black lines and symbols) from the literature (\citealp{kleiser11}; \citealp{pastorello05}; \citealp{pastorello12}; \citealp{taddia12}; and references therein).\label{color}}
 \end{figure}

Adopting the light curves and the extinctions used to compute the absolute magnitudes,
we also compared the colors of our SNe to those of other long-rising SNe. In Fig.~\ref{color} we plot three different optical colors: $B-V$, $g-r/R$, and $r/R-i/I$. 
The $B-V$ and $g-r/R$ color evolutions are similar for all of the SNe, with bluer colors at early epochs which progressively become redder. Among the reddest SNe we find SN~2005ci, similar to SN~1987A.  SNe~2004ek, 2004em, PTF09gpn, and PTF12kso are bluer, similar to SNe~2006V and 2006au. 

The $r/R-i/I$ color looks flatter for all of the SNe, as its value is more influenced by the different strengths of the spectral lines (e.g., H$\alpha$ and Ca~II) than by the continuum.

\section{Supernova spectra}
\label{sec:spectra}
\subsection{PTF SN spectra}

 \begin{figure*}
 \centering
 \includegraphics[width=15cm,angle=0]{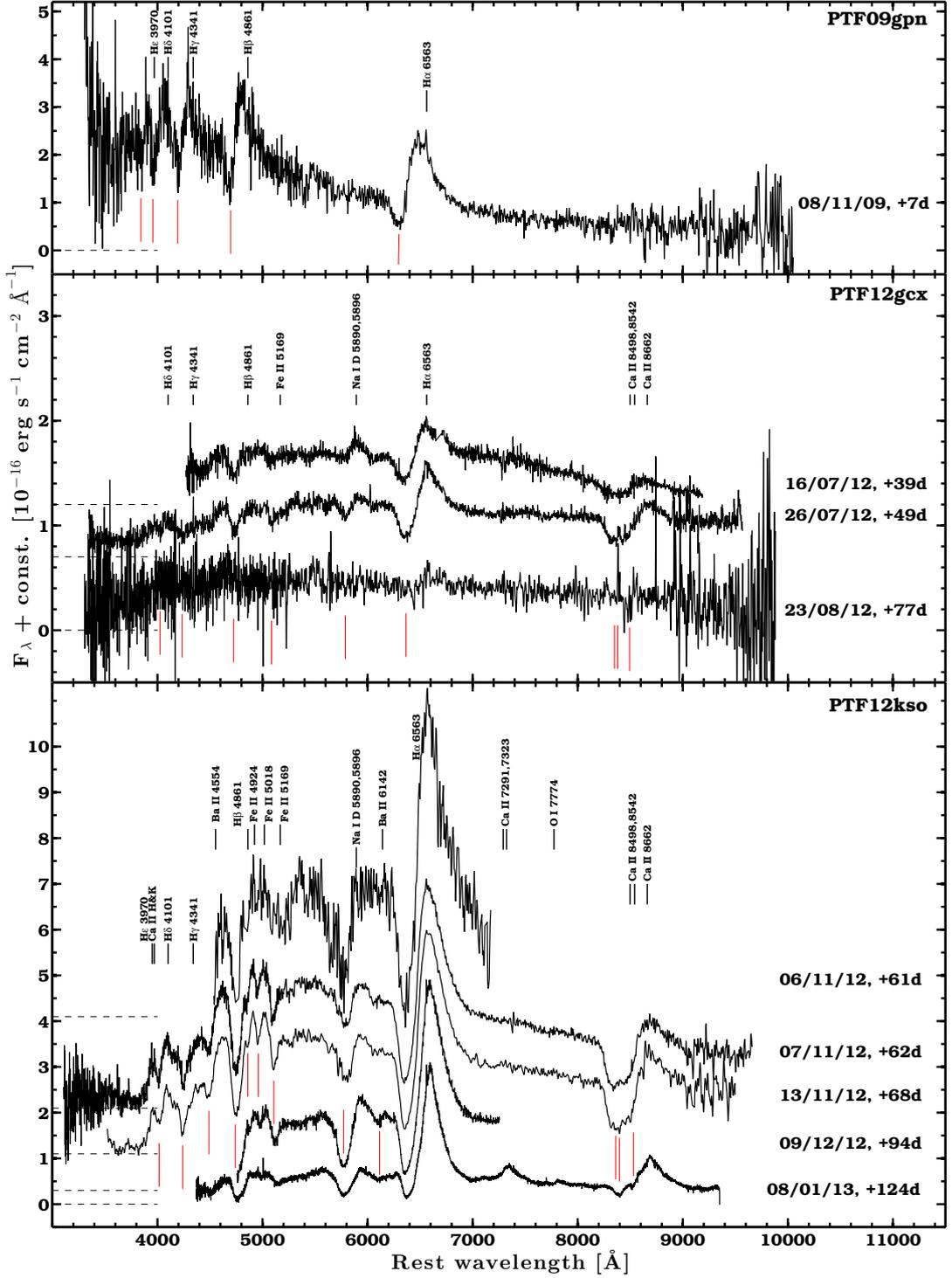}\\
  \caption{Spectral sequences for PTF09gpn, PTF12gcx, and PTF12kso. All spectra were calibrated against photometry and corrected for extinction. The main spectral lines, the dates, and the phases are reported. The absorption minimum of each line is marked by a red segment, the rest wavelength by a black segment. The zero-flux level is marked by a dashed horizontal line for each spectrum. \label{specPTF}}
 \end{figure*}

  \begin{figure*}
 \centering
 \includegraphics[width=15cm,angle=0]{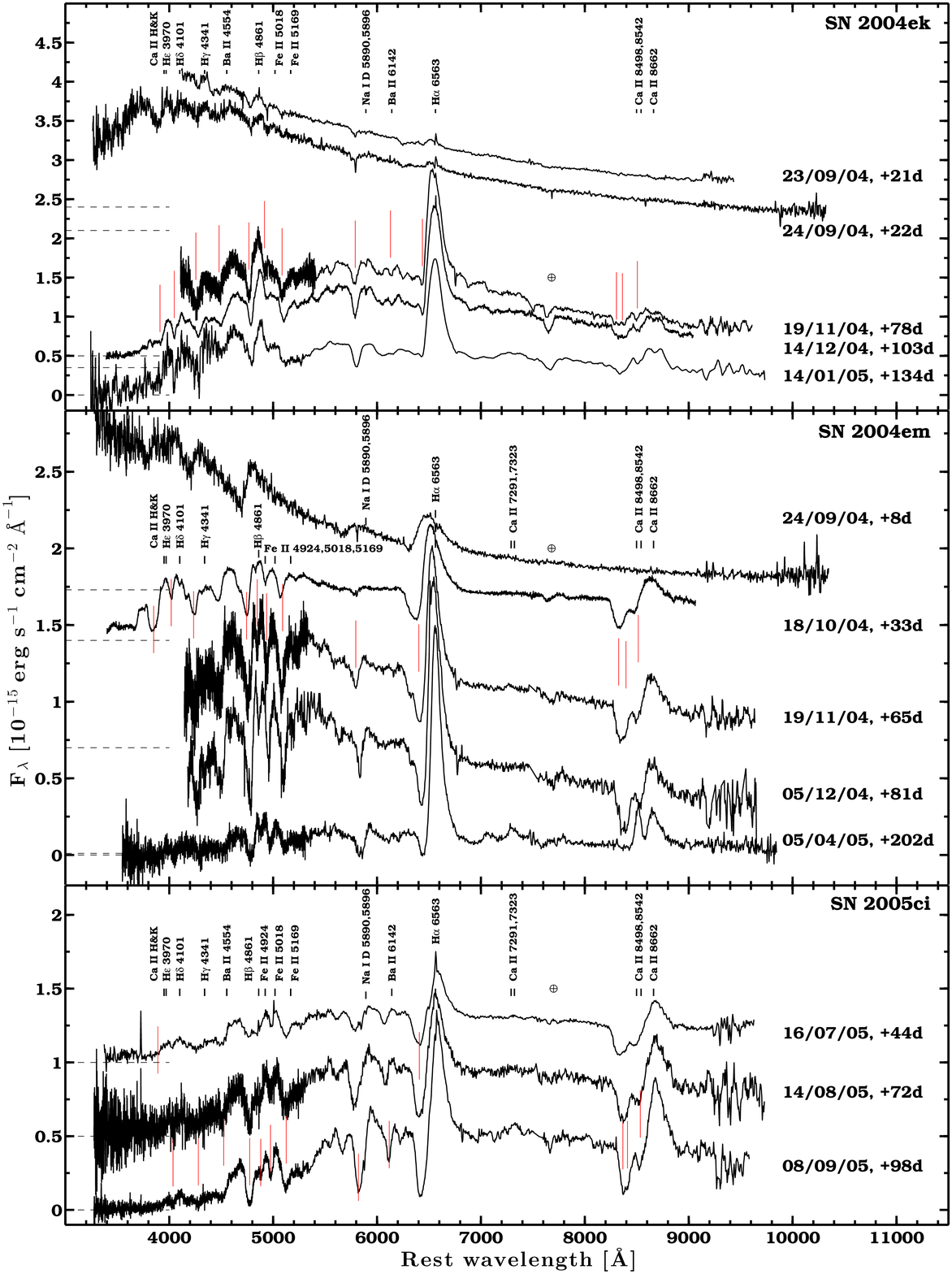}\\
  \caption{Spectral sequences for SNe~2004ek, 2004em, and 2005ci. All spectra were calibrated against photometry and corrected for extinction. The main spectral lines, the dates, and the phases are reported. The absorption minimum of each line is marked by a red segment, the rest wavelength by a black segment. The zero-flux level is marked by a dashed horizontal line for each spectrum. Telluric features are marked by $\oplus$. \label{specCCCP}}
 \end{figure*}

In Fig.~\ref{specPTF} we plot the spectral sequences of PTF09gpn, PTF12gcx, and PTF12kso. The spectra are made publicly available via WISeREP \citep{yaron12}. All of the spectra were calibrated against $r$-band photometry and corrected for reddening.

In Fig.~\ref{specPTF} (top panel) we show the only spectrum obtained of PTF09gpn, which was taken at $\sim7$~d, during the early peak of the $r$-band light curve, as indicated by the vertical dashed line in Fig.~\ref{lcPTF} (left-hand panel). The Balmer lines (H$\alpha$, H$\beta$, H$\gamma$, H$\delta$, H$\epsilon$) dominate the spectrum of PTF09gpn and are labeled in the figure. All of them exhibit a broad P-Cygni profile. The continuum appears quite blue, as the spectrum was obtained at early times.

In the central panel of Fig.~\ref{specPTF} three spectra of PTF12gcx are reported. The first two spectra, obtained on the rise of the light curve at 39 and 49~d, show clear P-Cygni profiles for the Balmer lines (from H$\alpha$ to H$\delta$),
for \ion{Na}{i}~D, and for the \ion{Ca}{ii} triplet. Redward of \ion{Na}{i}~D there may be traces of high-velocity H$\alpha$ absorption or \ion{Si}{ii}~$\lambda$6355. There are also signatures of
\ion{Fe}{ii}~$\lambda$5169 in absorption. The continuum spectral energy distribution (SED) peaks in the $V$ band.
 The last spectrum, obtained after peak, has low signal-to-noise ratio and only reveals a faint H$\alpha$ feature.

The spectral sequence of PTF12kso is shown in the bottom panel of Fig.~\ref{specPTF}. Five spectra were obtained, three of them before peak (at 61 to 68~d) and two after peak (at 94 and 124~d). In the early-time spectra, besides the Balmer lines down to H$\epsilon$, we detect P-Cygni features associated with \ion{Na}{i}~D (possibly blended with \ion{He}{i}~$\lambda$5875), \ion{Ca}{ii} triplet, \ion{Fe}{ii} triplet, and probably \ion{Ca}{ii} H\&K. There are also traces of \ion{Ba}{ii} at 6142~\AA\ and 4554~\AA. \ion{Ba}{ii}~$\lambda$6142 may be blended with high-velocity H$\alpha$ absorption or \ion{Si}{II}~$\lambda$6355. The continuum SED peaks in the $V$ band.
In the last spectrum, the absorption part of the P-Cygni features decreases its strength in all the lines except for \ion{Na}{i}~D. At these late epochs, \ion{Ca}{ii}~$\lambda\lambda$7291, 7323 emerges in emission, the continuum is faint and almost constant, and the lines dominate the SED.

\subsection{CCCP SN spectra}

 \begin{figure}
 \centering
\includegraphics[width=9cm,angle=0]{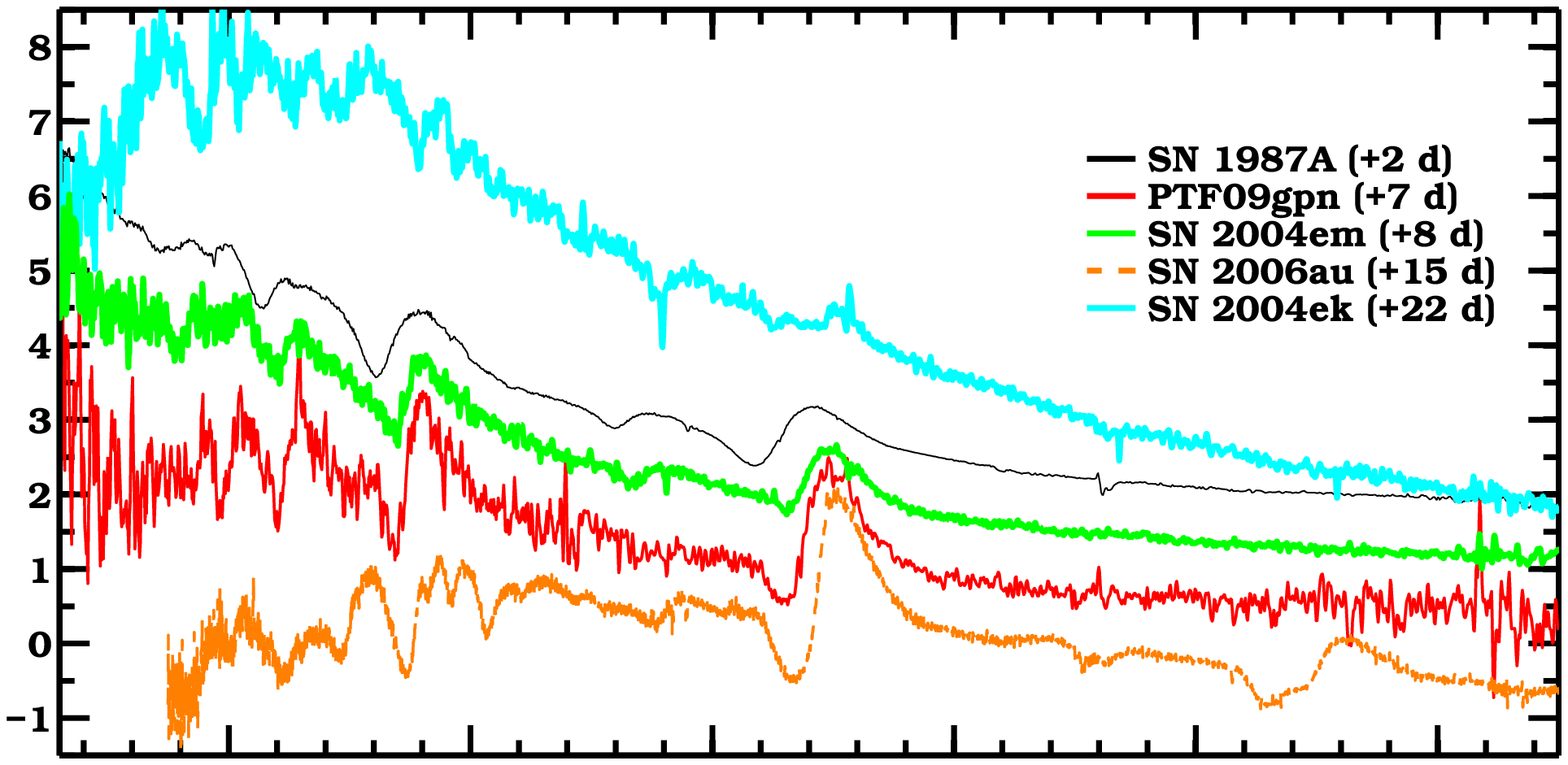}
\includegraphics[width=9cm,angle=0]{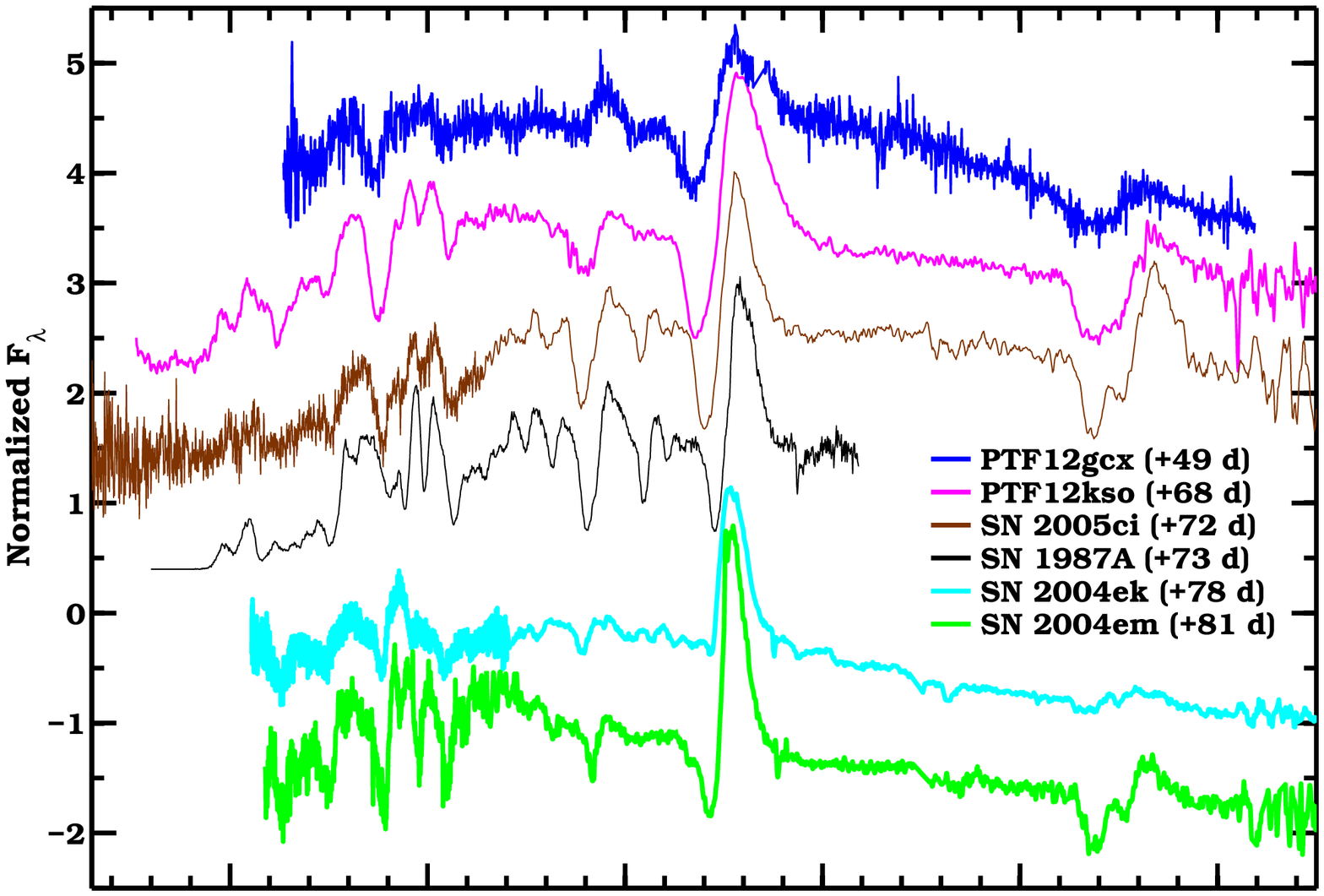}
\includegraphics[width=9cm,angle=0]{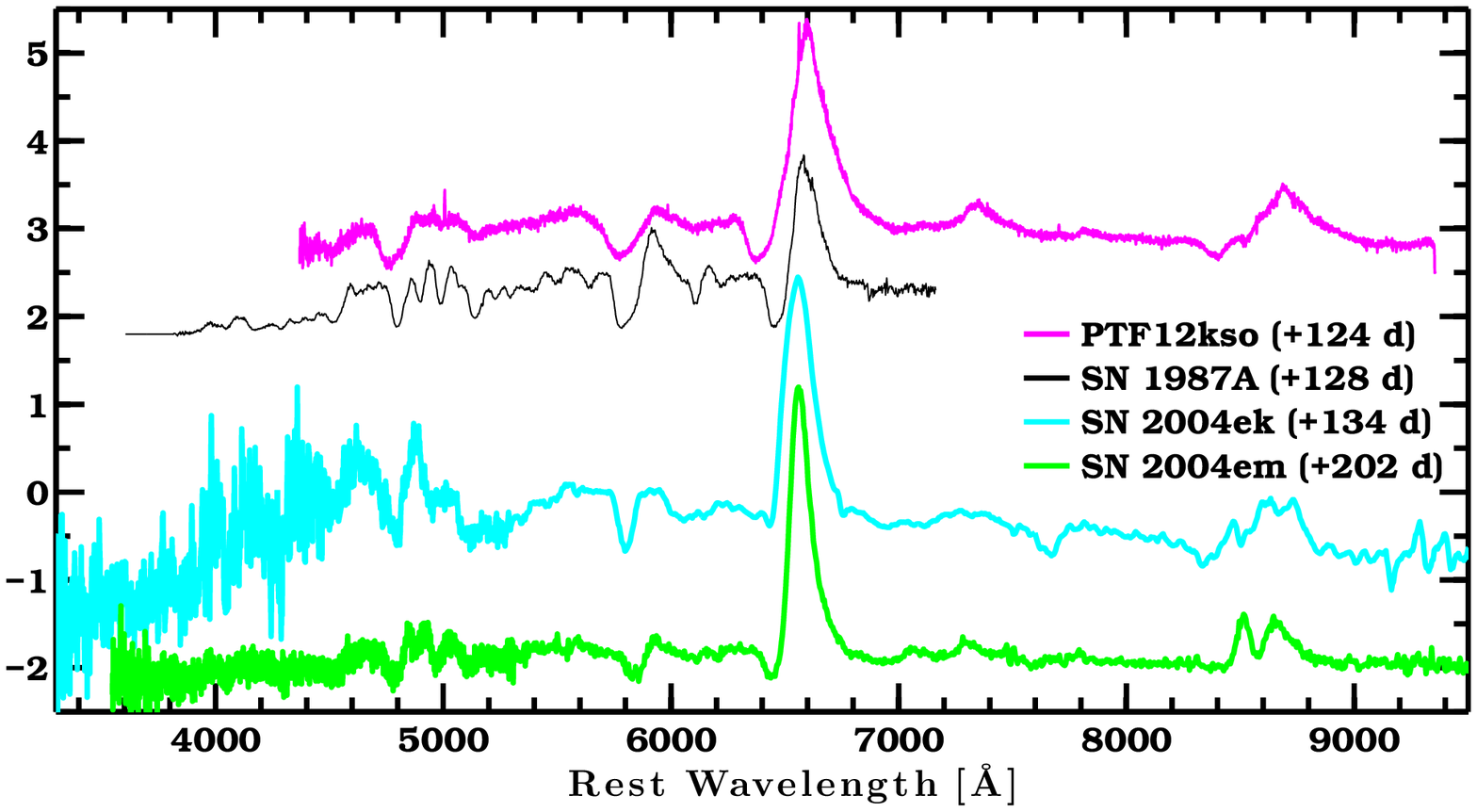}
  \caption{Spectral comparison among our PTF and CCCP long-rising SNe~II, SN~1987A (from \citealp{phillips88}) and SN~2006au (from \citealp{taddia12}). The top panel includes spectra taken at early epochs, in the central panel we show spectra obtained around peak. The bottom panel presents spectra taken on the tail of the light curves.\label{spec_comp}} 
 \end{figure}

  \begin{figure}
 \centering
\includegraphics[width=9cm,angle=0]{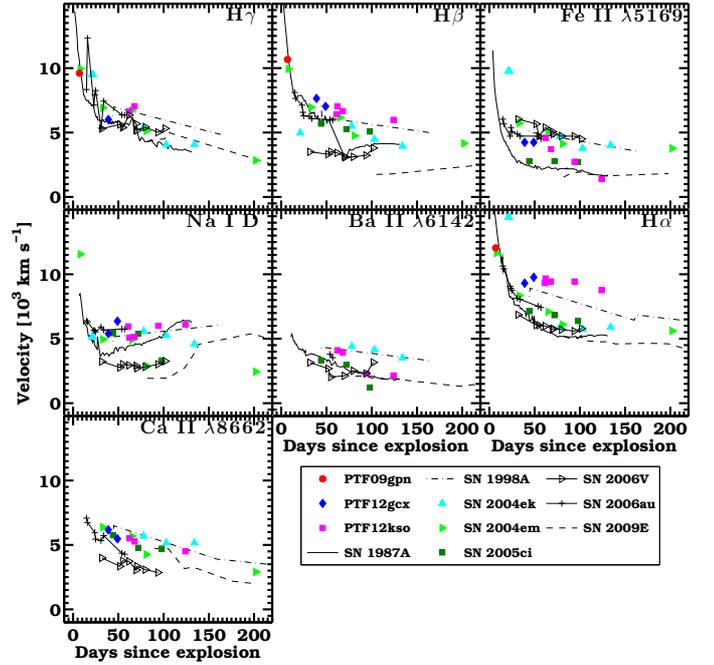}
  \caption{Velocity evolution from the P-Cygni absorption minima of the most important spectral lines for our PTF and CCCP SNe (colored symbols) as compared to SN~1987A and other long-rising SNe~II (black symbols, from the spectra presented by \citealp{phillips88}, \citealp{pastorello05}, \citealp{pastorello12}, \citealp{taddia12}). Typical uncertainties are of 500--1000~km~s$^{-1}$.\label{velocity}} 
 \end{figure}

In Fig.~\ref{specCCCP} we plot the spectral sequences of SNe~2004ek, 2004em, and 2005ci. The spectra are made publicly available via WISeREP \citep{yaron12}. All the spectra were calibrated against $R$-band photometry and corrected for reddening.

The spectra of SN~2004ek were obtained at 21, 22, 78 (pre-peak), 103 and 134 (post-peak) days.
 The first two spectra were taken at the epoch of the early cooling phase, and they show a blue continuum where diluted P-Cygni profiles of Balmer lines are barely visible. \ion{Fe}{ii}~$\lambda$5169 and possibly \ion{He}{i}~$\lambda$5875 in absorption are also visible. 
The later spectra  peak in the continuum around 5000--5500~\AA, and show strong broad H$\alpha$ emission with a marginal and narrow  P-Cygni absorption. The other Balmer lines and in particular H$\beta$ show prominent P-Cygni features.
\ion{Na}{i}~D emerges and \ion{Fe}{ii}~$\lambda$5169 (along with \ion{Fe}{ii}~$\lambda$5018) becomes more evident. The \ion{Ca}{ii} triplet emerges and traces of \ion{Ba}{ii}~$\lambda\lambda$4554, 6142 are observable. Possibly \ion{Ca}{ii} H\&K is also detected.

The first spectrum of SN~2004em was obtained at 8~d, at the cooling phase. Because of the early phase, it appears much bluer than the other spectra, and shows broad Balmer P-Cygni profiles. The next three spectra were also obtained before maximum, at 33, 65, and 81~d. Although the second spectrum was taken on the decline of the $B$-band light curve, it does not appear as blue as the first two spectra of SN~2004ek or as the previous spectrum of SN~2004em. These spectra have a continuum that peaks in the $V$ band, and are dominated by the typical SN~II P-Cygni Balmer profiles. As in the spectra of the SNe that were previously discussed, they exhibit broad P-Cygni profiles for \ion{Na}{i}~D and for the \ion{Ca}{ii} and \ion{Fe}{ii} triplets. 
In the last spectrum, taken at 202~d, \ion{Ca}{ii}~$\lambda\lambda$7291, 7323 emerges in emission, the \ion{Ca}{ii} triplet does not show absorption, the continuum becomes weak, and H$\alpha$ dominates the SED with a strong broad emission and residual absorption.

SN~2005ci has three spectra obtained at 44, 
72, and 98~d (only one after the peak and before the tail).
These spectra are particularly rich in features and are dominated by P-Cygni Balmer lines. All of the other lines mentioned for the other SNe are visible in emission and absorption, including \ion{Na}{i}~D, \ion{Ca}{ii} triplet, \ion{Fe}{ii}~$\lambda\lambda$4924, 5018, 5169, \ion{Ca}{ii}~$\lambda\lambda$7291, 7323 (in emission), and \ion{Ba}{ii}~$\lambda\lambda$4554, 6142. The spectral continua are quite red, peaking around 6000~\AA.

\subsection{Spectral comparison and expansion velocities}

For a more direct spectral comparison among our SNe and other SN~1987A-like events, we plot a sequence of spectra taken at early epochs in the top panel of Fig.~\ref{spec_comp}, one taken around peak in the central panel, and three spectra obtained at late epochs in the bottom panel.

In the top panel it is evident that at early epochs these long-rising SNe~II show Balmer and (later) Fe features diluted by a strong black-body continuum. The case of SN~2004ek is particularly extreme, with the spectral lines almost completely diluted even after 3 weeks from the explosion. As the ejecta expand, the continuum becomes less prominent and more features emerge. These lines characterize the spectra taken around peak, shown in the central panel. Here we can see the SNe exhibit different line broadening, corresponding to different expansion velocities.

The late-time spectra shown in the bottom panel highlight the transition to the nebular phase, with a faint continuum and strong emission lines, in particular H$\alpha$ and \ion{Ca}{ii}.

We measured the expansion velocities of our SNe from the P-Cygni minima of different lines at all available epochs, and display the results in Fig.~\ref{velocity}. Balmer lines, \ion{Fe}{ii}~$\lambda$5169, \ion{Na}{i}~D, \ion{Ba}{ii}~$\lambda$6142,  and \ion{Ca}{ii}~$\lambda$8662 are analyzed. 

Starting with the \ion{Fe}{ii}~$\lambda$5169 line velocities, which better represent the photospheric velocities \citep{dessart05}, we can see how our SNe are faster than SN~1987A and SN~2009E, and similar to the more energetic SNe~1998A, 2006V and 2006au. In particular, SNe~2004ek and 2004em exhibit the fastest photospheres. SN~2005ci is the slowest in our sample.  These results are confirmed when we look at the \ion{Ba}{ii} lines, another good indicator of the photospheric velocity.

Looking at the bright H$\alpha$ line, we observe that 
PTF12kso and PTF12gcx have larger velocities compared to the other SNe, which show similar expansion velocities as did SN~1987A. This is also observed in H$\beta$, but the difference is less significant.

\ion{Na}{i}~D and especially the \ion{Ca}{ii} lines show similar velocities in all of the SNe.

\section{Bolometric properties}
\label{sec:boloprop}

 \begin{figure}
 \centering
\includegraphics[width=9cm,angle=0]{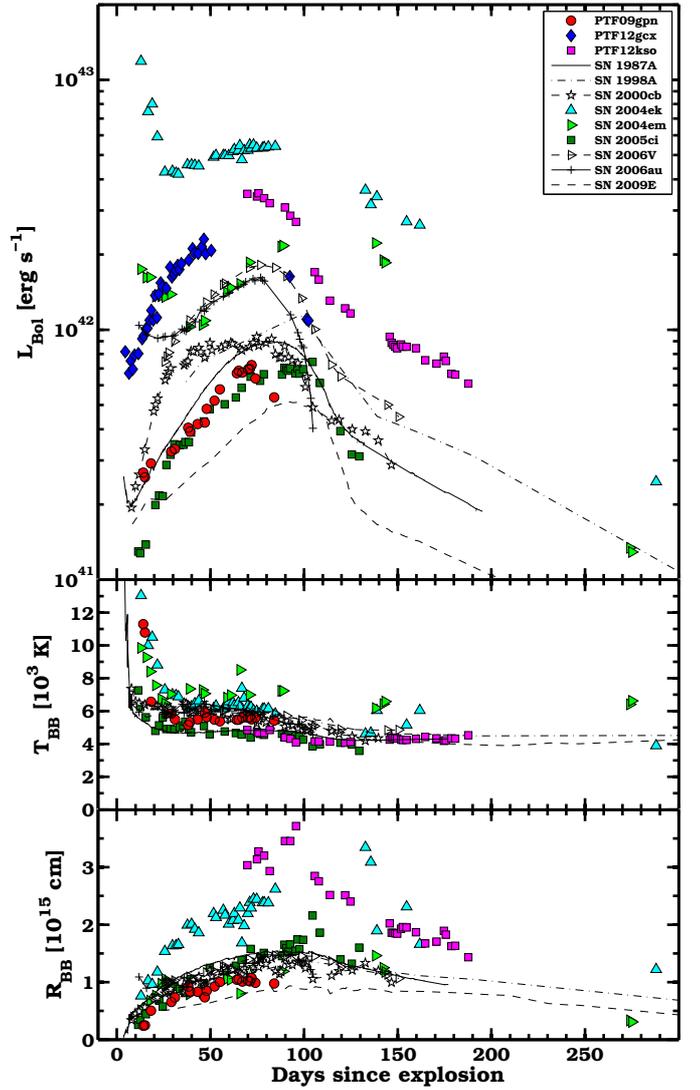}
  \caption{Bolometric luminosity ($L_{\rm Bol}$), radius ($R_{\rm BB}$), and temperature ($T_{\rm BB}$) evolution from the BB fit and integration of the SEDs of our PTF/CCCP SNe (colored symbols) and other similar events in the literature (black lines and symbols). \label{bolofig}}
 \end{figure}

For the CCCP and PTF SNe (except for PTF12gcx), we constructed the SEDs after having interpolated the $BVRI/Bgri$ light curves to the same epochs and converted the extinction-corrected magnitudes into fluxes at the effective wavelength of each filter. The bolometric light curves
were computed following the prescriptions by \citet{lyman14}, to account for the missing near-infrared (NIR) and UV flux: (a) 
the optical flux was obtained as the trapezoidal integral of the $BVRI/Bgri$ fluxes.  (b) The UV flux was computed as the integral of the black-body (BB) fit to the entire SED blueward of the effective $B$-band wavelength during the early cooling phase, and as the trapezoidal integral blueward of the effective $B$-band wavelength with $F_{\lambda}(\lambda \leq 2000$~\AA) = 0 
at later epochs. The cooling phase coincides with the epochs when the early $B$-band light curve is observed declining.
(c) the NIR flux was obtained as the integral of the best BB fit in the region redward of the effective $I/i$-band wavelength. The BB was fit to the entire SED during the cooling phase, and only to $VRI/gri(z)$ at later epochs, as the $B$-band flux is potentially affected by line blanketing. 
The sum of optical, UV, and NIR fluxes was then multiplied by $4\pi D^2$, with $D$ being the distance of the SN, to obtain the bolometric luminosity ($L_{\rm Bol}$). This is shown in Fig.~\ref{bolofig} (top panel). 
In the case of SN 2005ci, we verified that obtaining the bolometric light curve by simply integrating the best BB fit to the entire ($BVRI$) SED at all epochs would not have changed the result by much, with an average difference in luminosity of $\sim2$\%. For this reason, when the $B$ band was missing at late epochs (e.g., SN~2004em), we adopted the integral of the BB fit to $VRI$ as the bolometric flux. For PTF12gcx, where we only have a well-sampled $r$-band light curve, the bolometric light curve was obtained using the bolometric corrections from \citet{lyman14} assuming $g-r$ equal to that of SN~1987A (see Fig.~\ref{color}).

With the BB fit to the SED we also determined the BB radius ($R_{\rm BB}$) and the temperature ($T_{\rm BB}$), which are shown in the central and bottom panels of Fig.~\ref{bolofig}. Again, fitting only $VRI/gri$ or the entire SED does not have a large impact on the final results, with $T_{\rm BB}$ of SN~2005ci differing by 3\% on average and by 12\% at most. The radius differs by 6\% on average and by 27\% at most. 

With the same method outlined for CCCP and PTF SNe, we also derived $L_{\rm Bol}$, $T_{\rm BB}$, and $R_{\rm BB}$ for other well-observed SN~1987A-like SNe from the literature (data from \citealp{pastorello05,kleiser11,taddia12,pastorello12}). The results are shown in Fig.~\ref{bolofig}, where the top panel confirms that among the brightest long-rising SNe~II we find SN~2004ek, PTF12kso, and SN~2004em. The faintest are SN~2009E, PTF09gpn, and SN 2005ci. Long rising SNe~II from the literature (black lines and symbols) are on average fainter than the SNe in our sample (colored symbols), with SNe~2006V and 2006au being the most luminous. At late epochs, all of the light curves show a decline rate identical to that of SN~1987A, which is consistent with the $^{56}$Co decay rate, with the exception of SNe~2004ek and 2004em (which decline faster). SNe~2004ek, 2004em, 2005ci, and PTF09gpn have a decreasing luminosity and a decreasing temperature (central panel of Fig.~\ref{bolofig}) at early epochs. 
The temperature profiles are rather similar for all the SNe, 
with an early cooling (up to $\sim20$~d) followed by an almost constant temperature between $\sim4000$ and $\sim7000$~K depending on the object. The BB radii (bottom panel of Fig.~\ref{bolofig}) are on the order of $10^{15}$~cm,  with PTF12kso and SN~2004ek exhibiting the largest values. After a fast expansion, they peak around the epoch of maximum luminosity and then slowly decrease. 
In summary, our SN sample of long-rising SNe~II increases the spread of the observed bolometric properties and exhibits large variety.

\section{Modeling}
\label{sec:models}

\subsection{$^{56}$Ni mass from the linear decaying tail of the  bolometric light curve}
\label{sec:model_tail}
We can use the late part of the bolometric light curves to estimate the amount of $^{56}$Ni ejected in the SN. The light-curve decline at this phase is linear in magnitude and $\sim1$~mag every 100 days (the decay rate of $^{56}$Co). As the decay times of $^{56}$Ni and $^{56}$Co ($\tau_{^{56}{\rm Ni}}$ and $\tau_{^{56}{\rm Co}}$) are known, as well as their specific energy generation rates ($\epsilon_{^{56}{\rm Ni}}$ and $\epsilon_{^{56}{\rm Co}}$), it is possible to constrain the $^{56}$Ni mass ($M_{^{56}{\rm Ni}}$) by fitting the following expression to the bolometric light curve:

\begin{equation}
\begin{split}
L_{\rm Bol}=M_{^{56}{\rm Ni}}[\epsilon_{^{56}{\rm Ni}}{\rm exp}(-t/\tau_{^{56}{\rm Ni}}) +
\epsilon_{^{56}{\rm Co}}\tau_{^{56}{\rm Co}}/(\tau_{^{56}{\rm Co}}-\\\tau_{^{56}{\rm Ni}})({\rm exp}(-t/\tau_{^{56}{\rm Co}})-{\rm exp}(-t/\tau_{^{56}{\rm Ni}}))].
\end{split}
\end{equation}

For PTF12kso, SN~2004em, and SN~2005ci, we can fit at least two epochs on the linear decline, while SN~2004ek was observed only once at late epochs. For PTF09gpn and PTF12gcx, we can set only an upper limit on the $^{56}$Ni mass, as our last bolometric epochs coincide with the beginning of the light-curve tails. 

For PTF09gpn we also have $r$-band measurements at very late epochs. If we compare the absolute magnitude at these epochs with that of SN~2009E, the two SNe turn out to have similar $^{56}$Ni masses. \citet{pastorello12} estimated a  $^{56}$Ni mass of 0.04~M$_{\odot}$ from the quasi-bolometric light curve of SN~2009E, which we adopt as the best value for PTF09gpn.

The objects with the largest and smallest $^{56}$Ni masses are PTF12kso (0.23~M$_{\odot}$) and PTF09gpn (0.04~M$_{\odot}$), respectively. All of the results are reported in Table~\ref{tab:param}.

The uncertainties in these values are mainly affected by the uncertainty in the distance and in the explosion date. There could also be uncertainty related to the host-galaxy extinction. Therefore, we estimate the total error in the $^{56}$Ni mass to be at least 10\%.

\begin{figure}
 \centering
\includegraphics[width=9cm,angle=0]{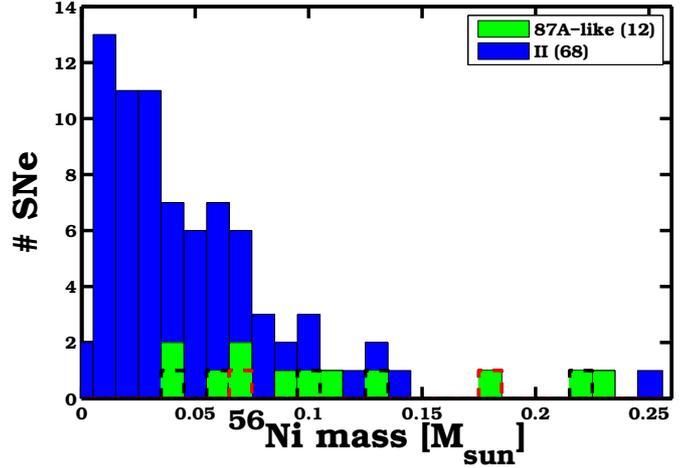}
  \caption{$^{56}$Ni mass distribution for SNe~II (IIP and IIL, estimates from \citealp{hamuy03}, \citealp{inserra13}, \citealp{anderson14}, and \citealp{rubin15}) in blue, and our SN~1987A-like SNe in green. $^{56}$Ni mass estimates that are based on at most two points on the radioactive tail are highlighted by dashed edges (red lines indicate upper limits, black lines mark lower limits).\label{56Nihisto}}
 \end{figure}

Our sample of long-rising SNe~II extends the range of $^{56}$Ni masses inferred for SN~1987A-like SNe (see Table 5 of \citealp{pastorello12}), with PTF12kso being the most $^{56}$Ni rich, about four times more than for SN~1987A. When compared to the $^{56}$Ni mass distribution of normal SNe~II from \citet{hamuy03}, \citet{inserra13}, \citet{anderson14}, and \citet{rubin15}, it appears that long-rising SNe~II produce more $^{56}$Ni on average (see Fig.~\ref{56Nihisto}). However, finding SN~1987A-like SNe with low $^{56}$Ni masses is more difficult than discovering SNe~IIP with the same $^{56}$Ni masses,
as their lower cooling-envelope luminosity at early epochs makes them fainter on the rise. Even accounting for this bias, long-rising SNe~II seem to have a different distribution of $^{56}$Ni masses as compared to SNe~IIP. A likely explanation for this result is that the progenitors of SN~1987A-like events are on average more massive than those of SNe~IIP, and therefore they also produce more $^{56}$Ni, as these two quantities show a correlation (see, e.g., \citealp{utrobin11}). 
We also notice that for seven SN~1987A-like events the $^{56}$Ni mass is derived from at most two points on the radioactive tail. This might introduce a bias in the $^{56}$Ni mass estimates if gamma-rays are escaping.

Based on the $^{56}$Ni mass estimates from the tail of the bolometric light curves of our SNe and other SN~1987A-like SNe, we can conclude that the amount of synthesized $^{56}$Ni in long-rising SNe~II can vary widely, between 0.04~M$_{\odot}$ and 0.23~M$_{\odot}$. The lower value
is consistent with those of normal SNe~IIP such as SN~1999em, the higher one with those of luminous SNe~IIP (e.g., SN~1992af; \citealp{nadyozhin03}) or moderately faint SNe~Ibc \citep[e.g.,][]{taddia15}. In Fig.~\ref{param} we can see how our sample (colored symbols) extends the range of $^{56}$Ni masses up to 0.23~M$_{\odot}$ compared to the previously known long-rising SNe~II.

\subsection{Explosion energy and ejecta mass from scaling relations}
\label{sec:explomass}

 \begin{figure}
 \centering
\includegraphics[width=9cm,angle=0]{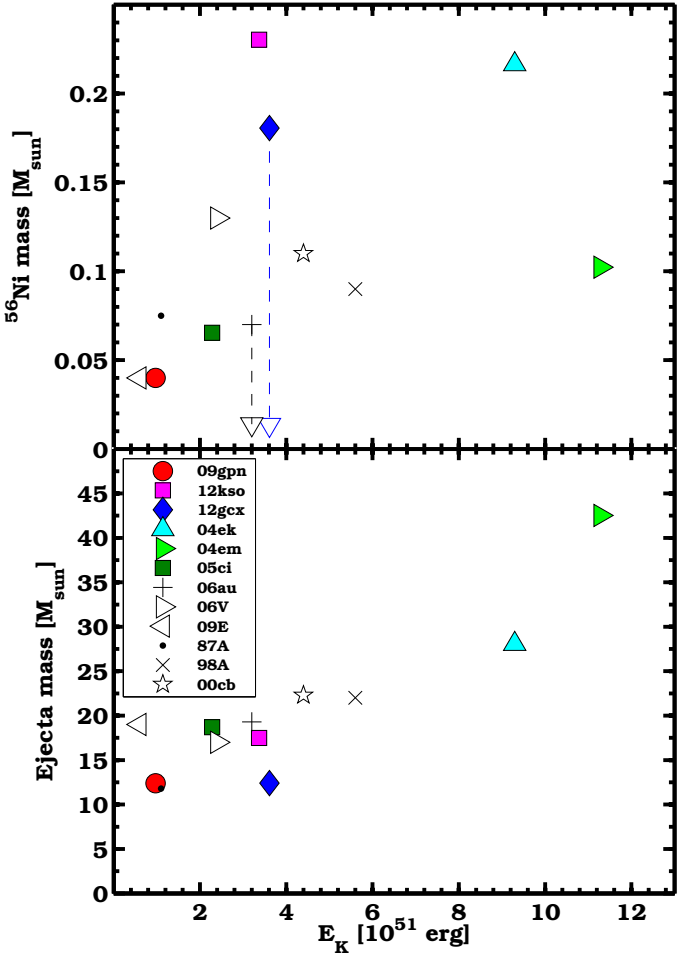}
  \caption{$^{56}$Ni mass and ejecta mass versus explosion energy for our PTF and CCCP SNe and for other well-observed SN~1987A-like SNe (\citealp{taddia12}). In the top panel, dashed vertical arrows are limits on the $^{56}$Ni mass.\label{param}}
 \end{figure}

 \begin{figure}
 \centering
\includegraphics[width=9cm,angle=0]{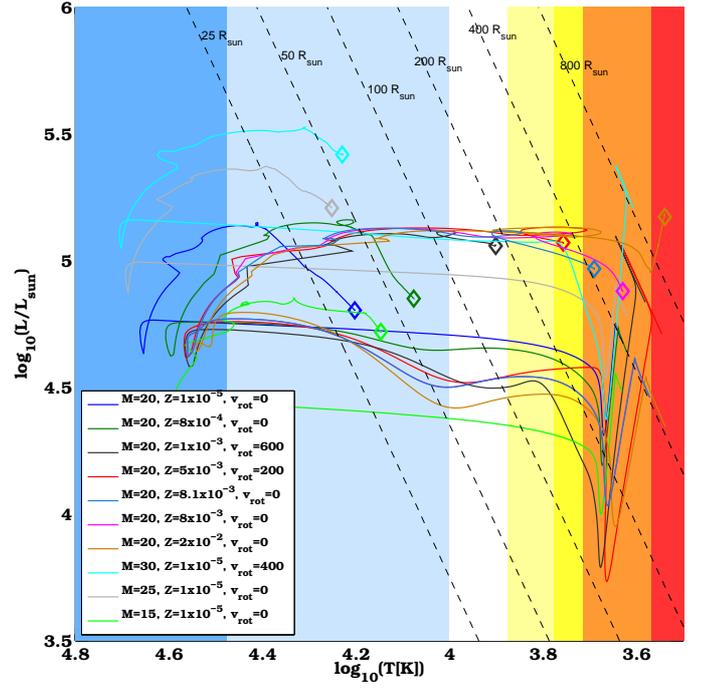}
  \caption{HR diagram for a series of stellar models obtained with MESA. Different metallicities and rotation velocities (see the legend) are used to create a sequence of SN progenitors with progressively larger radii (see the diamond symbols as compared to the dashed isoradius lines). \label{HR}}
 \end{figure}

In order to estimate the explosion energy ($E$) and the ejecta mass ($M_{\rm ej}$), we need to combine the information about the bolometric rise time and the photospheric velocity, as done by \citet{taddia12}.
For radioactively powered SNe, the diffusion time given by \citet{arnett79} is $t_d =(\kappa~M_{\rm ej}/v_{\rm ph})^{1/2}$, and $E\propto M_{\rm ej}~v_{\rm ph}^2$.
We assume the same mean opacity $\kappa$ for all our SNe, and adopt the \ion{Fe}{ii}~$\lambda$5169 velocity (see Fig.~\ref{velocity}) around peak as $v_{\rm ph}$ (for PTF09gpn we adopted the velocity of SN~1987A) and the bolometric rise time as $t_d$. Using values from SN~1987A, $E$(87A)~$= 1.1\times10^{51}$~erg, $M_{\rm ej}$(87A) = 14~M$_{\odot}$ \citep{blinnikov00}, $t_d$(87A) = 84~d, and $v_{\rm ph}$(87A) = 2200 km~s$^{-1}$, we obtain estimates of $E$ and $M_{\rm ej}$ for the other SNe by scaling. These are listed in Table~\ref{tab:param}.
 
The long rise times and fast expansion velocities of SNe~2004ek and 2004em imply large ejecta masses ($\sim30$--40 M$_{\odot}$) and energies ($\sim10^{52}$~erg). Lower values characterize the other SNe, which however show larger ejecta masses and energies than SN~1987A, with the exception of PTF09gpn. The SNe observed in our sample are found to be more energetic and have larger masses (on average) than the old sample, as shown in Fig.~\ref{param}.
It appears from the same figure that the ejecta mass and the explosion energy are correlated, as well as the $^{56}$Ni mass and the explosion energy. The estimates of $E$ and $M_{\rm ej}$ are based on simple scaling relations and therefore must be taken with caution. We will show in Sec.~\ref{sec:hydro} that the rise times of hydrodynamical models show a more complicated dependence on the progenitor parameters as compared to what we used here, including a dependence on the degree of $^{56}$Ni mixing and on the $^{56}$Ni mass.

\subsection{Hydrodynamical modeling}
\label{sec:hydro}

 \begin{figure*}
 \centering
 \includegraphics[width=6cm,angle=0]{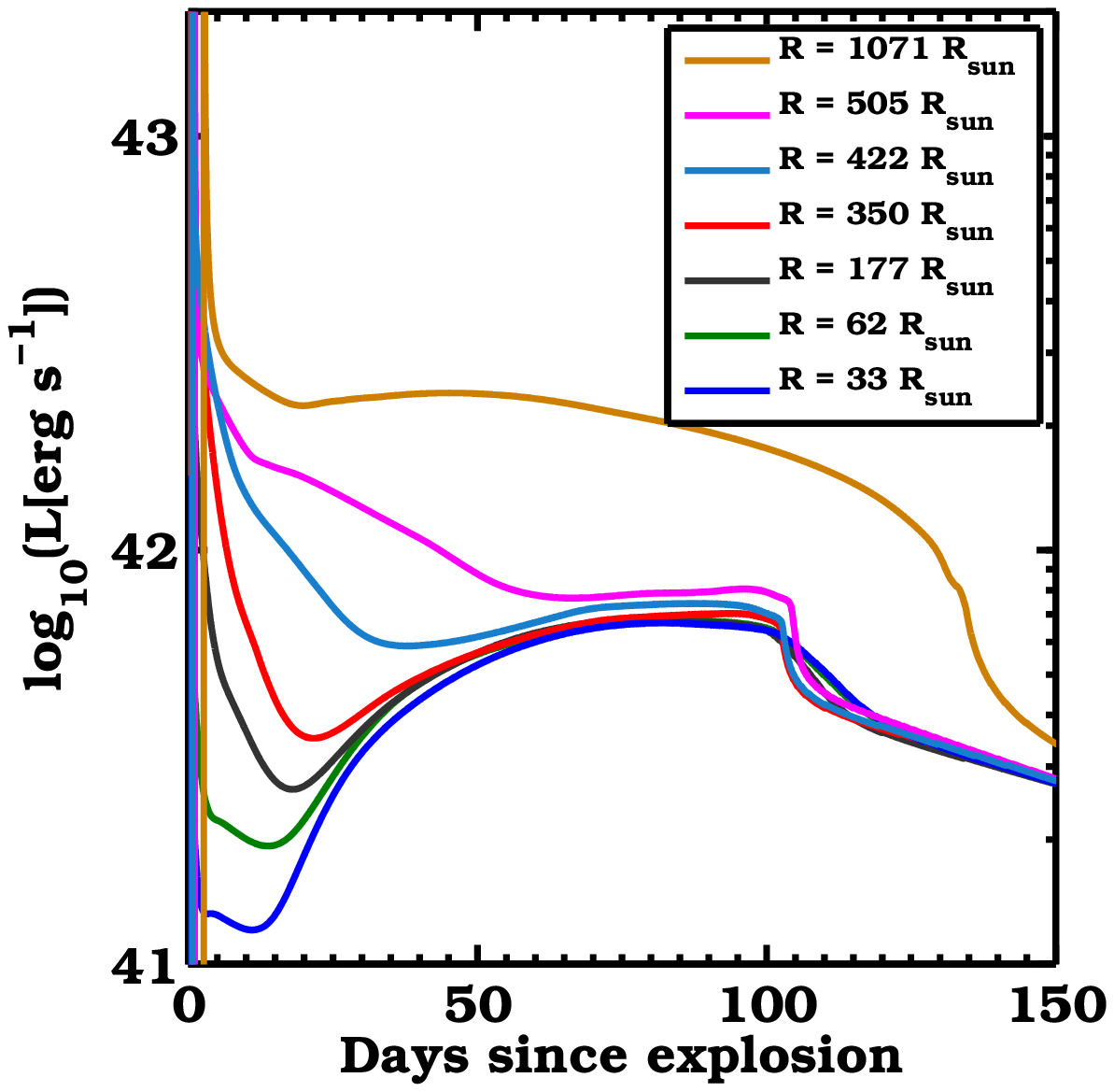}
\includegraphics[width=6cm,angle=0]{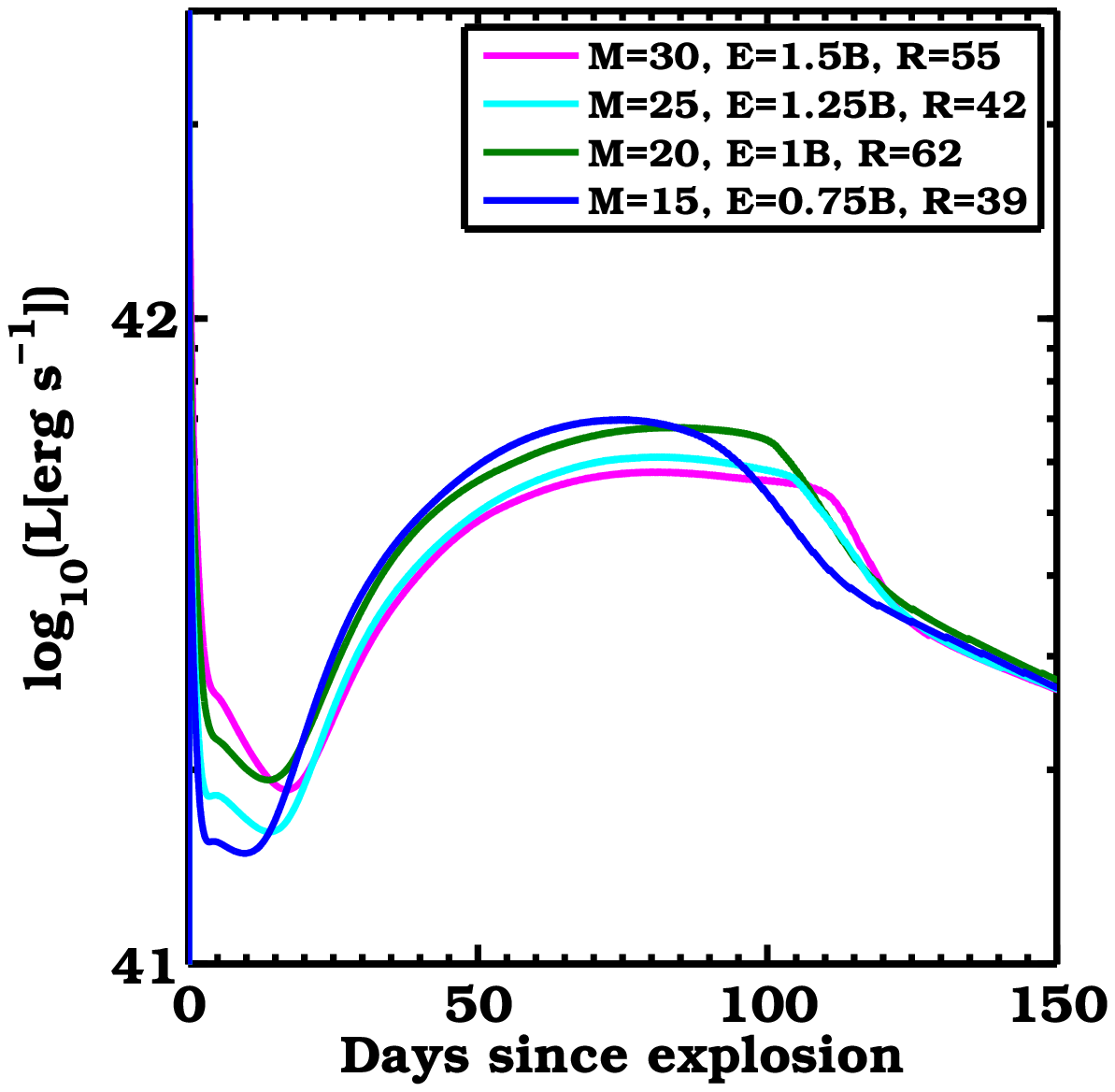}
\includegraphics[width=6cm,angle=0]{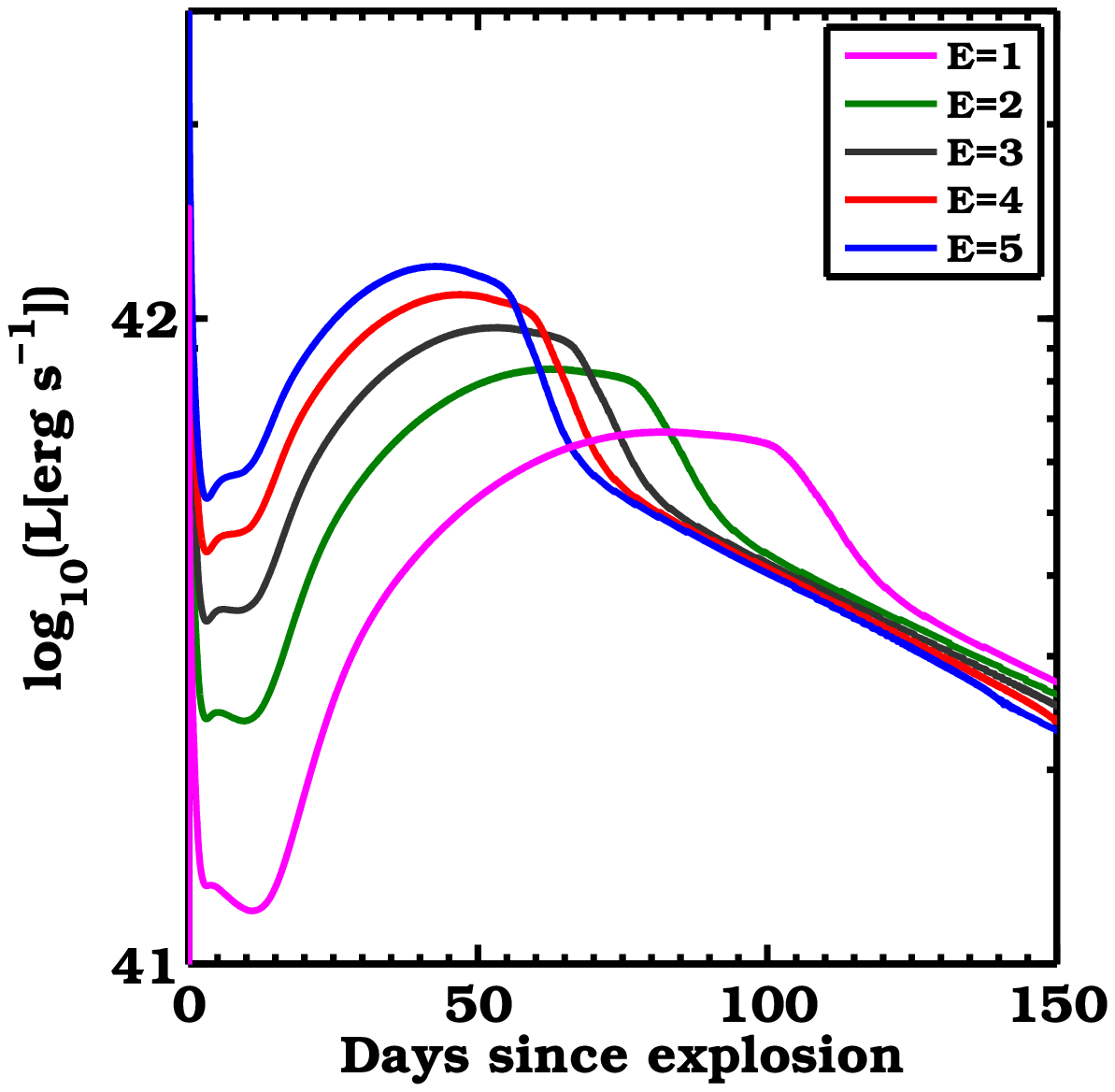}
\includegraphics[width=6cm,angle=0]{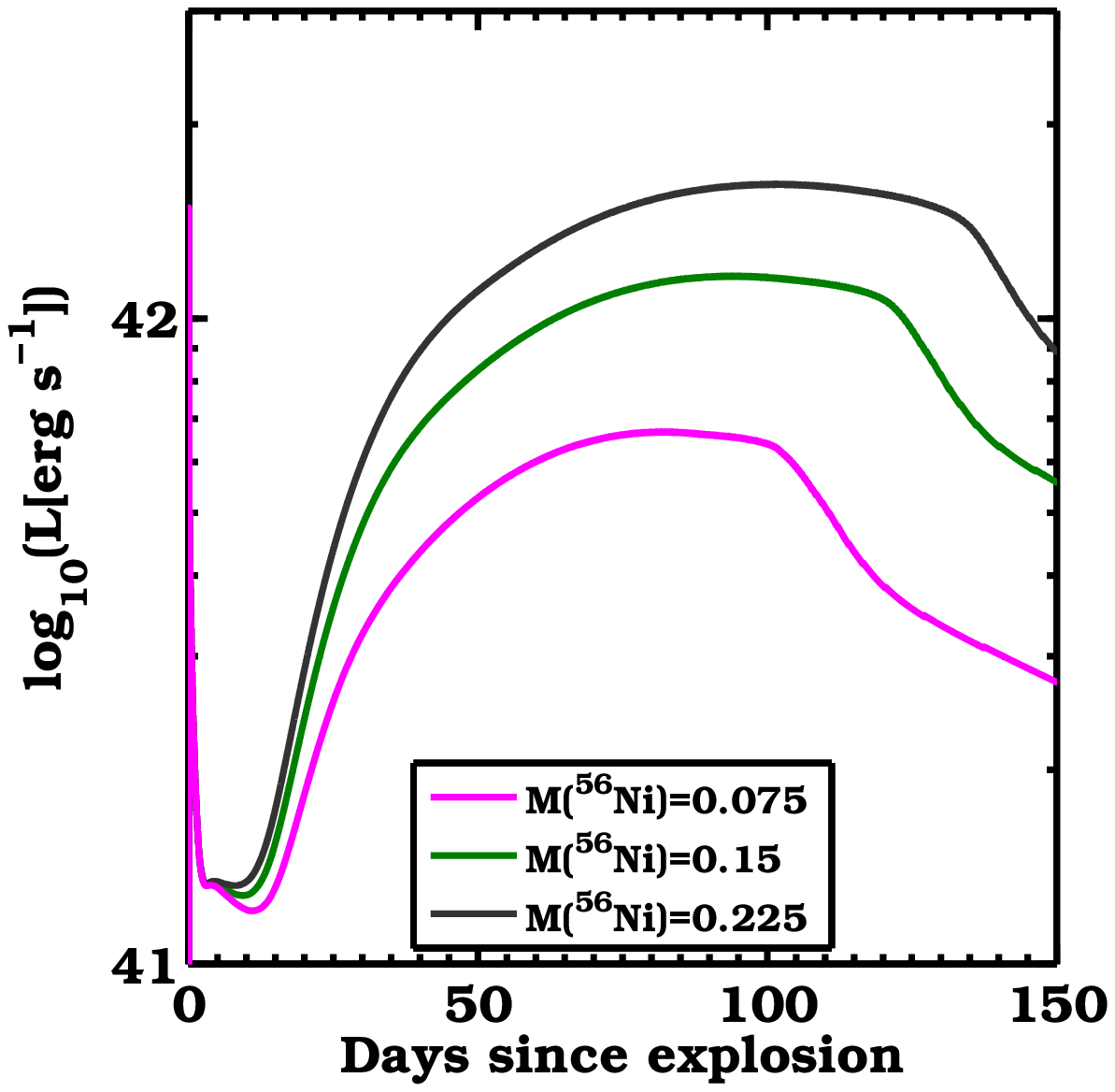}
\includegraphics[width=6cm,angle=0]{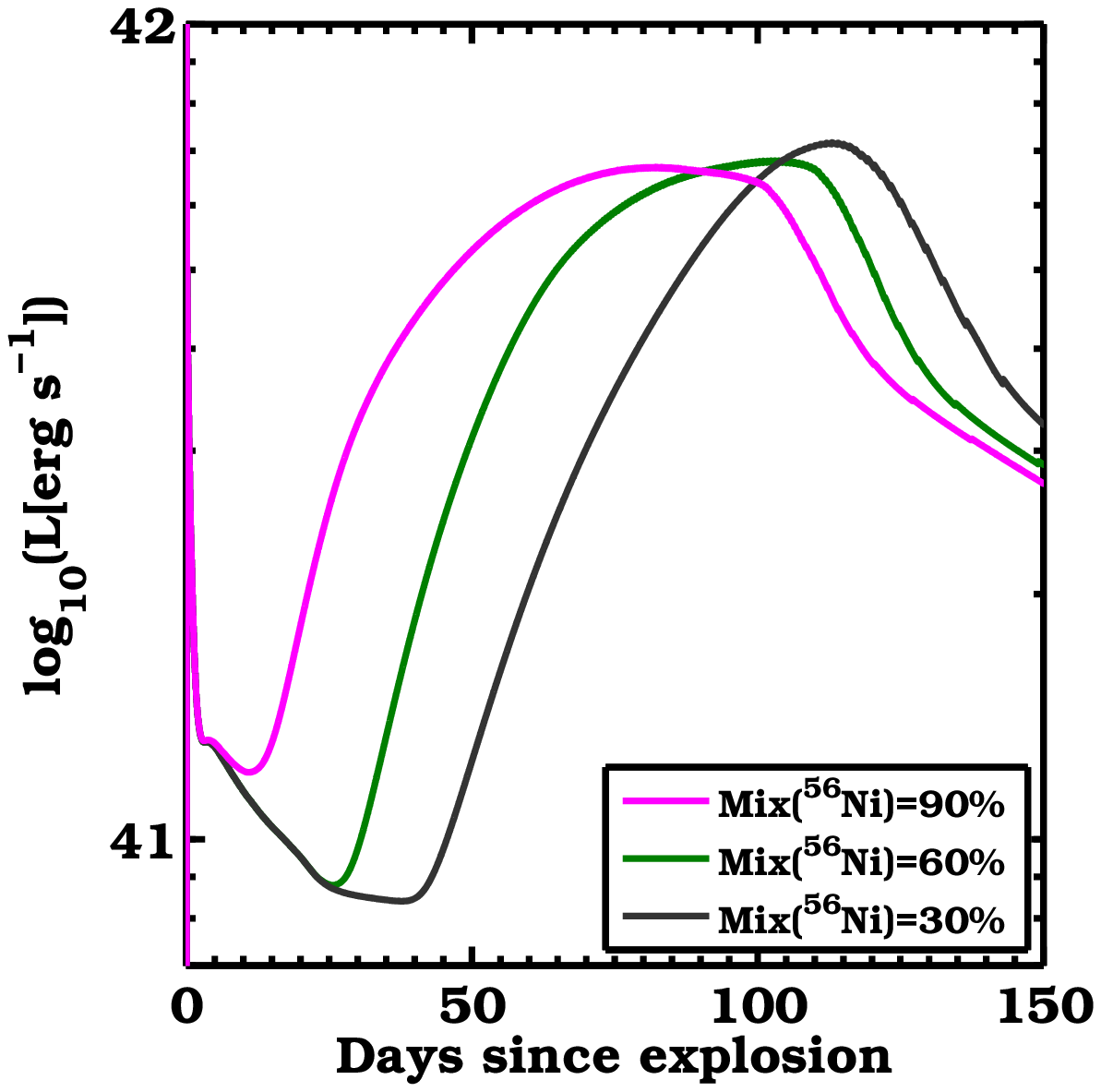}
\includegraphics[width=6cm,angle=0]{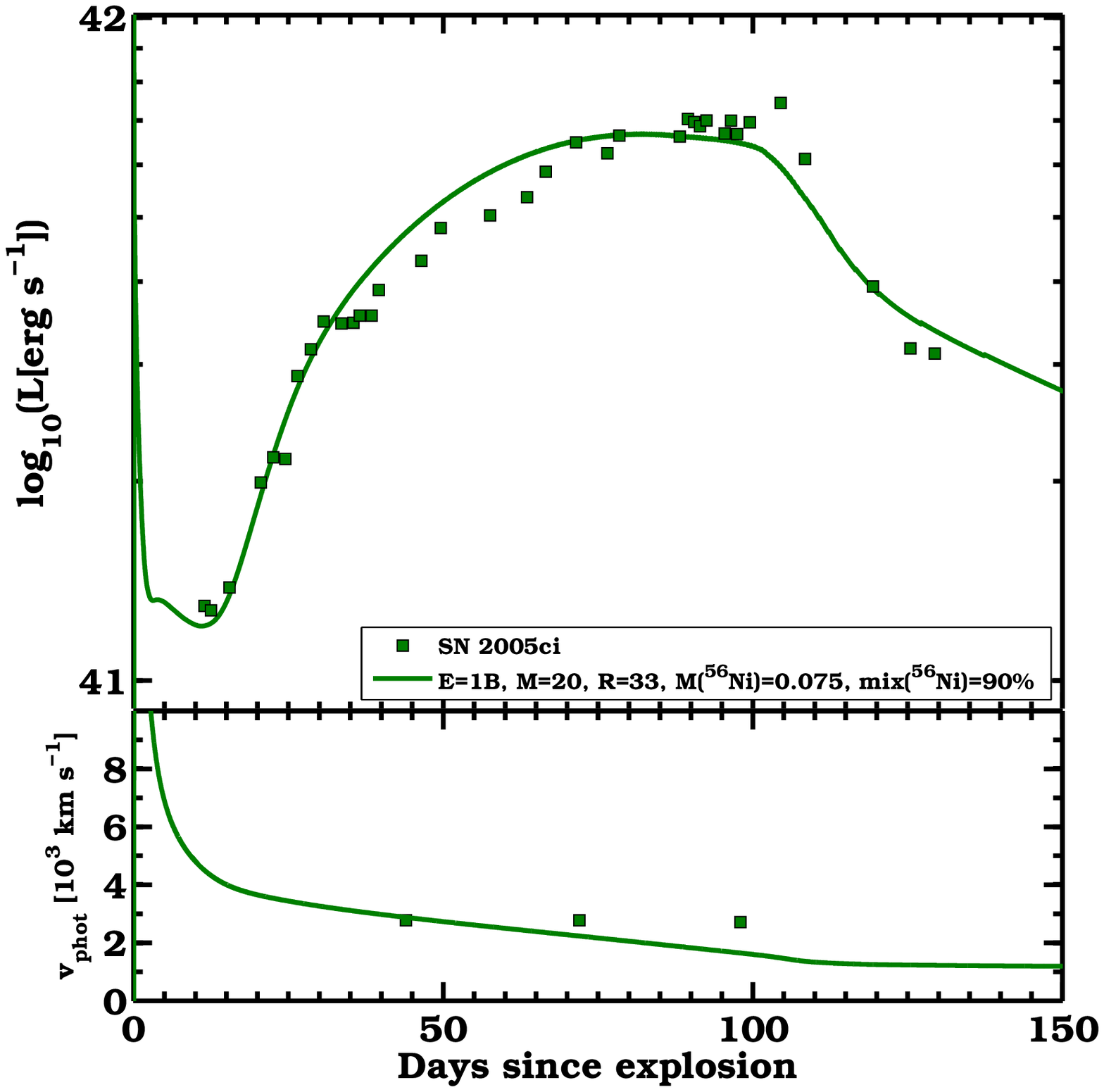}
\caption{SNEC bolometric light curves of the SNe from the progenitor stars obtained from the stellar models shown in Fig.~\ref{HR}. We tested how the light curves change depending on the different progenitor and explosion parameters. In the bottom-right panel we show one of our models fitting the bolometric light curve and the velocity profile of SN~2005ci. \label{hydro}}
 \end{figure*}

In Figs.~\ref{rise} and \ref{bolofig} we showed how the early-time shape
of our long-rising SNe~II is different for the different events. It is known, also from analytic modeling (see, e.g., \citealp{chevalier08}), that SN progenitor radii influence the early SN emission, with the more luminous early light curves corresponding to the SNe with larger progenitors.

Here we aim to demonstrate that we need progenitor radii between that of a RSG ($\gtrsim 500$~R$_{\odot}$, typical of SN~IIP progenitors) and that of a compact BSG ($\lesssim 100$~R$_{\odot}$, progenitor of a canonical SN~1987A-like SN) to explain the early light-curve dispersion of these objects. To do this, we made use of open-source codes to evolve massive and hydrogen-rich progenitor stars and explode them, producing bolometric light curves that were compared to our observations. 

We used the Modules for Experiments in Stellar Astrophysics (MESA; \citealp{paxton11}) to evolve $M_{\rm ZAMS}=15$, 20, 25, and 30~M$_{\odot}$ stars with different metallicities (from $Z=10^{-5}$ to $Z=5\times10^{-3}$) and rotation velocities ($v_{\rm rot}=0$--600~km~s$^{-1}$), until they end their lives at different positions in the HR diagram. We thus constructed a sequence of models with stellar radii from $\sim33$ to $\sim1071$~R$_{\odot}$ for stars with $M_{\rm ZAMS}=20$~M$_{\odot}$. We decided to produce stars with compact radii for the other initial masses (15, 25, and 30~M$_{\odot}$). We prohibited mass loss to preserve a large H envelope and did not include overshooting in our computations. The HR diagram including our models is shown in Fig.~\ref{HR}.
Low metallicity and fast rotation favor the production of BSG stars in the HR diagram, whereas high metallicity and low rotation velocity produce RSG stars as SN progenitors \citep[e.g.,][]{pod92}. 

The aim of this exercise was to construct a sequence of similar $M_{\rm ZAMS}$ stars of different final radii to use as input for the SuperNova Explosion Code (SNEC; \citealp{morozova15}), in order to produce bolometric light curves and analyze the effect of the different parameters on the SN light curves. We produced a set of hydrodynamical models with a range of progenitor radii ($R$), explosion energies ($E$), final masses ($M~=~M_{\rm ZAMS}=M_{\rm ej}+1.4$~M$_{\odot}$, as we inhibited the mass loss and consider the formation of a central compact object of 1.4~M$_{\odot}$), $^{56}$Ni masses ($M_{\rm{^{56}}Ni}$), and degrees of $^{56}$Ni mixing (${\rm mix}_{\rm{^{56}}{\rm Ni}}$). The degree of $^{56}$Ni mixing is expressed as a fraction of the final mass, and the $^{56}$Ni is uniformly distributed in the region where it is present.
For example, a ${\rm mix}_{\rm{^{56}}{\rm Ni}}$ of 0.9 implies that the $^{56}$Ni mass is uniformly distributed up to the radius that includes 90\% of the final progenitor mass.

If we keep the same mass ($M=20$~M$_{\odot}$), the
same $^{56}$Ni mass (0.075~M$_{\odot}$), and the same explosion energy ($10^{51}$~erg = 1~B), we can see in Fig.~\ref{hydro} (top-left panel) that for compact radii, the light curve is very similar to that of SN~1987A. For larger radii, the luminosity of the early part (before the light-curve rise) becomes larger, until the SN light curve reaches that of a normal SN~IIP.  This occurs for radii~$\gtrsim420$~R$_{\odot}$ for our given parameter set. The same result was found by \citet{young04}. 
For large radii, the decline in luminosity is also slower than for small radii. 

In  Fig.~\ref{hydro} (top-central panel) we show that, for a range of final masses from 15 to 30~M$_{\odot}$, if we keep the same ratio between $E$ and $M$, as well as approximately the same $R$, then the rise time and the early luminosity of the light curves are generically very similar. 

The top-right panel of Fig.~\ref{hydro} shows that for the same $M=20$~M$_{\odot}$, $M_{\rm ^{56}{\rm Ni}}=0.075$~M$_{\odot}$, and $R=33$~R$_{\odot}$, we obtain a brighter and earlier peak for larger energies (see also \citealp{young04}).

For larger $^{56}$Ni mass (keeping the same $M=20$~M$_{\odot}$, $R=33$~R$_{\odot}$, and $E=10^{51}$~erg = 1~B), the rise time is longer; for larger $^{56}$Ni mixing it is shorter (bottom panels of Fig.~\ref{hydro}). The effects of $^{56}$Ni mixing on the peak was also investigated by \citet{young04}.

As we could not produce all the models needed to scan the entire space of the progenitor parameters, we made use of our set of models to derive scaling relations for the time of the peak, luminosity at $+$10~d, and time of the rise. This allowed us to constrain the degree of $^{56}$Ni mixing, $E/M$, and the progenitor radius for our SNe. 

We measured the peak epochs of these models and we investigated how they change depending on each progenitor parameter. We found that
\begin{equation}
t_{\rm peak}\propto (E/M)^{-0.4}M_{^{56}{\rm Ni}}^{0.2}{\rm mix}_{^{56}\rm Ni}^{-0.25}
\label{eq:trise}
\end{equation}
\noindent
The luminosity of the early part (measured at 10~d after explosion) scales approximately like
\begin{equation}
L_{+10~{\rm d}}\propto (E/M)^{0.9} R^{0.7}
\label{eq:earlylum}
\end{equation}

\noindent
for $R\lesssim350$~R$_{\odot}$.
\noindent
The time at which the light curve starts to rise ($t_{\rm sr}$) after the early cooling depends mainly on the degree of $^{56}$Ni mixing as
\begin{equation}
t_{\rm sr}\propto{\rm mix}_{^{56}{\rm Ni}}^{-0.9}.
\label{eq:trs_mix}
\end{equation}

We also obtained estimates of the progenitor radius from these relations 
by scaling with the parameters of the hydrodynamical model of SN~2005ci, shown in the bottom-left panel of Fig.~\ref{hydro}. The bolometric light curve and the photospheric velocity of SN~2005ci are well reproduced by a model with $M=20$~M$_{\odot}$, $R=33$~R$_{\odot}$, $E=1$~B, $M_{\rm ^{56}{\rm Ni}}=0.075$~M$_{\odot}$, and  ${\rm mix}_{^{56}{\rm Ni}}=90\%$. The light curve of SN~2005ci is rather similar to that of SN~2000cb \citep{kleiser11}, which was modeled by \citet{utrobin11}, who found a similar mass ($M_{\rm env}=22.3$~M$_{\odot}$) and progenitor radius ($R=35$~R$_{\odot}$), but higher energy ($E=4.4$~B).
 
For each of our SNe we make use of the $M_{\rm ^{56}Ni}$ measured in Sec.~\ref{sec:model_tail}. We measure the epoch of the light-curve rise ($t_{\rm sr}$) to assess the degree of ${\rm mix}_{^{56}\rm Ni}$ via Eq.~\ref{eq:trs_mix} and scaling to SN~2005ci. By measuring the rise time we make use of Eq.~\ref{eq:trise} to estimate $E/M$ by scaling to SN~2005ci. Then we can use Eq.~\ref{eq:earlylum} and the obtained values of $E/M$ to estimate the radius by scaling to SN~2005ci.
 
In Table~\ref{tab:param} we report the parameter estimates obtained from these scaling relations. We obtain radii of about 320--350~R$_{\odot}$ for SN~2004em and PTF12gcx, of $\sim120$~R$_{\odot}$ for PTF09gpn, and of a few thousands R$_{\odot}$ for SN~2004ek. Therefore, we suggest that we have observed H-rich SNe with progenitor radii that are intermediate between those of RSGs and BSGs, implying that H-rich stars in the A to K spectral classes can also explode and become SNe~II. Moreover, we found that there is a large variety in the degree of $^{56}$Ni mixing, ranging from 25\% to 100\% of the final mass.

\section{Discussion}
\label{sec:discussion}

In this work we have added three new and well-observed long-rising SNe~II from PTF and analyzed three more SNe from CCCP. These objects are an important addition to this SN class as long-rising SNe~II are extremely rare events. \citet{pastorello12} estimated that SN~1987A-like SNe make up 1--3\% of CC~SNe based on the data from the Asiago Supernova Catalogue. Also, \citet{kleiser11} derived a similar value based on the data from the Lick Observatory SN Search \citep{filippenko01, leaman11, li11a, li11b}. If we consider the CCCP SN program, which observed nearby CC~SNe, three SN~1987A-like SNe (from a total of 62 CC SNe) were followed, suggesting a fraction of 5\%. The PTF/iPTF survey is better suited to estimate the rate as it is not a targeted search. Within a distance modulus of $\mu \approx 35$~mag, we found only PTF09gpn out of a total of 108 CC~SNe observed within the same volume. This suggests a fraction of 1\%, consistent with the value suggested by \citet{pastorello12}.

The progenitor parameters suggest stars with a radius smaller than that of a typical RSG, which is $\gtrsim500$~R$_{\odot}$ (with the exception of SN~2004ek).


We estimated the progenitor radii by using hydrodynamical models.
In particular, we obtained a good fit to the bolometric luminosity and photospheric velocity of SN~2005ci. Thereafter, using the hydrodynamical models, we derived scaling relations for the peak epoch, the early luminosity, and the epoch of the light-curve rise, which depends on $R$, $E/M$, $M_{^{56}{\rm Ni}}$, and ${\rm mix}_{^{56}{\rm Ni}}$.
We confirmed that the early luminosity depends on the ratio $E/M$ and on the $R$ with similar power-law indexes. However, compared to the scaling relation based on the diffusion time from the model of Arnett, we found important differences. For instance, we found that $t_{\rm peak}$ depends on $M/E$ instead of $M^3/E$. This was also found by \citet{utrobin05} for his set of hydrodynamical models in the envelope-mass range 15--21~M$_{\odot}$. We also considered the dependence of $t_{\rm peak}$ on $M_{^{56}{\rm Ni}}$ and ${\rm mix}_{^{56}{\rm Ni}}$, which were previously ignored.

With the relations estimated from this series of hydrodynamical models, we obtained progenitor radii of $\sim30$~R$_{\odot}$ for SN~2005ci, 
$\sim120$~R$_{\odot}$ for PTF09gpn, and $\sim320$--350~R$_{\odot}$ for SN~2004em and PTF12gcx. 
Considering that the ejecta masses range from $\sim10$ to $\sim40$~M$_{\odot}$, the radii of these stars imply a range of progenitor stars from BSGs (SN~2005ci and PTF09gpn) to yellow supergiants (YSGs; e.g., SN~2004em and PTF12gcx), as can be seen in the HR diagram of \citet{taddia12} and in Fig.~\ref{HR}.


For SN~2004ek, we obtained a large progenitor radius of a few thousand~R$_{\odot}$, typical of a RSG. Its bolometric light curve shows a late-time peak, but the rise occurring after the initial luminosity drop is very shallow compared to that of SN~1987A, and the early-time light curve is very bright. 
We can interpret the light curve of SN~2004ek as the combination of radioactive emission from its large $^{56}$Ni mass (important only at peak and even more at later epochs) and cooling emission from its large envelope (dominating the early epochs). 


For two host galaxies of our PTF SNe (PTF09gpn and PTF12kso), we have measured very low metallicity ($Z\approx1/10$~Z$_{\odot}$), lower than the metallicity measured in the Small Magellanic Cloud (SMC). A low metallicity ($\lesssim1/3$~Z$_{\odot}$) can favor a blue solution in the HR diagram for a single star, with the star exploding as a BSG (see \citealp{pod92} and \citealp{weiss89}). In this case, the SN progenitors could have spent their entire lives in the BSG stage, and did not have a RSG phase such as that of SN~1987A. These two SNe confirm the trend of long-rising SNe~II at low metallicity found by \citet{taddia13}.

\section{Conclusions}
\label{sec:conclusion}

\begin{itemize}

\item[\textbullet]{We have added and analyzed six new long-rising SNe~II to this rare family, three from PTF and three from CCCP. These objects present similarities in the light curves, but also interesting differences, especially at early epochs.}

\item[\textbullet]{Our SN sample expands the range of $^{56}$Ni masses, explosion energies, and ejecta masses for the long-rising SNe~II.}

\item[\textbullet]{The radii and the large ejecta masses found for our SNe suggest a BSG to YSG origin for these events. Canonical SN~1987A-like SNe arise from compact BSGs (e.g., SN~2005ci), while SNe with more luminous early phases arise from stars with a radius of a few hundred R$_{\odot}$ (e.g., SN~2004em). We also found that SN~2004ek comes from an even more extended RSG progenitor star, but shows a long-rising light curve owing to the large amount of ejected $^{56}$Ni.}

\item[\textbullet]{We found PTF09gpn and PTF12kso at the lowest 
host metallicities ever estimated for long-rising SNe~II, confirming that low host metallicity is an important ingredient for producing peculiar SNe II, whose radii are typically shorter than those of normal SN~IIP progenitors.}

\end{itemize}

\begin{acknowledgements}
We thank the staffs of the various observatories (Palomar, Lick, Keck, etc.) where data for this study were obtained.
The Oskar Klein Centre is funded by the Swedish Research Council. 
We gratefully acknowledge the support from the Knut and Alice Wallenberg Foundation. A.G.-Y. is supported by the EU/FP7 via ERC grant No. 307260, the Quantum Universe I-Core program by the Israeli Committee for Planning and Budgeting and the ISF; by Minerva and ISF grants; by the Weizmann-UK ``making connections'' program; and by Kimmel and ARCHES awards. A.V.F.'s research is supported by the Christopher R. Redlich Fund, the 
TABASGO Foundation, and NSF grant AST-1211916.
 D.C.L. and S.F.A. acknowledge support from NSF grants AST-1009571 and AST-1210311,
under which part of this research (photometry data collected at MLO) was carried out. We
thank Joseph Fedrow, Alyssa Del Rosario, Chuck Horst, and David Jaimes for assistance
with the MLO observations. 
J.M.S. is supported by an NSF Astronomy and Astrophysics Postdoctoral Fellowship under award AST-1302771. 
This research has made use of the APASS database, located at
the AAVSO web site; funding for APASS has been provided by the Robert Martin Ayers
Sciences Fund. 
Research at Lick Observatory is partially supported by a generous gift from Google. 
Some of the data presented herein were obtained at the W. M. Keck Observatory, which is operated as a scientific partnership among the California Institute of Technology, the University of California, and NASA; the observatory was made possible by the generous financial support of the W. M. Keck Foundation. 
We acknowledge contributions to CCCP by S. B. Cenko, D. Fox, D. Sand, and A. Soderberg. We acknowledge M. Sullivan and K. Sharon for helping with the CCCP spectral observations. LANL participation in iPTF is supported by the US Department of Energy as part of the Laboratory Directed Research and Development program. 
\end{acknowledgements}

\bibliographystyle{aa}

\onecolumn

\clearpage
\begin{deluxetable}{lllccllccc}
\tabletypesize{\scriptsize}
\tablewidth{0pt}
\tablecaption{PTF and CCCP Sample of SN~1987A-like SNe.\label{tab:sample}}
\tablehead{
\colhead{SN} &
\colhead{$\alpha$(J2000)} &
\colhead{$\delta$(J2000)} &
\colhead{Galaxy} &
\colhead{Galaxy } &
\colhead{Redshift} &
\colhead{Distance} &
\colhead{$A_V^{\rm MW}$} &
\colhead{$A_V^{\rm host}$} &
\colhead{EW$(\rm \ion{Na}{i}~D)_{\rm h}$}   \\
\colhead{} &
\colhead{(hh:mm:ss)} &
\colhead{(dd:mm:ss)} &
\colhead{} &
\colhead{Type} &
\colhead{} &
\colhead{(Mpc)} &
\colhead{(mag)} &
\colhead{(mag)} &
\colhead{(\AA)}}
\startdata
2004ek   & 01:09:58.51 & $+$32:22:47.7 &  UGC~724     &   S       &    0.017275    &    75.3$\pm$2.1               &0.183         & 0.347$\pm$0.015              & 0.70$\pm$0.03 \\
2004em   &19:31:31.11 & $+$35:52:15.7  & IC~1303    &  SAc        &   0.014894    &   	59.6$\pm$1.7               & 0.298        & \ldots              &n.d.       \\
2005ci  & 14:34:44.88 & $+$48:40:19.8 & NGC~5682     &   SB(s)b       & 0.007581    &    36.3$\pm$13.8                &  0.089       & \ldots              &n.d.             \\
PTF09gpn   & 03:43:43.26 & $-$17:08:43.1 & Anon.     &   \ldots       &   0.0154    &    66.3$\pm$1.2                & 0.453        & \ldots              &n.d.              \\
PTF12gcx   & 15:44:17.32 & $+$09:57:43.1 & SDSS~J154417.02$+$095743.8     &  \ldots        &   0.045041     &      198.3$\pm$3.7         & 0.140        & \ldots              &n.d.       \\
        PTF12kso  & 04:36:10.89 & $-$05:04:53.3 & Anon.     &   \ldots       &   0.030    &        130.6$\pm$2.4            &  0.184        & \ldots              &n.d.               \\

\enddata
\tablecomments{MW extinctions are obtained via NED \citep{fitzpatrick99,schlafly11}. Redshifts are from NED except for PTF09gpn and PTF12kso, where we measured it from the narrow lines of their host-galaxy spectra. Distances are redshift-independent for the three CCCP SNe, and obtained from the redshift for the three PTF SNe (assuming WMAP5 cosmological parameters; \citealp{komatsu09}). PTF12kso is also known as SN~2012gg, CSS J043611.05-050452.1, and LSQ12fjm \citep{lsq12fjm}.}
\end{deluxetable}

\begin{deluxetable}{lll|cc|cc|l}
\tabletypesize{\scriptsize}
\tablewidth{0pt}
\tablecolumns{7}
\tablecaption{Discovery Data for the PTF and CCCP Sample of SN~1987A-like SNe.\label{tab:disco}}
\tablehead{
\colhead{SN} &
\colhead{Discovery\tablenotemark{a}} &
\colhead{Classification\tablenotemark{a}} &
\colhead{Last nondetection} &
\colhead{Limit} &
\colhead{Discovery} &
\colhead{Discovery} &
\colhead{Explosion date}  \\
\colhead{} &
\colhead{source} &
\colhead{source} &
\colhead{JD$-$2,450,000} &
\colhead{mag} &
\colhead{JD$-$2,450,000}&
\colhead{mag} &
\colhead{JD$-$2,450,000}}
\startdata
2004ek    & IAUC 8405   & CBET 86  &  3242.5   &  19.5 & 3258.4730  & 17.1   &    3250.5$\pm$8.0                \\
2004em    & CBET 82     & IAUC 8411 &   3214.5  &  19.0 & 3263.4680  & 17.5  &       3263.3$^{+0.168}_{-7.10}$             \\
2005ci    &    CBET 168    & CBET 169/\citet{kleiser11} &  3523.81  & 19.0   & 3523.86 &  20.0 &        3523.69$^{+0.17}_{-12.03}$       \\   
PTF09gpn  &      PTF \citep{arcavi10}    &    PTF \citep{arcavi10}   &     5127.91 (P48)    &  21.00 (P48)   & 5137.77 (P48)  &  20.30 (P48)  &    5137.23$^{+0.05}_{-0.05}$                 \\
PTF12gcx &       PTF  (ATel 4293)             &   PTF (ATel 4293)                 &    6085.39 (P48)           & 20.93 (P48)   &
 6085.84 (P48)  &   21.14 (P48) &       6085.72$^{+0.12}_{-0.40}$    \\
PTF12kso &        PTF / CBET 3296,1          &      PTF / CBET  3296.2         &    6153.00 (P48)          &  20.85 (P48,$g$)   & 6198.98 (P48) &  18.41 (P48)  &   6176.0$\pm$23.0        \\
\enddata
\tablenotetext{a}{References: \\
IAUC 8405: \citet{disco04ek}; \\
CBET 86: \citet{class04ek}; \\
CBET 82: \citet{disco04em};\\
IAUC 8411: \citet{class04em}; \\  
CBET 168: \citet{disco05ci};\\
CBET 169: \citet{class05ci};\\
ATel 4293: \citet{class12gcx};\\
CBET 3296,1: \citet{disco12kso};\\
CBET 3296,2: \citet{class12kso};}

\tablecomments{Magnitudes are unfiltered or $r$ band except for the limiting magnitude of PTF12kso. For PTF12kso there is a nondetection at JD = 2,456,169.5 by LSQ (Aug. 30), which is later than the last PTF nondetection, but the limiting magnitude is unknown. LSQ discovered it at JD = 2,456,202.7324, later than PTF.
For SN~2005ci the explosion date (3522.3$^{+1.5}_{-3.0}$) derived from the expanding photospheric method (EPM; see \citealp{jones09}) using the $VI$ bands and the first two spectra matches that derived from the PL fit which we adopted. We notice that the explosion date of SN 2005ci occurs before the last nondetection, but the magnitude limit at this phase was not deep enough (19~mag) to detect SN~2005ci.}
\end{deluxetable}

\begin{deluxetable}{lcccc}
\tabletypesize{\scriptsize}
\tablewidth{0pt}
\tablecaption{Optical Photometry of Three CCCP Long-Rising SNe~II.\label{tab:photCCCP}}
\tablehead{
\colhead{JD$-$2,400,000} &
\colhead{$B$} &
\colhead{$V$} &
\colhead{$R/unf.$} &
\colhead{$I$} \\
\colhead{(days)} &
\colhead{(mag)} &
\colhead{(mag)} &
\colhead{(mag)} &
\colhead{(mag)}}
\startdata
\multicolumn{5}{c}{{\bf SN~2004ek}}\\
53263.47* & 16.870(0.030) & 16.770(0.030) & 16.610(0.020) & 16.380(0.040) \\  
53267.20 & 17.105(0.016) & 16.866(0.018) & 16.622(0.007) & 16.277(0.032) \\  
53269.41* & 17.060(0.030) & 16.900(0.040) & 16.570(0.040) & 16.300(0.050) \\  
53272.17 & 17.305(0.016) & 16.966(0.019) & 16.712(0.010) & 16.258(0.033) \\  
53276.18 & 17.505(0.038) & 17.036(0.030) & 16.723(0.010) & 16.358(0.037) \\  
53280.24 & 17.495(0.016) & 17.066(0.020) & 16.742(0.009) & 16.308(0.032) \\  
53282.16 & \ldots        & \ldots        & \ldots        & 16.328(0.033) \\  
53283.40 & 17.525(0.016) & 17.116(0.019) & 16.782(0.007) & 16.328(0.033) \\  
53288.22 & 17.485(0.019) & \ldots        & 16.663(0.018) & 16.238(0.034) \\  
53290.41 & \ldots        & 17.116(0.066) & \ldots        & \ldots        \\ 
53292.17 & 17.515(0.020) & \ldots        & 16.642(0.010) & \ldots        \\ 
53294.26* & 17.570(0.030) & 17.030(0.030) & 16.610(0.030) & 16.250(0.030) \\  
53302.18 & 17.555(0.025) & 17.016(0.021) & \ldots        & \ldots        \\ 
53303.46 & 17.525(0.019) & 16.966(0.027) & 16.492(0.007) & 16.117(0.037) \\  
53307.41* & 17.620(0.060) & 16.960(0.040) & 16.490(0.030) & 16.040(0.030) \\  
53308.37 & 17.595(0.070) & \ldots        & \ldots        & \ldots        \\ 
53310.24 & \ldots        & 16.896(0.023) & 16.483(0.015) & 16.087(0.044) \\  
53312.51* & 17.520(0.040) & 16.870(0.030) & 16.430(0.030) & 16.070(0.030) \\  
53313.24 & \ldots        & 16.876(0.018) & \ldots        & 16.008(0.033) \\  
53315.45* & 17.500(0.040) & 16.860(0.020) & 16.430(0.020) & 16.050(0.050) \\  
53316.24 & 17.505(0.019) & 16.856(0.021) & 16.363(0.007) & 15.998(0.032) \\  
53317.30* & 17.650(0.030) & 16.780(0.020) & 16.500(0.020) & 16.090(0.050) \\  
53318.50* & 17.580(0.040) & 16.790(0.030) & 16.370(0.040) & \ldots        \\ 
53320.45* & 17.530(0.020) & 16.890(0.020) & 16.420(0.020) & 16.040(0.040) \\  
53321.47* & 17.640(0.030) & 16.820(0.020) & 16.310(0.010) & 15.950(0.040) \\  
53322.17 & 17.595(0.018) & 16.856(0.019) & 16.392(0.007) & 15.967(0.033) \\  
53323.49* & 17.570(0.030) & 16.870(0.020) & 16.350(0.020) & 15.950(0.030) \\  
53325.26 & 17.605(0.015) & 16.906(0.018) & 16.402(0.006) & \ldots        \\ 
53328.35 & 17.645(0.015) & \ldots        & \ldots        & 15.957(0.032) \\  
53329.25 & \ldots        & 16.856(0.019) & 16.363(0.007) & \ldots        \\ 
53331.47* & 17.650(0.040) & 16.840(0.020) & 16.360(0.020) & 15.930(0.030) \\  
53335.16 & \ldots        & 16.906(0.027) & 16.312(0.026) & \ldots        \\ 
53383.27* & 18.790(0.180) & \ldots        & 16.760(0.040) & \ldots        \\ 
53386.12 & 18.995(0.051) & 17.776(0.029) & 16.992(0.014) & 16.317(0.046) \\  
53389.37* & \ldots        & 17.400(0.200) & \ldots        & \ldots        \\ 
53405.22* & \ldots        & 17.840(0.140) & 17.070(0.070) & \ldots        \\ 
53412.26* & \ldots        & 17.710(0.190) & 17.190(0.140) & \ldots        \\ 
53538.45 & \ldots        & 21.006(0.657) & 19.973(0.091) & 19.087(0.540) \\  
53539.41 & \ldots & \ldots & 20.023(0.122)& \ldots \\
                                                             
\hline   
\multicolumn{5}{c}{{\bf SN~2004em}}\\                                                 
53276.17 & 17.870(0.082) &  17.647(0.055) &  17.503(0.079) &  17.382(0.073)     \\
53279.14 & 17.930(0.074) &  17.647(0.060) &  17.483(0.126) &  17.432(0.081)     \\
53281.13 & 17.900(0.074) &  17.577(0.058) &  17.353(0.032) &  17.082(0.064)     \\
53284.14 & 18.140(0.075) &  17.617(0.054) &  17.453(0.104) &  17.252(0.059)     \\
53287.10 & \ldots        &  \ldots        &  17.303(0.053) &  \ldots            \\
53289.17 & 18.270(0.112) &  17.607(0.057) &  \ldots        &  16.872(0.071)     \\
53292.12 & 18.170(0.077) &  17.547(0.074) &  17.403(0.026) &  17.042(0.054)     \\
53302.17 & 18.560(0.086) &  17.707(0.066) &  17.453(0.039) &  17.172(0.069)     \\
53309.11 & 18.630(0.079) &  17.677(0.057) &  17.383(0.031) &  17.172(0.082)     \\
53310.18 & 18.550(0.113) &  17.667(0.106) &  17.403(0.072) &  17.052(0.155)     \\
53322.11 & \ldots        &  17.387(0.054) &  \ldots        &  \ldots            \\
53324.11 & \ldots        &  \ldots        &  \ldots        &  16.592(0.081)     \\
53329.14 & 18.210(0.079) &  17.117(0.054) &  17.013(0.041) &  16.682(0.046)     \\
53334.14 & \ldots        &  \ldots        &  16.713(0.151) &  16.542(0.088)     \\
53351.14 & \ldots        &  \ldots        &  16.593(0.035) &  \ldots            \\
53352.11 & 17.830(0.073) &  16.897(0.053) &  16.583(0.020) &  16.402(0.068)     \\
53401.55 & \ldots        &  17.127(0.154) &  16.753(0.179) &  16.312(0.106)     \\
53405.56 & \ldots        &  17.297(0.069) &  \ldots        &  \ldots            \\
53406.54 & \ldots        &  \ldots        &  16.943(0.054) &  16.662(0.065)     \\
53537.32 & \ldots        &  \ldots        &  19.453(0.717) &  \ldots            \\
53538.37 & \ldots        &  20.296(0.153) &  19.543(0.689) &  20.362(0.983)     \\
                             
\hline
\multicolumn{5}{c}{{\bf SN~2005ci}}\\                                         
53523.86** & \ldots        &  \ldots        &  20.160(0.150) &  \ldots            \\
53531.81** & \ldots        &  \ldots        &  18.780(0.130) &  \ldots            \\
53532.85** & \ldots        &  \ldots        &  18.750(0.130) &  \ldots            \\
53533.83** & \ldots        &  \ldots        &  18.760(0.130) &  \ldots            \\
53534.76** & \ldots        &  \ldots        &  18.860(0.130) &  \ldots            \\
53535.18 & 19.340(0.299) &  \ldots        &  18.742(0.107) &  18.417(0.329)     \\
53536.21 & 19.480(0.280) &  \ldots        &  18.662(0.152) &  18.117(0.383)     \\

53536.80** &  \ldots        &\ldots        & 18.510(0.160) & \ldots\\

53539.23 & \ldots        &  19.052(0.138) &  \ldots        &  17.877(0.417)     \\
53540.82** & \ldots        &  \ldots        &  18.310(0.130) &  \ldots            \\
53544.26 & 19.540(0.482) &  18.702(0.188) &  18.152(0.071) &  17.527(0.214)     \\
53545.79** & \ldots        &  \ldots        &  18.180(0.130) &  \ldots            \\
53546.28 & 19.610(0.182) &  18.432(0.186) &  17.992(0.079) &  17.377(0.259)     \\
53548.23 & 19.380(0.277) &  18.322(0.177) &  17.982(0.049) &  17.417(0.258)     \\
53549.78** & \ldots        &  \ldots        &  17.940(0.130) &  \ldots            \\
53550.18 & 19.000(0.233) &  \ldots        &  17.822(0.026) &  17.087(0.191)     \\
53552.34 & \ldots        &  18.172(0.156) &  17.632(0.050) &  \ldots            \\
53553.79** & \ldots        &  \ldots        &  17.750(0.130) &  \ldots            \\
53554.41 & 18.870(0.244) &  18.112(0.133) &  17.452(0.037) &  \ldots            \\
53557.30 & 19.000(0.227) &  \ldots        &  \ldots &  16.987(0.152)     \\
53557.78** & \ldots &  \ldots        &  17.630(0.130) &  \ldots     \\
53559.22 & 19.010(0.192) &  17.972(0.115) &  \ldots        &  16.967(0.115)     \\
53560.26 & 19.070(0.181) &  17.822(0.107) &  17.422(0.047) &  16.847(0.138)     \\
53561.76** & \ldots        &  \ldots        &  17.720(0.150) &  \ldots            \\
53562.18 & 19.160(0.222) &  17.982(0.127) &  17.472(0.028) &  16.847(0.133)     \\
53563.30 & 19.000(0.190) &  17.972(0.072) &  \ldots        &  16.747(0.110)     \\
53565.75** & \ldots        &  \ldots        &  17.530(0.130) &  \ldots            \\
53567.68** & \ldots        &  \ldots        &  17.680(0.150) &  \ldots            \\
53570.19 & 18.990(0.245) &  17.682(0.113) &  17.262(0.033) &  16.597(0.116)     \\
53573.30 & \ldots        &  17.762(0.134) &  17.222(0.051) &  16.477(0.113)     \\
53581.29 & 18.990(0.212) &  17.612(0.103) &  17.162(0.043) &  16.387(0.107)     \\
53587.24 & 18.970(0.251) &  17.632(0.099) &  17.012(0.036) &  16.407(0.113)     \\
53590.29 & 18.770(0.192) &  17.572(0.094) &  \ldots        &  16.287(0.093)     \\
53595.16 & \ldots        &  17.522(0.091) &  \ldots        &  16.137(0.149)     \\
53600.21 & 18.840(0.129) &  17.452(0.075) &  16.922(0.035) &  16.177(0.097)     \\
53602.18 & 18.990(0.493) &  17.422(0.073) &  16.802(0.033) &  16.137(0.098)     \\
53611.93 & 18.760(0.139) &  17.422(0.074) &  16.792(0.028) &  16.177(0.103)     \\
53613.23 & 18.930(0.215) &  17.442(0.103) &  \ldots        &  16.077(0.100)     \\
53614.24 & \ldots        &  \ldots        &  16.782(0.023) &  \ldots            \\
53615.17 & 18.720(0.152) &  17.422(0.093) &  \ldots        &  16.127(0.116)     \\
53616.21 & 18.840(0.252) &  17.472(0.098) &  16.802(0.033) &  16.097(0.102)     \\
53619.16 & 18.900(0.197) &  17.462(0.082) &  \ldots        &  16.147(0.103)     \\
53620.18 & \ldots        &  17.552(0.056) &  \ldots        &  16.117(0.174)     \\
53621.19 & \ldots        &  17.492(0.092) &  \ldots        &  16.157(0.107)     \\
53623.23 & \ldots        &  \ldots        &  16.822(0.047) &  \ldots            \\
53628.20 & 18.920(0.476) &  \ldots        &  \ldots        &  16.037(0.125)     \\
53632.13 & 19.260(0.356) &  17.842(0.203) &  17.102(0.033) &  \ldots            \\
53643.11 & 19.240(0.655) &  18.332(0.203) &  17.532(0.084) &  16.787(0.157)     \\
53649.15 & \ldots        &  18.612(0.297) &  17.842(0.074) &  \ldots            \\
53653.11 & 19.740(0.430) &  19.072(0.349) &  17.932(0.078) &  17.137(0.202)     \\
                                                                
\enddata
\tablecomments{Besides CCCP photometry from \citet{arcavi12}, the table includes $BVRI$ photometry from \citet{tse08} (marked by ``*" and shifted by +0.1 ($B$), +0.2($V$), +0.2($R$), and +0.1($I$) mag with respect to the original values of \citealp{tse08} to match the CCCP photometry) of SN~2004ek and unfiltered photometry of SN~2005ci from \citet{kleiser11} (marked by ``**").}
\end{deluxetable}

\begin{deluxetable}{lcccccc}
\tabletypesize{\scriptsize}
\tablewidth{0pt}
\tablecaption{Optical Photometry of Three PTF SN~1987A-like SNe.\label{tab:photPTF}}
\tablehead{
\colhead{JD-2,400,000}&
\colhead{$B$}&
\colhead{$g$}&
\colhead{$r$}&
\colhead{$i$}&
\colhead{$z$}\\
\colhead{(days)}&
\colhead{(mag)}&
\colhead{(mag)}&
\colhead{(mag)}&
\colhead{(mag)}&
\colhead{(mag)}}
\startdata

\multicolumn{6}{c}{{\bf PTF09gpn}}\\   

55137.77* & \ldots        & \ldots        & 20.300(0.017) & \ldots       & \ldots          \\
55142.82* & \ldots        & \ldots        & 19.560(0.170) & \ldots       & \ldots          \\
55144.89* & \ldots        & \ldots        & 19.491(0.094) & \ldots       & \ldots          \\
55151.80  & 20.302(0.143) & \ldots        & 19.181(0.048) & 19.248(0.086) & \ldots         \\ 
55152.71  & \ldots        & \ldots        & 19.238(0.027) & 19.290(0.043) & \ldots         \\
55152.72* & \ldots        & \ldots        & 19.510(0.004) & \ldots       & \ldots          \\
55155.89  & 20.585(0.065) & 19.873(0.044) & 19.286(0.027) & 19.238(0.036) & 19.446(0.110)  \\
55166.78  & \ldots        & 19.947(0.129) & 19.227(0.054) & 19.061(0.065) & 19.268(0.133)  \\
55168.72  & \ldots        & 20.097(0.086) & 19.188(0.045) & 19.127(0.051) & 19.139(0.083)  \\
55168.74* & \ldots        & \ldots        & 19.557(0.045) & \ldots       & \ldots          \\
55175.80  & 20.804(0.119) & 20.004(0.056) & 19.058(0.036) & 18.839(0.034) & \ldots         \\
55176.73  & 20.576(0.111) & 19.997(0.049) & 19.046(0.043) & 18.949(0.050) & 18.704(0.198)  \\
55180.80  & 20.584(0.126) & \ldots        & 18.997(0.042) & 18.772(0.042) & 18.808(0.096)  \\
55181.62* & \ldots        & \ldots        & 19.149(0.260) & \ldots       & \ldots          \\
55183.64* & \ldots        & \ldots        & 19.153(0.046) & \ldots       & \ldots          \\
55184.64  & \ldots        & 19.615(0.059) & 18.997(0.079) & 18.697(0.069) & \ldots         \\
55185.65  & 20.336(0.043) & \ldots        & 18.801(0.020) & 18.669(0.020) & 18.847(0.062)  \\
55189.60* & \ldots        & \ldots        & 18.798(0.137) & \ldots       & \ldots          \\
55189.83  & 20.243(0.121) & 19.616(0.057) & 18.727(0.032) & 18.621(0.035) & 18.608(0.185)  \\
55192.59* & \ldots        & \ldots        & 18.910(0.227) & \ldots       & \ldots          \\
55192.67  & \ldots        & \ldots        & 18.612(0.017) & 18.530(0.032) & 18.617(0.072)  \\
55201.78  & 19.878(0.078) & 19.363(0.035) & 18.454(0.014) & 18.376(0.017) & 18.446(0.063)  \\
55202.60* & \ldots        & \ldots        & 18.827(0.318) & \ldots       & \ldots          \\
55202.78  & \ldots        & 19.321(0.035) & 18.444(0.019) & 18.321(0.025) & 18.374(0.103)  \\
55203.74* & \ldots        & \ldots        & 18.772(0.056) & \ldots       & \ldots          \\
55204.71  & \ldots        & \ldots        & 18.434(0.011) & 18.354(0.013) & 18.487(0.035)  \\
55207.69* & \ldots        & \ldots        & 18.613(0.059) & \ldots       & \ldots          \\
55207.79  & 19.869(0.056) & \ldots        & 18.397(0.017) & 18.312(0.022) & 18.353(0.037)  \\
55208.67  & 19.840(0.159) & 19.244(0.087) & 18.391(0.016) & 18.320(0.034) & 18.393(0.108)  \\
55209.65* & \ldots        & \ldots        & 18.600(0.008) & \ldots       & \ldots          \\
55209.67  & 19.745(0.029) & 19.260(0.014) & 18.406(0.009) & 18.259(0.012) & 18.325(0.082)  \\
55221.70  & 20.392(0.141) & 19.519(0.051) & \ldots        & 18.437(0.023) & \ldots         \\                                                                                          
55429.98* & \ldots        & \ldots        & 21.972(0.189) & \ldots       & \ldots          \\
55441.98* & \ldots        & \ldots        & 21.741(0.012) & \ldots       & \ldots          \\
55450.96* & \ldots        & \ldots        & 21.706(0.365) & \ldots       & \ldots          \\
55452.98* & \ldots        & \ldots        & 22.307(0.458) & \ldots       & \ldots          \\
55455.96* & \ldots        & \ldots        & 22.103(0.336) & \ldots       & \ldots          \\
55456.92* & \ldots        & \ldots        & 21.767(0.323) & \ldots       & \ldots          \\
55457.97* & \ldots        & \ldots        & 22.475(0.451) & \ldots       & \ldots          \\
55459.01* & \ldots        & \ldots        & 21.998(0.386) & \ldots       & \ldots          \\
55473.84* & \ldots        & \ldots        & 21.598(0.345) & \ldots       & \ldots          \\
55480.90* & \ldots        & \ldots        & 21.864(0.490) & \ldots       & \ldots          \\
55481.92* & \ldots        & \ldots        & 21.524(0.000) & \ldots       & \ldots          \\

\multicolumn{6}{c}{{\bf PTF12gcx}}\\   

56085.85* & \ldots  & \ldots &     21.145(0.343) & \ldots & \ldots \\
56086.93* & \ldots  & \ldots &     20.653(0.371) & \ldots & \ldots \\
56087.91* & \ldots  & \ldots &     20.811(0.098) & \ldots & \ldots \\
56089.84* & \ldots  & \ldots &     20.322(0.137) & \ldots & \ldots \\
56091.89* & \ldots  & \ldots &     20.462(0.147) & \ldots & \ldots \\
56092.79* & \ldots  & \ldots &     20.337(0.080) & \ldots & \ldots \\
56093.88* & \ldots  & \ldots &     20.416(0.157) & \ldots & \ldots \\
56094.79* & \ldots  & \ldots &     20.276(0.082) & \ldots & \ldots \\
56096.81* & \ldots  & \ldots &     20.273(0.090) & \ldots & \ldots \\
56098.83* & \ldots  & \ldots &     20.142(0.072) & \ldots & \ldots \\
56099.74* & \ldots  & \ldots &     20.124(0.015) & \ldots & \ldots \\
56100.80* & \ldots  & \ldots &     20.073(0.085) & \ldots & \ldots \\
56101.73* & \ldots  & \ldots &     20.074(0.153) & \ldots & \ldots \\
56102.80* & \ldots  & \ldots &     20.025(0.062) & \ldots & \ldots \\
56103.73* & \ldots  & \ldots &     19.962(0.078) & \ldots & \ldots \\
56104.77* & \ldots  & \ldots &     20.048(0.072) & \ldots & \ldots \\
56105.76* & \ldots  & \ldots &     19.827(0.058) & \ldots & \ldots \\
56106.75* & \ldots  & \ldots &     20.050(0.162) & \ldots & \ldots \\
56107.78* & \ldots  & \ldots &     19.827(0.075) & \ldots & \ldots \\
56108.78* & \ldots  & \ldots &     19.732(0.233) & \ldots & \ldots \\
56109.68* & \ldots  & \ldots &     19.785(0.083) & \ldots & \ldots \\
56111.76* & \ldots  & \ldots &     19.830(0.147) & \ldots & \ldots \\
56113.79* & \ldots  & \ldots &     19.643(0.062) & \ldots & \ldots \\
56114.73* & \ldots  & \ldots &     19.747(0.064) & \ldots & \ldots \\
56115.84* & \ldots  & \ldots &     19.723(0.045) & \ldots & \ldots \\
56116.74* & \ldots  & \ldots &     19.697(0.090) & \ldots & \ldots \\
56117.82* & \ldots  & \ldots &     19.658(0.039) & \ldots & \ldots \\
56118.75* & \ldots  & \ldots &     19.700(0.078) & \ldots & \ldots \\
56119.80* & \ldots  & \ldots &     19.649(0.065) & \ldots & \ldots \\
56123.78* & \ldots  & \ldots &     19.611(0.051) & \ldots & \ldots \\
56125.80* & \ldots  & \ldots &     19.510(0.088) & \ldots & \ldots \\
56126.80* & \ldots  & \ldots &     19.565(0.049) & \ldots & \ldots \\
56128.78* & \ldots  & \ldots &     19.556(0.084) & \ldots & \ldots \\
56130.75* & \ldots  & \ldots &     19.489(0.047) & \ldots & \ldots \\
56131.77* & \ldots  & \ldots &     19.413(0.075) & \ldots & \ldots \\
56132.77* & \ldots  & \ldots &     19.545(0.045) & \ldots & \ldots \\
56135.73* & \ldots  & \ldots &     19.513(0.036) & \ldots & \ldots \\
56177.68  & \ldots & 21.343(0.167) & 19.754(0.054) & 19.281(0.028) & \ldots \\  
56178.67 & \ldots  &  21.136(0.125) & \ldots & \ldots & \ldots \\
56186.65  & \ldots & \ldots & 20.259(0.051) & 19.833(0.067) & \ldots \\  
56187.68  & \ldots & \ldots & 20.289(0.068) & 19.825(0.074) & \ldots \\

\multicolumn{6}{c}{{\bf PTF12kso}}\\   
2456198.983*  & \ldots        & \ldots        &  18.406(0.018) & \ldots        & \ldots       \\
2456228.944*  & \ldots        & \ldots        &  18.041(0.025) & \ldots        & \ldots       \\
2456238.002*  & \ldots        & \ldots        &  17.974(0.002) & \ldots        & \ldots       \\
2456238.97 & \ldots       & \ldots       & 18.020(0.008) & \ldots                 & \ldots       \\
2456239.992*  & \ldots        & \ldots        &  17.984(0.027) & \ldots        & \ldots       \\
2456240.84**  & \ldots        & 18.931(0.028) & 18.007(0.013) & 17.837(0.038) & \ldots           \\
2456241.80**  & \ldots        & 19.016(0.047) & 18.036(0.060) & 17.779(0.067) & \ldots           \\
2456245.71 & 19.652(0.060) & 18.950(0.024) & 17.993(0.014) & 17.782(0.012)        & \ldots       \\
2456245.77**  & \ldots        & 19.054(0.036) & 18.010(0.048) & 17.791(0.046) & 17.532(0.020)    \\
2456246.81**  & \ldots        & 18.991(0.018) & \ldots       & \ldots       & \ldots             \\
2456247.74**  & \ldots        & 19.049(0.151) & 17.964(0.023) & 17.909(0.029) & 17.843(0.093)    \\
2456248.78**  & \ldots        & 19.101(0.060) & 18.084(0.044) & 17.822(0.059) & 17.639(0.122)    \\
2456250.77**  & \ldots        & 19.135(0.070) & 17.937(0.074) & 17.807(0.028) & 17.539(0.092)    \\
2456250.88 & 19.744(0.039) & 19.080(0.019) & 18.011(0.011) & 17.853(0.014)        & \ldots       \\
2456251.70 & 19.882(0.076) & 19.034(0.026) & 18.026(0.020) & 17.765(0.016)        & \ldots       \\
2456252.76**  & \ldots        & 19.130(0.032) & 18.066(0.117) & 17.813(0.045) & 17.640(0.074)    \\

2456252.99$\dagger$ & \ldots        & \ldots        & 18.057(0.129) &  \ldots           & \ldots          \\

2456253.81**  & \ldots        & 19.055(0.102) & 17.996(0.068) & 17.897(0.137) & \ldots           \\
2456254.69 & 19.903(0.079) & 19.136(0.036) & 18.029(0.014) & 17.872(0.015)        & \ldots       \\
2456254.716*  & \ldots        & \ldots        &  18.000(0.035) & \ldots        & \ldots       \\
2456254.76**  & \ldots        & 19.109(0.030) & 18.018(0.006) & 17.867(0.061) & 17.748(0.104)    \\
2456256.76**  & \ldots        & 19.176(0.032) & 18.144(0.028) & 17.863(0.076) & 17.672(0.101)    \\
2456256.915*  & \ldots        & \ldots        &  18.033(0.005) & \ldots        & \ldots       \\
2456257.68 & \ldots       & 19.036(0.078) & 18.047(0.029) & 17.875(0.024)         & \ldots       \\
2456259.76**  & \ldots        & 19.226(0.104) & 18.069(0.063) & 17.875(0.123) & 17.544(0.083)    \\
2456261.75**  & \ldots        & \ldots       & 18.140(0.048) & \ldots       & \ldots             \\
2456264.77**  & \ldots        & 19.365(0.112) & 18.140(0.100) & 17.916(0.015) & 17.588(0.081)    \\
2456265.73**  & \ldots        & 19.470(0.110) & 18.280(0.120) & 18.038(0.124) & 17.649(0.064)    \\
2456265.76 & 20.179(0.139) & 19.433(0.042) & 18.233(0.021) & 17.970(0.023)        & \ldots       \\
2456266.73**  & \ldots        & 19.509(0.020) & 18.303(0.036) & 17.938(0.068) & 17.799(0.155)    \\
2456267.73**  & \ldots        & 19.666(0.082) & 18.349(0.085) & 18.041(0.067) & 17.873(0.051)    \\
2456268.65 & 20.394(0.069) & 19.618(0.029) & 18.309(0.012) & 18.086(0.023)        & \ldots       \\
2456268.73**  & \ldots        & 19.623(0.145) & 18.491(0.035) & 18.032(0.017) & 17.834(0.039)    \\
2456269.67**  & \ldots        & 19.764(0.043) & 18.362(0.048) & 18.110(0.066) & 17.809(0.049)    \\
2456270.72**  & \ldots        & 19.761(0.075) & 18.496(0.055) & 18.177(0.064) & 17.773(0.146)    \\

2456270.88$\dagger$ & 20.401(0.076) & 18.897(0.057) & 18.203(0.097) &  18.250(0.170)    & \ldots          \\

2456271.67 & 20.628(0.077) & 19.910(0.033) & 18.467(0.017) & 18.183(0.034)        & \ldots       \\
2456272.72**  & \ldots        & 19.928(0.121) & 18.516(0.055) & 18.222(0.092) & 17.847(0.142)    \\
2456273.72**  & \ldots        & 19.911(0.119) & 18.655(0.043) & 18.294(0.053) & 17.987(0.123)    \\
2456274.71**  & \ldots        & 19.997(0.075) & 18.604(0.010) & 18.319(0.102) & 17.993(0.075)    \\
2456275.67**  & \ldots        & 20.160(0.052) & 18.816(0.107) & 18.393(0.037) & 18.091(0.068)    \\
2456276.68**  & \ldots        & 20.085(0.045) & 18.783(0.060) & 18.432(0.042) & 17.989(0.090)    \\
2456277.67**  & \ldots        & 20.250(0.080) & 18.741(0.039) & 18.479(0.069) & 18.112(0.101)    \\
2456278.67**  & \ldots        & 20.215(0.046) & 18.825(0.079) & 18.470(0.097) & 18.250(0.060)    \\
2456279.67**  & \ldots        & \ldots       & 18.869(0.034) & 18.554(0.153) & 18.186(0.108)     \\
2456280.77**  & \ldots        & \ldots       & 18.922(0.065) & 18.693(0.020) & 18.283(0.055)     \\
2456281.71 & 20.977(0.135) & 20.381(0.114) & 18.914(0.038) & 18.704(0.035)        & \ldots       \\
2456281.77**  & \ldots        & \ldots       & 18.939(0.021) & 18.592(0.032) & \ldots            \\
2456282.67**  & \ldots        & \ldots       & 18.931(0.053) & 18.617(0.120) & \ldots            \\
2456283.67**  & \ldots        & 20.349(0.019) & 19.008(0.117) & 18.810(0.072) & 18.233(0.127)    \\
2456283.88 & 20.986(0.157) & \ldots       & \ldots       & \ldots                 & \ldots       \\
2456284.67**  & \ldots        & 20.480(0.093) & 19.076(0.093) & 18.912(0.038) & 18.519(0.099)    \\
2456285.72**  & \ldots        & \ldots       & 18.968(0.151) & 18.898(0.109) & 18.214(0.000)     \\
2456286.72**  & \ldots        & \ldots       & 19.094(0.109) & 18.798(0.135) & \ldots            \\
2456287.67**  & \ldots        & \ldots       & \ldots       & 18.931(0.172) & \ldots             \\
2456288.71**  & \ldots        & \ldots       & 19.078(0.101) & 18.846(0.110) & \ldots            \\
2456289.67**  & \ldots        & \ldots       & 19.240(0.007) & 18.986(0.124) & \ldots            \\
2456289.86 & \ldots       & \ldots       & 19.263(0.078) & 18.961(0.089)          & \ldots       \\
2456290.67**  & \ldots        & \ldots       & 19.128(0.073) & 18.868(0.105) & \ldots            \\
2456291.67**  & \ldots        & \ldots       & 19.180(0.111) & 18.860(0.136) & \ldots            \\
2456295.72**  & \ldots        & \ldots       & 19.246(0.040) & 18.843(0.106) & \ldots            \\

2456295.80$\dagger$ & 20.768(0.136) & 19.758(0.078) & 18.904(0.111) &  19.159(0.171)    & \ldots          \\

2456297.68**  & \ldots        & \ldots       & \ldots       & 18.949(0.139) & \ldots             \\
2456297.87 & \ldots       & 20.889(0.128) & \ldots       & \ldots                 & \ldots       \\
2456298.72**  & \ldots        & \ldots       & 19.250(0.076) & 19.008(0.105) & \ldots            \\
2456299.68**  & \ldots        & \ldots       & 19.223(0.064) & 19.107(0.038) & \ldots            \\
2456300.76 & 21.460(0.179) & \ldots       & 19.304(0.029) & 19.157(0.038)         & \ldots       \\
2456300.77**  & \ldots        & \ldots       & 19.256(0.149) & 19.124(0.093) & \ldots            \\
2456301.68**  & \ldots        & \ldots       & 19.262(0.059) & 19.132(0.064) & \ldots            \\
2456308.63**  & \ldots        & \ldots       & 19.349(0.050) & 19.213(0.099) & \ldots            \\
2456309.68**  & \ldots        & \ldots       & 19.350(0.065) & 19.235(0.077) & \ldots            \\
2456310.62**  & \ldots        & \ldots       & 19.402(0.027) & 19.259(0.134) & \ldots            \\
2456311.63**  & \ldots        & \ldots       & 19.306(0.066) & 19.088(0.089) & \ldots            \\
2456312.68**  & \ldots        & \ldots       & 19.384(0.011) & 19.385(0.034) & \ldots            \\
2456314.57**  & \ldots        & \ldots       & 19.497(0.016) & 19.346(0.108) & \ldots            \\
2456315.61**  & \ldots        & \ldots       & 19.529(0.032) & 19.450(0.051) & \ldots            \\
2456316.61**  & \ldots        & \ldots       & 19.405(0.161) & 19.228(0.127) & \ldots            \\
2456317.57**  & \ldots        & \ldots       & 19.516(0.074) & 19.354(0.061) & \ldots            \\
2456318.61**  & \ldots        & \ldots       & 19.435(0.071) & 19.269(0.115) & \ldots            \\
2456321.74 & \ldots       & \ldots       & 19.465(0.110) & 19.366(0.102)          & \ldots       \\
2456322.71 & \ldots       & 20.962(0.145) & 19.422(0.050) & 19.474(0.055)         & \ldots       \\
2456323.68**  & \ldots        & \ldots       & 19.389(0.029) & 19.402(0.014) & \ldots            \\
2456323.71 & 21.752(0.171) & \ldots       & \ldots       & \ldots                 & \ldots       \\
2456324.69 & 21.763(0.191) & \ldots       & \ldots       & \ldots                 & \ldots       \\
2456325.57**  & \ldots        & \ldots       & 19.500(0.039) & 19.425(0.060) & \ldots            \\
2456325.69 & \ldots       & 21.134(0.155) & 19.464(0.119) & 19.587(0.172)         & \ldots       \\
2456326.57**  & \ldots        & \ldots       & 19.435(0.101) & 19.410(0.167) & \ldots            \\
2456327.60**  & \ldots        & \ldots       & 19.463(0.040) & 19.434(0.061) & \ldots            \\
2456327.77 & 21.431(0.185) & 21.152(0.112) & \ldots       & \ldots                & \ldots       \\
2456328.60**  & \ldots        & \ldots       & 19.358(0.031) & 19.301(0.063) & \ldots            \\
2456328.68 & 22.014(0.202) & \ldots       & 19.517(0.028) & 19.511(0.038)         & \ldots       \\
2456329.64**  & \ldots        & \ldots       & 19.560(0.016) & 19.441(0.053) & \ldots            \\
2456330.64**  & \ldots        & \ldots       & 19.491(0.025) & 19.379(0.030) & \ldots            \\
2456330.68 & \ldots       & 21.143(0.086) & \ldots       & \ldots                 & \ldots       \\
2456331.64**  & \ldots        & \ldots       & 19.563(0.009) & 19.531(0.051) & \ldots            \\
2456333.60**  & \ldots        & \ldots       & 19.550(0.059) & 19.600(0.030) & \ldots            \\
2456334.60**  & \ldots        & \ldots       & 19.470(0.036) & 19.573(0.021) & \ldots            \\
2456335.72 & 21.540(0.191) & 21.066(0.121) & 19.564(0.053) & 19.508(0.053)        & \ldots       \\
2456338.61**  & \ldots        & \ldots       & 19.637(0.001) & 19.587(0.001) & \ldots            \\
2456339.60**  & \ldots        & \ldots       & 19.617(0.041) & \ldots       & \ldots             \\
2456340.61**  & \ldots        & \ldots       & 19.584(0.039) & 19.644(0.006) & \ldots            \\
2456340.70 & \ldots       & 21.120(0.127) & 19.594(0.055) & 19.670(0.057)         & \ldots       \\
2456341.61**  & \ldots        & \ldots       & 19.609(0.054) & 19.626(0.051) & \ldots            \\
2456342.57**  & \ldots        & \ldots       & 19.543(0.006) & 19.685(0.068) & \ldots            \\
2456343.57**  & \ldots        & \ldots       & 19.692(0.091) & 19.648(0.040) & \ldots            \\
2456344.57**  & \ldots        & \ldots       & 19.627(0.031) & 19.585(0.087) & \ldots            \\
2456345.57**  & \ldots        & \ldots       & 19.695(0.049) & 19.616(0.082) & \ldots            \\
2456346.57**  & \ldots        & \ldots       & 19.584(0.058) & 19.653(0.041) & \ldots            \\
2456346.73 & \ldots       & \ldots       & 19.635(0.078) & 19.741(0.122)          & \ldots       \\
2456347.56**  & \ldots        & \ldots       & 19.585(0.151) & \ldots       & \ldots             \\
2456348.57**  & \ldots        & \ldots       & 19.622(0.069) & 19.759(0.083) & \ldots            \\
2456349.57**  & \ldots        & \ldots       & 19.716(0.090) & 19.662(0.040) & \ldots            \\
2456350.67 & \ldots       & \ldots       & 19.636(0.105) & 19.671(0.117)          & \ldots       \\
2456351.66 & \ldots       & 21.409(0.176) & \ldots       & \ldots                 & \ldots       \\
2456354.66 & \ldots       & \ldots       & 19.693(0.048) & 19.903(0.083)          & \ldots       \\
2456356.72 & \ldots       & 21.488(0.152) & \ldots       & \ldots                 & \ldots       \\
2456363.66 & 21.822(0.170) & 21.296(0.161) & 19.746(0.044) & 19.953(0.059)        & \ldots       \\
2456364.65 & 21.946(0.168) & \ldots       & \ldots       & \ldots                 & \ldots       \\
2456366.64 & 21.933(0.199) & \ldots       & \ldots       & \ldots                 & \ldots       \\
2456367.65 & 21.460(0.142) & \ldots       & 19.821(0.062) & 20.109(0.143)         & \ldots       \\
2456377.66 & \ldots       & \ldots       & 19.795(0.128) & \ldots                 & \ldots       \\

\enddata
\tablecomments{Photometric epochs obtained with P48 are marked with ``*", those obtained by LCOGT are marked with ``"**". For PTF12kso the epochs marked with ``$\dagger$" correspond to $BVRI$ photometry from MLO, which was standardized to the Johnson-Cousins
  filter system (no correction has been made to convert the obtained $VRI$ magnitudes to the $gri$ system). The rest of the data were obtained by P60.} 
\end{deluxetable}

\begin{deluxetable}{ccccccc}
\tabletypesize{\scriptsize}
\tablewidth{0pt}
\tablecaption{Optical Spectroscopy of PTF and CCCP SN~1987A-like SNe and their Host Galaxies.\label{tab:spectra}}
\tablehead{
\colhead{Date (UT)}&
\colhead{JD-2,450,000}&
\colhead{Phase\tablenotemark{a}}&
\colhead{Target}&
\colhead{Telescope}&
\colhead{Instrument}&
\colhead{Range}\\
\colhead{}&
\colhead{(days)}&
\colhead{(days)}&
\colhead{}&
\colhead{}&
\colhead{}&
\colhead{(\AA)}}
\startdata
08 Nov. 2009 & 5143.9 &   $+$6.7 & PTF09gpn & P200 & DBSP  &  3357$-$10207   \\
26 Dec. 2011 & 5921.5  & \ldots &  PTF09gpn host galaxy & Keck-1 & LRIS  & 3580$-$10198   \\
\hline

06 Nov. 2012  &   
6237.6        &   $+$61.6 & PTF12kso  &    Copernico 1.82m    & AFOSC      & 4673$-$7386 \\
07 Nov. 2012 & 6238.9  &   $+$62.9 & PTF12kso & P200 & DBSP &  3202$-$9946   \\%
13 Nov. 2012 &  6242.5 &   $+$68.5 & PTF12kso & P200 & DBSP &   3630$-$9782   \\
09 Dec. 2012 & 6270.5 &   $+$94.5 & PTF12kso &  Keck-2 & DEIMOS  &  4902$-$7467 \\
08 Jan. 2013 & 6301.8 &   $+$124.8 & PTF12kso & Keck-2 & DEIMOS  &  4500$-$9632 \\
27 Sep. 2015                     &7292.5  & \ldots &  PTF12kso host galaxy & Keck-2  & DEIMOS  &   4650$-$9500   \\%
                     
\hline
16 Jul. 2012 & 6124.5     &   $+$38.8 & PTF12gcx & Keck-2 & DEIMOS &  4402$-$9593  \\
26 Jul. 2012 &     6134.5    &   $+$48.8 & PTF12gcx & P200 & DBSP &    3502$-$9992  \\
23 Aug. 2012 & 6162.7   &   $+$77.0 & PTF12gcx & Lick 3-m & Kast &  3458$-$10332 \\
15 Jun. 2007   &  4266.5 & \ldots &  PTF12gcx host galaxy &  SDSS 2.5-m &  SDSS spectrograph &   3801$-$9221 \\
\hline

23 Sep. 2004 &  3271.5    &   $+$21.0 & SN~2004ek & P200 & DBSP &   4190$-$9600  \\
24 Sep. 2004 & 3272.5   &   $+$22.0 & SN~2004ek& Lick 3-m & Kast &   3320.00$-$10500  \\
19 Nov. 2004 &  3328.8  &   $+$78.3 & SN~2004ek & P200 & DBSP &   4000$-$9776  \\%
14 Dec. 2004 &  3353.5    &   $+$103.0 & SN~2004ek & Keck-1 & LRIS &    3450$-$9200  \\
14 Jan. 2005 &  3385.1     &   $+$133.6 & SN~2004ek & P200 & DBSP &  3300$-$9900  \\%

\hline
24 Sep. 2004 & 3272.5   &   $+$8.2 & SN~2004em & Lick 3-m & Kast &  3320.00$-$10500  \\
18 Oct. 2004 &   3296.5    &   $+$32.2 & SN~2004em & Keck-1 & LRIS &   3450$-$9200   \\
19 Nov. 2004 &  3328.6     &   $+$65.3 & SN~2004em & P200 & DBSP &   3500$-$9780   \\
05 Dec. 2004 &  3344.6     &   $+$81.3 & SN~2004em &P200 & DBSP &   3500$-$9796   \\
05 Apr. 2005 &    3466.0    &   $+$201.7 & SN~2004em & P200 & DBSP &  3600$-$9991   \\%

\hline
16 Jul. 2005 &    3567.5   &   $+$43.8 & SN~2005ci & P200 & DBSP &   3400$-$9700  \\
14 Aug. 2005 &  3596.7  &   $+$73.0 & SN~2005ci &P200 & DBSP &   3500$-$9801  \\
08 Sep. 2005 &  3621.6    &   $+$97.9 & SN~2005ci & P200 & DBSP &   3300$-$9654  \\
\enddata
\tablenotetext{a}{From explosion date.}
\end{deluxetable}

\begin{deluxetable}{lc||cc}
\tabletypesize{\scriptsize}
\tablewidth{0pt}
\tablecaption{Relative Line Fluxes and Host-Galaxy Metallicity Measurements.\label{line}}
\tablehead{
\colhead{Line}&
\colhead{Flux}&
\colhead{Method}&
\colhead{12$+$log(O/H)}\\
\colhead{}&
\colhead{($F/F_{{\rm H}\alpha}$)}&
\colhead{}&
\colhead{(dex)}}
\startdata
\multicolumn{4}{c}{\bf PTF09gpn host}\\

$[\ion{O}{ii}]~\lambda$3727    &0.4093$\pm$0.2367  &  R23          &  7.77$\pm$0.15     \\
H$\delta$~$\lambda$4101        &0.0507$\pm$0.0259  &  N2           &  8.06$\pm$0.18  \\
H$\gamma$~$\lambda$4341     &0.1227$\pm$0.0335  & O3N2         & 8.06$\pm$0.14   \\   
$[\ion{O}{iii}]~\lambda$4363   &0.0487$\pm$0.0186  &               &    \\
H$\beta$~$\lambda$4861         &0.3683$\pm$0.0244 &            & \\
$[\ion{O}{iii}]~\lambda$4959   &0.4341$\pm$0.0703  &              &    \\
$[\ion{O}{iii}]~\lambda$5007   &1.2848$\pm$0.0249  &              &    \\
H$\alpha$~$\lambda$6563       &1.0000$\pm$0.0192\tablenotemark{*}  & & \\
$[\ion{N}{ii}]~\lambda$6584    &0.0289$\pm$0.0113  &              &    \\
$[\ion{S}{ii}]~\lambda$6717   &  0.1225 $\pm$0.0140  &              &    \\
$[\ion{S}{ii}]~\lambda$6731   &  0.0547$\pm$ 0.0178 &              &    \\
\multicolumn{4}{c}{\bf PTF12gcx host}\\

$[\ion{O}{ii}]~\lambda$3727    &2.0219$\pm$0.0111 &R23          &  8.27$\pm$0.15  \\
H$\gamma$~$\lambda$4341   & 0.1198$\pm$0.0075 &N2           &  8.52$\pm$0.18  \\
H$\beta$~$\lambda$4861       &0.3485$\pm$0.0055 &O3N2         & 8.56$\pm$0.14   \\   
$[\ion{O}{iii}]~\lambda$4959   &0.0883$\pm$0.0093 & &\\
$[\ion{O}{iii}]~\lambda$5007   &0.2860$\pm$0.0073 & &\\
$[\ion{N}{ii}]~\lambda$6548    &0.083$\pm$0.0099   & &\\
H$\alpha$~$\lambda$6563      & 1.0000$\pm$0.0089\tablenotemark{\#} & &\\
$[\ion{N}{ii}]~\lambda$6584    &0.2408$\pm$0.0041 & &\\
$[\ion{S}{ii}]~\lambda$6717    &0.2166$\pm$0.0062 & &\\
$[\ion{S}{ii}]~\lambda$6731    &0.1590$\pm$0.0079 & &\\
\multicolumn{4}{c}{\bf PTF12kso host}\\
$[\ion{O}{iii}]~\lambda$4959   &0.2429$\pm$0.0256& N2           &  $\leq$~8.04$\pm$0.18\\
H$\alpha$~$\lambda$6563   & 1.0000$\pm$0.0028\tablenotemark{**} & &  \\
$[\ion{N}{ii}]~\lambda$6584    &$\leq$0.0257 & &\\

\enddata
\tablenotetext{*}{H$\alpha$ flux is 2.42$\times$10$^{-14}$ erg~s$^{-1}$~cm$^{-2}$.}
\tablenotetext{\#}{H$\alpha$ flux is 6.72$\times$10$^{-14}$ erg~s$^{-1}$~cm$^{-2}$.}
\tablenotetext{**}{H$\alpha$ flux is 1.36$\times$10$^{-15}$ erg~s$^{-1}$~cm$^{-2}$.}
\end{deluxetable}

\begin{deluxetable}{l|cc|cc|cc|}
\tabletypesize{\scriptsize}
\tablewidth{0pt}
\tablecaption{\label{tab:rise}Rise Times and Peak Magnitudes of Long-Rising SNe~II.}
\tablehead{
\colhead{}&
\colhead{$t_{\rm rise}$}&
\colhead{$m_{\rm max}$}&
\colhead{$t_{\rm rise}$}&
\colhead{$m_{\rm max}$}&
\colhead{$t_{\rm rise}$}&
\colhead{$m_{\rm max}$}\\
\colhead{}&
\colhead{(days)}&
\colhead{(mag)}&
\colhead{(days)}&
\colhead{(mag)}&
\colhead{(days)}&
\colhead{(mag)}}
\startdata
Band/SN               & SN~2004ek   &      &  SN~2004em     &     &  SN~2005ci       &          \\
$B$         &  55 & 17.55 & \ldots &  \ldots &     74 & 18.85  \\
$V$         &  75 & 16.84 & 111    &  16.61  &     86 & 17.43  \\
$R$         &  83 & 16.35 & 108    &  16.47  &     87 & 16.77  \\
$I$         &  94 & 15.90 & 115    &  15.93  &     89 & 16.09  \\
\hline
Band/SN   &PTF09gpn&        &  PTF12gcx&  &       PTF12kso  &  \\    
$B$         &  77   &    19.74    &  \ldots   &  \ldots           &  \ldots &\ldots\\
$g$         &  77  &  19.26      &    \ldots &   \ldots           & \ldots &\ldots\\
$r$         & $>$72 & $<$18.35  & 68 & 19.32       & 68 & 18.00      \\
$i$         &   77 & 18.26       &   \ldots   &   \ldots           &  68  & 17.78           \\
$z$         & $>72$   & $<$18.32       &   \ldots  &    \ldots          &  79   &   17.60          \\
\enddata
\tablecomments{Rise times and peak magnitudes are obtained from low-order polynomial fits to the SN light curves.}
\end{deluxetable}

\begin{deluxetable}{c|cc|cc|ccc}
\tabletypesize{\scriptsize}
\tablewidth{0pt}
\tablecaption{\label{tab:param}Explosion and Progenitor Parameters.}
\tablehead{
\colhead{SN}&
\colhead{$E^{a}$}&
\colhead{$M_{\rm ej}^{a}$}&
\colhead{$M_{^{56}{\rm Ni}}^{t}$}&
\colhead{${\rm mix}_{^{56}{\rm Ni}}^{h}$}&
\colhead{$R^{h}$}\\
\colhead{}&
\colhead{($10^{51}$ erg)}&
\colhead{(M$_{\odot}$)}&
\colhead{(M$_{\odot}$)}&
\colhead{($\%/M_{\rm ej}$)}&
\colhead{(R$_{\odot}$)}}
\startdata
2004ek   & 9.3           & 28            &  0.217     & 41  & 2303  \\
2004em   & 11.3          & 43            &  0.102     & 25  & 316   \\
2005ci   & 2.3           & 19            &  0.065     & 90  & 33    \\
PTF09gpn & 1.0 & 12   &  0.040     & 35  & 120   \\
PTF12gcx & 3.6           & 12            &  $<$ 0.181 & 100 & 346   \\
PTF12kso & 3.4           & 17            &  0.230     & 90  & \ldots\\
\enddata
\tablenotetext{a}{From \citet{arnett89}.}
\tablenotetext{t}{From the modeling of the bolometric light curve tail.}
\tablenotetext{h}{From the hydrodynamical models.}
\tablecomments{Errors in the $^{56}$Ni mass are mainly due to the uncertainty in the distance and in the explosion epoch. We estimate the total uncertainty to be $\sim10$\% for each SN. The $^{56}$Ni mass for PTF09gpn is determined by the $r$-band light-curve comparison to SN~2009E at late epochs.}
\end{deluxetable}

\end{document}